\documentclass[final,5p,twocolumn]{elsarticle}

\usepackage{graphicx}
\usepackage{amssymb}
\usepackage{lineno}

\usepackage[T1]{fontenc}
\usepackage[utf8]{inputenc}
\usepackage{url}
\usepackage{amsmath}
\usepackage{algorithm}
\usepackage{algpseudocode}
\usepackage[caption=false,font=footnotesize,margin=0pt]{subfig}
\usepackage{multirow}
\usepackage{bm}
\usepackage{enumitem}

\DeclareMathOperator*{\argmax}{arg\,max}
\DeclareMathOperator*{\argmin}{arg\,min}

\journal{Robotics and Autonomous Systems}

\begin{document}

\begin{frontmatter}


\title{Lightweight and Optimized Sound Source Localization and Tracking Methods for Open and Closed Microphone Array Configurations}



\author{Fran\c{c}ois Grondin}
\author{Fran\c{c}ois Michaud}

\address{Department of Electrical Engineering and Computer Engineering, Interdisciplinary Institute for Technological Innovation (3IT), \\Universit\'{e} de Sherbrooke, Sherbrooke, Qu\'{e}bec, Canada\\
\{francois.grondin2,francois.michaud\}@usherbrooke.ca}

\begin{abstract}
Human-robot interaction in natural settings requires filtering out the different sources of sounds from the environment. 
Such ability usually involves the use of microphone arrays to localize, track and separate sound sources online.
Multi-microphone signal processing techniques can improve robustness to noise but the
processing cost increases with the number of microphones used, limiting response time and widespread use on different types of mobile robots.
%% FG: Reviewer 3: Ajout d'une justification pourquoi la localization est complexe
Since sound source localization methods are the most expensive in terms of computing resources as they involve scanning a large 3D space, minimizing the amount of computations required would facilitate their implementation and use on robots.
%% FG: Reviewer 3: Correction typo
The robot's shape also brings constraints on the microphone array geometry and configurations.
In addition, sound source localization methods usually return noisy features that need to be smoothed and filtered by tracking the sound sources.
This paper presents a novel sound source localization method, called SRP-PHAT-HSDA, that scans space with coarse and fine resolution grids to reduce the number of memory lookups.
A microphone directivity model is used to reduce the number of directions to scan and ignore non significant pairs of microphones. 
A configuration method is also introduced to automatically set parameters that are normally empirically tuned according to the shape of the microphone array.
For sound source tracking, this paper presents a modified 3D Kalman (M3K) method capable of simultaneously tracking in 3D the directions of sound sources.
Using a 16-microphone array and low cost hardware, results show that SRP-PHAT-HSDA and M3K perform 
at least as well as other sound source localization and tracking methods while using up to 4 and 30 times less computing resources respectively.
\end{abstract}


\end{frontmatter}

\section{Introduction}
\label{sec:intro}

Distant Speech Recognition (DSR) occurs when speech is acquired with one or many microphone(s) moved away from the mouth of the speaker, making recognition difficult because of background noise, overlapping speech from other speakers, and reverberation \cite{woelfel2009distant,kumatari2012microphone}. 
DSR is necessary for enabling verbal interactions without the necessity of using intrusive body- or head-mounted devices.
But still, recognizing distant speech robustly remains a challenge \cite{vacher2015distant}.
Microphone arrays make it possible to capture sounds for DSR \cite{kumatari2012microphone} in human-robot interaction (HRI).
This requires the installation of multiple microphones on the robot platform, and process distant speech perceived by filtering out noise from fans and actuators on the robot and non-stationary background sound sources in reverberant environments, fast enough to support live interactions.
This process usually relies first on localizing and tracking the perceived sound sources, to then be able to separate them \cite{grondin2013manyears} for specific processing such as speech recognition \cite{brodeur2016integration,frechette2012integration}.
%% FG: Reviewer 1: Ajout d'une spécification pour justifier pourquoi la localisation est importante
Using sound source localization and tracking methods robust to noise and low computational cost is important \cite{hoshiba2017uav}, as it is usually the first step to engage speech based human-robot interaction (HRI).
A natural speech HRI requires the robot to be able to detect speech commands in noisy environments, while avoiding false detections.

%% FM Ramena la figure 1 au début afin de faire une introduction aux deux techniques avant d'aller dans le plus spécifique. Déplaça le paragraphe avant la partie SST ici dans le même objectif
%% FM: Enlever les caractères en gras dans le bloc de SST
%% FG: Le terme SST n'est plus en gras
%
  \begin{figure}[!ht]
  \centering
  \includegraphics[width=\columnwidth]{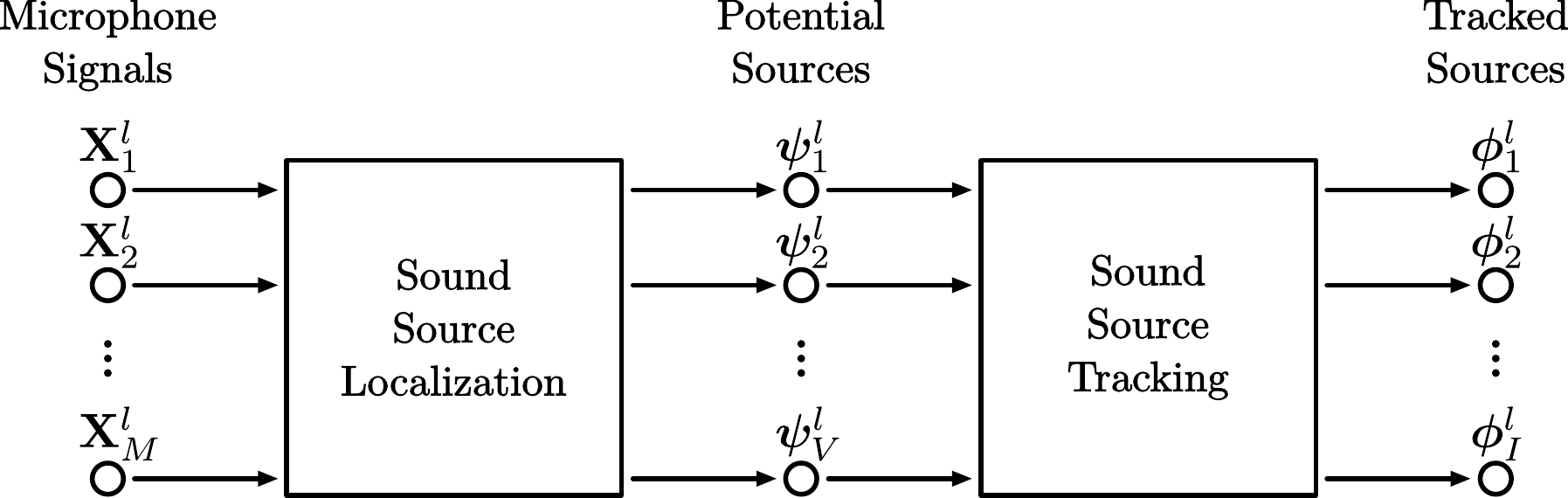}
  \caption{Block diagram of sound source localization and tracking}
  \label{fig:sst_cascade}
  \end{figure}

As illustrated by Fig. \ref{fig:sst_cascade}, sound source localization (SSL) and sound source tracking (SST) are done in sequence.
%% FM: Modifia la phrase pour introduire DoA et intégrer avec SST
%% FG: C'est parfait
For each frame $l$, SSL uses the captured signals from the $M$-microphone array $\mathbf{X}^l = \{\mathbf{X}^l_1, \mathbf{X}^l_2, \dots, \mathbf{X}^l_M\}$ and generates $V$ potential sources $\bm{\Psi}^l = \{\bm{\psi}^l_1, \bm{\psi}^l_2, \dots, \bm{\psi}^l_V \}$, where each potential source $\bm{\psi}^l_v$ consists of a direction of arrival (DoA) $\bm{\lambda}^l_v$
in Cartesian coordinates, and the steered beamformer energy level $\Lambda^l_v$.
SSL methods provide noisy observations of the DoAs of sound sources, caused by the sporadic activities of sound sources (e.g., the sparsity of speech), combined with the presence of multiple competing sound sources. 
SST then uses the potential sources and returns $I$ tracked sources $\bm{\Phi}^l = \{\phi^l_1,\phi^l_2,\dots,\phi^l_I\}$ to filter out this noise and provide a smooth trajectory of the sound sources.
%% FM: Enleva cette partie car il est question de ceci dans le paragraphe précédent
%, which can be used to derive distinct audio streams for each source (a process known as sound source separation (SSS) \cite{grondin2013manyears}).
%% FM: Changea pour "the most expensive", mais devrait-on commenter aussi pour SST? Ou devrait-on généraliser pour SSL et SST (car j'imagine que SST peut être expensive aussi en utilisant l'approche des filtres particulaires)? L'objectif serait de rester général en terme de SSL et SST, et aborder les points spécifiques en SSL et SST dans les deux autres paragraphes. J'ai rebrassé le tout.
%% FG: Je crois qu'on peut conserver le most expensive pour SSL, vu que c'est tout de même la méthode la plus exigeante en terme de puissance, et qu'elle continue d'augmenter la charge de calculs avec le nombre de microphones, contrairement à SST.
Improved capabilities for SSL and SST can be directly associated with the number of microphones used, which influences processing requirements \cite{valin2007robust}. 
%% FM: Est-ce que la phrase suivante est nécessaire?
%% FG: Vu qu'on le dit maintenant dans la phrase précédente, on peut la retirer
%Reducing the amount of computations to perform robust SSL can therefore be a benefit, either to increase the number of microphones in the array, to free on-board computing resources for other processes (e.g., speech processing, vision, planning, navigation), or to facilitate its implementation on various types of embedded computing platforms. 

%% FM: Changea la transition et le "For instance" de la phrase suivante
In this process, SSL is the most expensive in terms of computation, and a variety of SSL algorithms exists.
Rascon et al. \cite{rascon2015lightweight} present a lightweight SSL method that uses little memory and CPU resources, but is limited to three microphones and scans the DoA of sound source only in 2D.
Nesta and Omologo \cite{nesta2012generalized} describe a generalized state coherence transform to perform SSL, which is particularly effective when multiple sound sources are present.
However, this method relies on independent component analysis (ICA), which takes many seconds to converge.
Drude et al. \cite{drude2015doa} use a kernel function that relies on both phase and level differences, at the cost of increasing the computational load.
Loesch and Yang \cite{loesch2010blind} also introduce a localization method based on time-frequency sparseness, which remains sensitive to high reverberation levels.
Multiple Signal Classification based on Standard Eigenvalue Decomposition (SEVD-MUSIC) makes SSL robust to additive noise \cite{nakadai2010design}.
SEVD-MUSIC, initially used for narrowband signal \cite{schmidt1986multiple}, has been adapted for broadband sound sources such as speech \cite{ishi2009evaluation}, and is robust to noise as long as the latter is less powerful than the signals to be localized.
Multiple Signal Classification based on Generalized Eigenvalue Decomposition (GEVD-MUSIC) method \cite{nakamura2011intelligent} has been introduced to cope with this issue, but the latter method increases the computations.
Multiple Signal Classification based on Generalized Singular Value Decomposition (GSVD-MUSIC) reduces computational load of GEVD-MUSIC and improves localization accuracy \cite{nakamura2012real}, but still relies on eigen decomposition of a matrix. 
Other methods take advantage of specific array geometries (linear, circular or spherical) to improve robustness and reduce computational load \cite{danes2010information,pavlidi2012real,rafaely2010spherical}.
Even though interesting properties arise from these geometries, these configurations are less practical for a mobile robot due to physical constraints introduced by its specific shape.
SSL can also be performed using a Steered Response Power with Phase Transform (SRP-PHAT).
The SRP-PHAT is usually computed using weighted Generalized Cross-Correlation with Phase Transform (GCC-PHAT) at each pair of microphones \cite{grondin2013manyears,valin2007robust}.
SRP-PHAT requires less computations than MUSIC-based methods, but still requires a significant amount of computations when scanning the 3D-space for a large number of microphones.
Stochastic region contraction \cite{do2007real}, hierarchical search \cite{zotkin2004accelerated, do2009stochastic,nunes2014steered} and vectorization \cite{lee2010vectorized} have also been studied to speed up scanning with SRP-PHAT, but usually limit the search to a 2D surface and a single source.
Marti et al. \cite{marti2013steered} also propose a recursive search over coarse and fine grids.
This method divides space in rectangular volumes, and maps points from the fine grid to only one point on the coarse grid.
It however neglects microphone directivity, and also uses an averaging window over the GCC values, which may reduce the contribution of a peak during the coarse scan when neighboring values are negative. 

SST methods can be categorized into four types: 
\begin{itemize}

\item Viterbi search. Anguera et al. \cite{anguera2007acoustic} propose a post-processing Viterbi method to track a sound source over time.
This method introduces a significant latency when used online, making it appropriate only for offline processing.
Tracking is also performed on discrete states, which restrains the direction of the tracked source to a fixed grid. 

\item Sequential Monte Carlo (SMC) filtering. The SMC method, also called particle filtering, performs low latency tracking for a single sound source \cite{williamson2002particle,ward2003particle,vermaak2001nonlinear}.
Valin et al. \cite{grondin2013manyears,valin2007robust} adapt the SMC method to track multiple sound sources.
This method consists in sampling the space with finite particles to model the non-Gaussian state distribution. 
SMC allows tracking with continuous trajectories, but requires a significant amount of computations, and is undeterministic because it uses randomly generated particles.

\item Kalman filtering. Rascon et al. \cite{rascon2015lightweight} propose a lightweight method that relies on Kalman filters, which allows tracking with continuous trajectories and  reduce considerably the amount of computations.
This method is however limited to DoAs in spherical coordinates, using elevation and azimuth, which generates distortion as the azimuth resolution changes with elevation.
It also introduces azimuth wrapping.
Markovi\'{c} et al. \cite{markovic2016wrapping} present an extended Kalman filter on Lie groups (LG-EKF) to perform directional tracking with an 8-microphone array.
LG-EKF solves the azimuth wrapping phenomenon, but limits the tracking to a 2D circle, and is therefore unsuitable for tracking sources on a 3D spherical surface.

%% FG: J'ai ajouté cet item pour incorporé les travaux de Markovic et al. avec les distributinos von Mise, que je ne connaissais pas mais qui a été mentionné par le Reviewer #3
\item Joint probabilistic data association filter (JPDA). Markovi\'{c} et al.  \cite{markovic2016multitarget} introduces this tracking method for 3D spherical surface that relies on Bayesian von Mises-Fisher estimator.
This approach however requires prior knowledge of the number of active sources, and neglects the motion model for each tracked source, which leads to switched or merged trajectories when two sources cross each other.

\end{itemize}

To improve SSL and SST, this paper introduces a SRP-PHAT method referred to as SRP-PHAT-HSDA, for Hierarchical Search with Directivity model and Automatic calibration, and a tracking method based on a modified 3D Kalman filter (M3K) using Cartesian coordinates. 
SRP-PHAT-HSDA scans the 3D space over a coarse resolution grid, and then refines search over a specific area.
It includes a Time Difference of Arrival (TDOA) uncertainty model to optimize the scan accuracy using various grid resolution levels using open and closed microphone array configurations.
A microphone directivity model is also used to reduce the number of directions to scan and ignore non significant pairs of microphones.
M3K replaces the SMC filters used in Valin et al. \cite{valin2007robust} by Kalman filters, and introduces three new concepts: 1) normalization of the states to restrict the space to a unit sphere; 2) derivation of a closed-form expression for the likelihood of a coherent source to speed up computations; and 3) weighted update of the Kalman mean vector and covariance matrix for simultaneous tracking of sound sources.
These modifications provide efficient tracking of multiple sound sources, makes the method convenient for low-cost embedded hardware as it requires less computations than the SMC method, and solves the distortion and wrapping introduced by Kalman filtering with spherical coordinates. 
%% FG: J'ai enlevé la dernière phrase qui parlait du fait que le filtre était déterministic, pour éviter la confusion, suite aux réserves émises par le Reviewer 3.

The paper is organized as follows.
%% FM: modifia légèrement la transition
%% FG: OK
First, Section \ref{sec:complexity} characterizes the computing requirements of SRP-PHAT in comparison to SEVD-MUSIC, to justify and situate the improvements brought by SRP-PHAT-HSDA.
Sections \ref{sec:ssl} and \ref{sec:sst} then describe SRP-PHAT-HSDA and M3K, respectively.
Section \ref{sec:setup} presents the experimental setup involving 8 and 16-microphone circular and closed cubic arrays on a mobile robot, implementing SSL and SST methods on a Raspberry Pi 3.
%% FM: Sépara la phrase et ajouta des précisions
%% FG: OK
Section \ref{sec:results} presents the results obtained from experiments comparing SRP-PHAT with SRP-PHAT-HSDA, and M3K with SMC.
Finally, Section \ref{sec:conclusion} concludes this paper with final remarks and future work.

\section[Computing Requirements of SRP-PHAT versus SEVD-MUSIC]{Computing Requirements of SRP-PHAT versus \\SEVD-MUSIC}
\label{sec:complexity}

SSL is usually divided in two tasks: 1) estimation of TDOA, and 2) DoA search over the 3D space around the microphone array.
The main difference between SRP-PHAT and SEVD-MUSIC lies in Task 1: SRP-PHAT relies on the Generalized Cross-Correlation with Phase Transform method (GCC-PHAT), while SEVD-MUSIC uses Singular Eigenvalue Decomposition (SEVD).
The intend here is to demonstrate which method is the most efficient for Task 1, and then, using this method, how can Task 2 be further improved to reduce computing needs.

Both methods first capture synchronously the acoustic signals $x_m$ from the $M$ microphones in the array.
These signals are divided in frames of $N$ samples, spaced by $\Delta N$ samples and multiplied by a the sine window $w[n]$:
  \begin{equation}
    x^l_m[n] = w[n] x_m[n+l\Delta N]
    \label{eq:complexity_x}
  \end{equation}
with $l$, $i$ and $n$ representing the frame, microphone and sample indexes, respectively.
The methods then compute the Short-Time Fourier Transform (STFT) with a $N$-samples real Fast Fourier Transform (FFT), where the expression $X^l_m[k]$ stands for the spectrum at each frequency bin $k$, and the constant $j$ is the complex number $\sqrt{-1}$:
  \begin{equation}
    X_m^l[k] = \sum_{n=0}^{N-1}{x_m^l[n]\exp{\left(-j2\pi kn/N\right)}}
    \label{eq:complexity_X}
  \end{equation}

%% FG: Je donne ici une précision concernant la "sine window", pour répondre à une question soulevée par le Reviewer 3
The sine window allows reconstruction in the time-domain with a 50\% frame overlap, and thus the same STFT results can be used for both the localization and separation steps, which reduces the total amount of computations.
SRP-PHAT relies on the Generalized Cross-Correlation with Phase Transform (GCC-PHAT), which is computed for each pair of microphones $p$ and $q$ (where $p \neq q$).
The Inverse Fast Fourier Transform (IFFT) provides an efficient computation of the GCC-PHAT, given that the time delay $n$ is an integer:
  \begin{equation}
    r^l_{pq}[n] = \frac{1}{N}\sum_{k=0}^{N-1}{\frac{X_{p}^{l}[k]X_{q}^{l}[k]^*}{|X_{p}^{l}[k]||X_{q}^{l}[k]|+\epsilon}\exp{\left(j2\pi kn/N\right)}}
    \label{eq:complexity_r}
  \end{equation}

The IFFT complexity depends on the number of samples per frame $N$, which is usually a power of 2.
The order of complexity for a real IFFT is $\mathcal{O}(N \log N)$).
With $M(M-1)/2$ pairs of microphones, SRP-PHAT computing complexity reaches $\mathcal{O}(M^2 N \log N)$.

SEVD-MUSIC relies on singular eigenvalue decomposition of the cross-correlation matrix.
The $M \times M$ correlation matrix $\mathbf{R}[k]$ is defined as follows, where $\textrm{E}\{\dots\}$ and $\{\dots\}^H$ stand for the expectation and Hermitian operators, respectively:
  \begin{equation}
    \mathbf{R}[k] = \textrm{E}\{\mathbf{X}[k]\mathbf{X}[k]^H\}
    \label{eq:complexity_Rv}
  \end{equation}

The $M \times 1$ vector $\mathbf{X}^l[k]$ concatenates the spectra of all microphones for each frame $l$ and frequency bin $k$ (where the operator $\{\dots\}^T$ stands for the transpose): 
  \begin{equation}
    \mathbf{X}^l[k] = \left[ 
    \begin{array}{cccc}
    X^l_1[k] & X^l_2[k] & \dots & X^l_M[k]
    \end{array}
    \right]^T
    \label{ssl:eq:complexity_Xv}
  \end{equation}

In practice, the correlation matrix is usually computed at each frame $l$ with an estimator that sums vectors over time (a window of $L$ frames) for each frequency bin $k$:
  \begin{equation}
    \mathbf{R}^{l}[k] = \frac{1}{L}\sum_{\Delta L=0}^{L-1}{\mathbf{X}^{l+\Delta L}[k] \mathbf{X}^{l+\Delta L}[k]^H}
    \label{eq:complexity_Rvest}
  \end{equation}

SEVD-MUSIC complexity depends on the size of the matrix $\mathbf{R}^l[k]$, and is $\mathcal{O}(M^3)$ \cite{holmes2007fast}.
This operation is performed at each frequency bin $k$, for a total of $N/2$ bins, which leads to an overall complexity of $\mathcal{O}(M^3 N)$.

To better express the computing requirements of both methods, Table \ref{tab:complexity_srpphatsvdmusic} presents simulation results of the time (in sec) required to process one frame $l$, with various values of $N$ and $M$, on a Raspberry Pi 3.
SRP-PHAT compute $M(M-1)/2$ real $N$-sample FFTs using the FFTW C library \cite{frigo1998fftw}, and SVD-MUSIC evaluates $N/2$ SEVD of $M \times M$ matrices using the Eigen C++ library \cite{guennebaud2014eigen}.
Some methods (e.g., \cite{nakadai2010design, nakamura2011intelligent, nakamura2012real}) compute SEVD only in the lower frequency range (where speech is usually observed) to reduce the computational load.
However, this discards some useful spectral information in the higher frequencies (in speech fricative sounds for instance), which are considered with the SRP-PHAT method.
To ensure a fair comparison, both methods treat the whole spectral range.
For $N = 256$, when the number of microphones increase from $M=8$ to $M=32$, the processing time increases by a factor of $17.8 \approx (M_1/M_2)^2 = (32/8)^2 = 16$ for SRP-PHAT, and a factor of $36.6 < (M_1/M_2)^3 = (32/8)^3 = 64$ for SEVD-MUSIC.
The latter factor is less than the expected complexity of $M^3$, which is probably explained by truncated computation of SVD for small singular values.
Similarly, for a fixed number of microphones $M=8$, the complexity increases by a factor of $8.9 \approx (N_1/N_2) \log (N_1/N_2) = (2048/256) \log (2048/256) = 11$ for SRP-PHAT, and by a factor of $7.3 \approx (N_1/N_2) = (2048/256) = 8$ for SEVD-MUSIC.
SRP-PHAT requires from 206 ($M = 8$ and $N = 2048$) to 525 ($M = 32$ and $N = 512$) less computing time that the SEVD-MUSIC method.
This suggests that SRP-PHAT is more suitable for online processing, as it performs Task 1 effectively.
It is therefore desirable to use SRP-PHAT for Task 1, and optimize Task 2 to get an efficient SSL 
%% FM: Changea system pour method
%% FG: OK
method.
  \begin{table*}[!ht]
    \centering
    \caption{Processing time in sec/frame for \{SRP-PHAT, SEVD-MUSIC\} and ratio between both (between parentheses)}
    \renewcommand{\arraystretch}{1.2}
    \begin{tabular}{|c||c|c|c|c|}
    \hline
    M & $N = 256$ & $N = 512$ & $N = 1024$ & $N = 2048$ \\
    \hline
    \hline
    $8$ & $\{1.8\textrm{E}{-4},4.1\textrm{E}{-2}\}$ (228) & $\{3.3\textrm{E}{-4}, 8.1\textrm{E}{-2}\}$ (245) & $\{7.1\textrm{E}{-4}, 1.6\textrm{E}{-1}\}$ (225) & $\{1.6\textrm{E}{-3}, 3.3\textrm{E}{-1}\}$ (206) \\
    $16$ & $\{7.6\textrm{E}{-4}, 2.3\textrm{E}{-1}\}$ (303) & $\{1.4\textrm{E}{-3}, 4.5\textrm{E}{-1}\}$ (321) & $\{3.1\textrm{E}{-3}, 9.0\textrm{E}{-1}\}$ (290) & $\{6.9\textrm{E}{-3}, 1.8\textrm{E}{+0}\}$ (260) \\
    $24$ & $\{1.8\textrm{E}{-3}, 7.0\textrm{E}{-1}\}$ (389) & $\{3.3\textrm{E}{-3}, 1.4\textrm{E}{+0}\}$ (424) & $\{7.0\textrm{E}{-3}, 2.8\textrm{E}{+0}\}$ (400) & $\{1.6\textrm{E}{-2}, 5.6\textrm{E}{+0}\}$ (350) \\
    $32$ & $\{3.2\textrm{E}{-3}, 1.5\textrm{E}{+0}\}$ (469) & $\{5.9\textrm{E}{-3}, 3.1\textrm{E}{+0}\}$ (525) & $\{1.3\textrm{E}{-2}, 6.2\textrm{E}{+0}\}$ (477) & $\{2.9\textrm{E}{-2}, 1.2\textrm{E}{+1}\}$ (524) \\
    \hline
    \end{tabular}
    \label{tab:complexity_srpphatsvdmusic}
  \end{table*}

\section{SRP-PHAT-HSDA Method}
\label{sec:ssl}

\begin{figure}[!ht]
\centering
\includegraphics[width=\columnwidth]{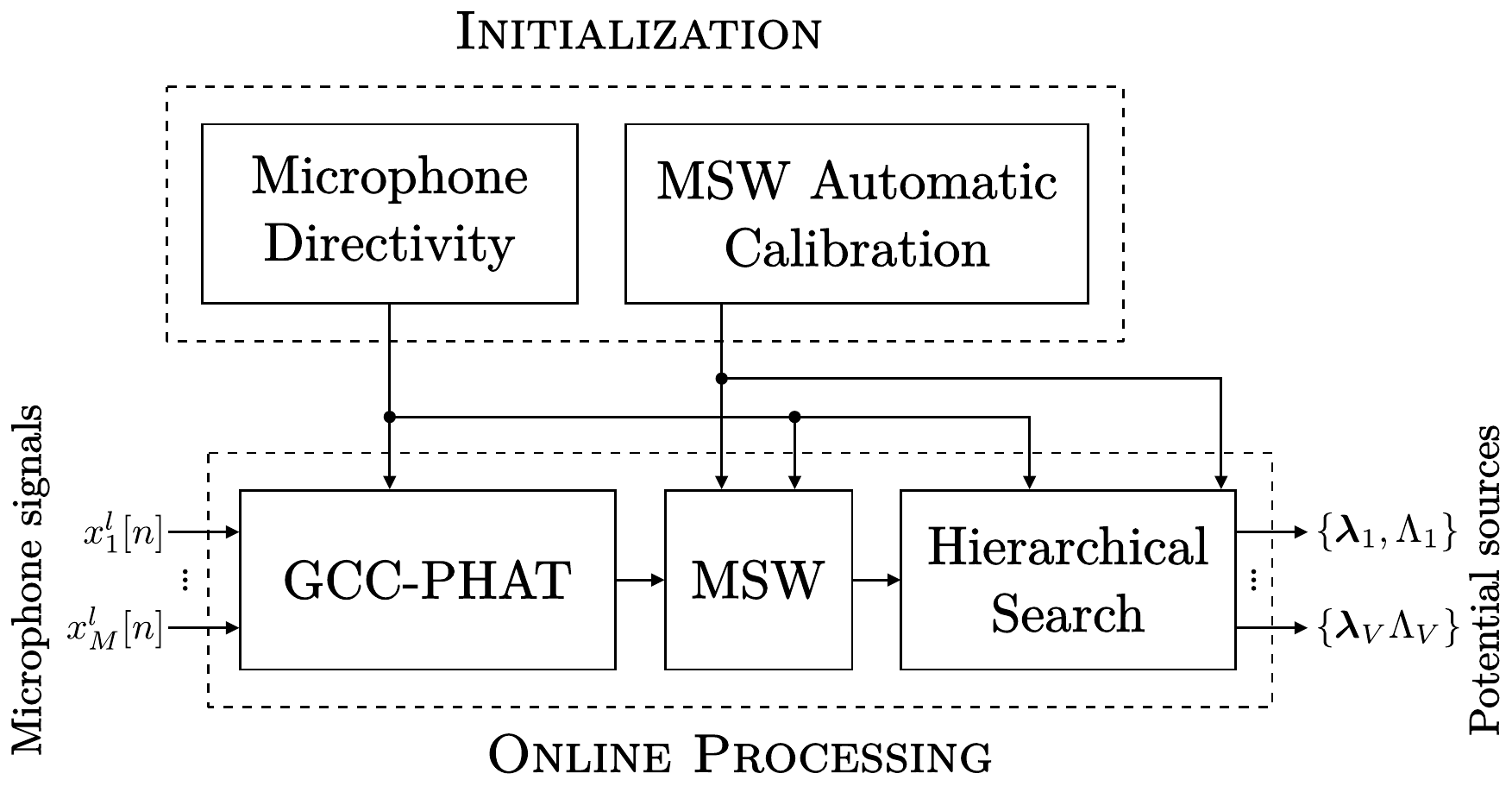}
\caption{Block diagram of SRP-PHAT-HSDA}
\label{fig:ssl_overview}
\end{figure}

To understand how SRP-PHAT-HSDA works, let us start by explaining SRP-PHAT, to then explain the added particularities of SRP-PHAT-HSDA.
Figure \ref{fig:ssl_overview} 
%% FM shows the overview of %% FG: OK
illustrates the SRP-PHAT-HSDA method, using $M$ microphone signals to localize $V$ potential sources.
The Microphone Directivity module and the MSW Automatic Calibration module are used at initialization, and provide parameters to perform optimized GCC-PHAT, Maximum Sliding Window (MSW) filtering and Hierarchical Search online.

%% FG: J'ai ajouté une référence aux travaux de dibiase et al.
The underlying mechanism of SRP-PHAT is to search for $V$ potential sources for each frame $l$ over a discrete space \cite{grondin2013manyears, dibiase2001robust}.
For each potential source, the computed GCC-PHAT frames are filtered using a Maximum Sliding Windows (MSW).
The sum of the filtered GCC-PHAT frames for all pairs of microphones provide the acoustic energy for each direction on the discrete space, and the direction with the maximum energy corresponds to a potential source.
Once a potential source is obtained, its contribution is removed from the GCC-PHAT frames by setting the amplitude of the corresponding TDOA to zero, and the space is scanned again.
This process is repeated $V$ times until the DoAs ($\bm{\lambda}_v, v = 1, \dots, V$) and energy levels ($\Lambda_v, v = 1, \dots, V$) of all potential sources are generated.

\begin{figure}[!ht]
\centering
\subfloat[$\mathcal{L} = 0$]{\includegraphics[width=0.27\columnwidth]{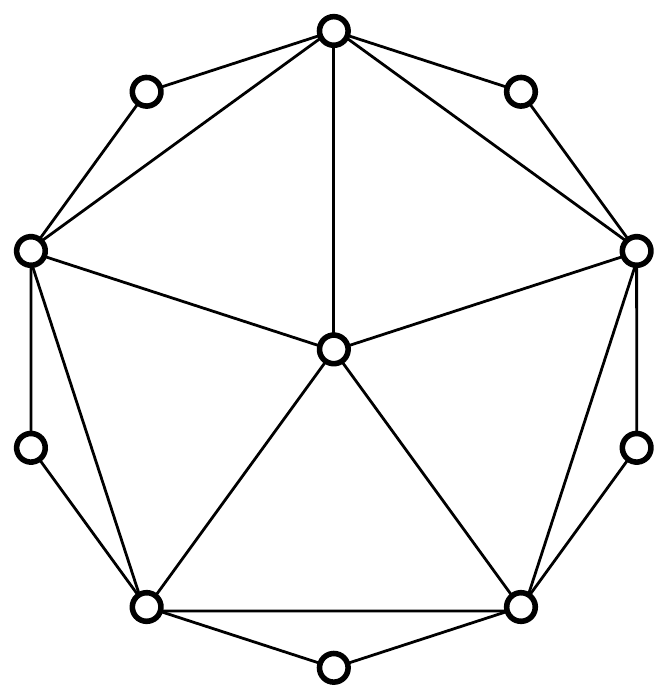}}
\hfil
\subfloat[$\mathcal{L} = 1$]{\includegraphics[width=0.27\columnwidth]{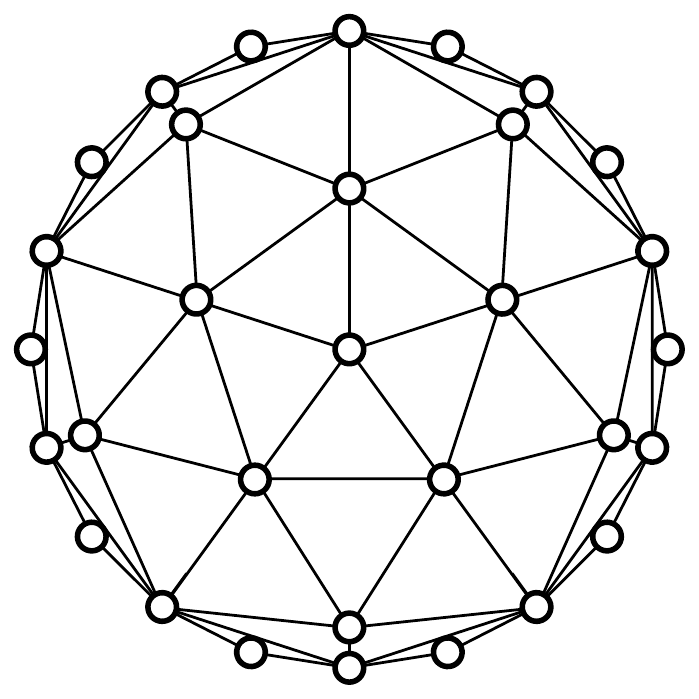}}
\hfil
\subfloat[$\mathcal{L} = 2$]{\includegraphics[width=0.27\columnwidth]{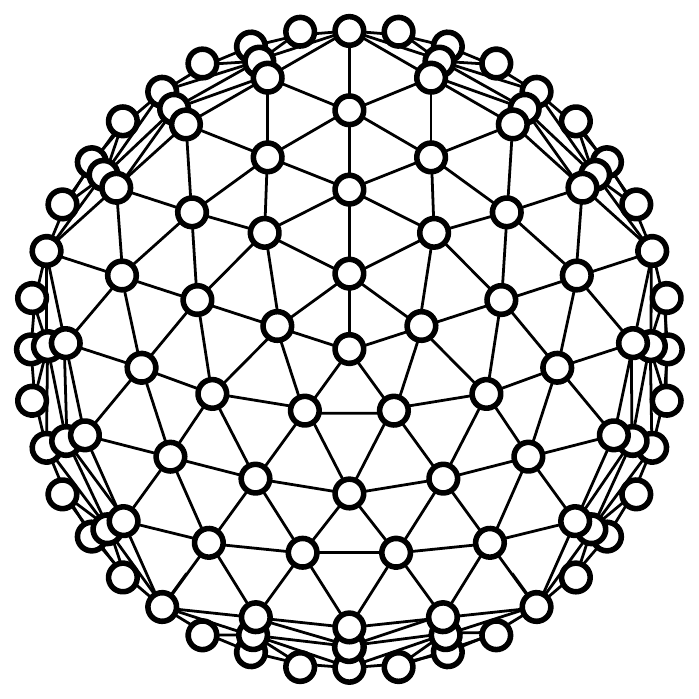}}
\caption{Discrete unit spheres}
\label{fig:ssl_spheres}
\end{figure}

A discrete unit sphere provides potential DoAs for sound sources.
As in \cite{grondin2013manyears} and \cite{valin2007robust}, a regular convex icosahedron made of 12 points defines the initial discrete space, and is refined recursively $\mathcal{L}$ times until the desired space resolution is obtained.
Figure \ref{fig:ssl_spheres} shows the regular icosahedron ($\mathcal{L} = 0$), and subsequent refining iterations levels ($\mathcal{L} = 1$ and $\mathcal{L} = 2$).

Each point on the discrete sphere corresponds to a unit vector $\mathbf{u}_k$, where $k$ stands for the point index where $k = 1, 2, \dots, K$, and $S = \{\mathbf{u}_1, \mathbf{u}_2, \dots, \mathbf{u}_K\}$ is the set that contains all vectors, where the number of points $K = 10 \times 4^{\mathcal{L}} + 2$ depends on the resolution level $\mathcal{L}$.
In the SRP-PHAT method proposed in \cite{valin2007robust}, the scan space is refined four times ($\mathcal{L} = 4$) to generate $2562$ points and obtain a spatial resolution of $3$ degrees.

To further reduce SRP-PHAT computations, and maintain a high localization accuracy regardless of the microphone array shape, SRP-PHAT-HSDA adds the following elements:

\begin{itemize}

\item \emph{Microphone Directivity (MD)}: When the number of microphone $M$ increases, the computational load also increases by a complexity of $\mathcal{O}(M^2)$. The proposed method assumes that microphone have a directivity pattern, and this introduces constraints that reduces the space to be scanned and the number of pairs of microphones to use, which in turn decreases the amount of computations.

\item \emph{Maximum Sliding Window Automatic Calibration (MSWAC)}: TDOA estimation is influenced by the uncertainty in the speed of sound and the microphones positions (which may be difficult to measure precisely with microphone arrays of complex geometry), and scan grid discretization, which should be modelled somehow.
The MSW size can be tuned manually by hand to maximize localization accuracy, but this remains a time consuming task which has to be repeated for each new microphone array geometry.
The TDOA uncertainty model solves this challenge as it automatically tunes the MSW size to maximize localization accuracy. 

\item \emph{Hierarchical Search (HS)}: Searching for potential sources involves scanning the 3D space according to a grid with a specific resolution. 
Finer resolution means better precision but higher computation. 
To reduce computations, a solution is to first do a scan with a grid at coarse resolution to identify a potential sound source, and then do another scan with a grid with a fine resolution using the location found during the first scan to pinpoint a more accurate direction.

\end{itemize}

\subsection{Microphone Directivity}
\label{subsec:ssl_micdir}

\begin{figure}[!ht]
    \centering
    \includegraphics[width=\columnwidth]{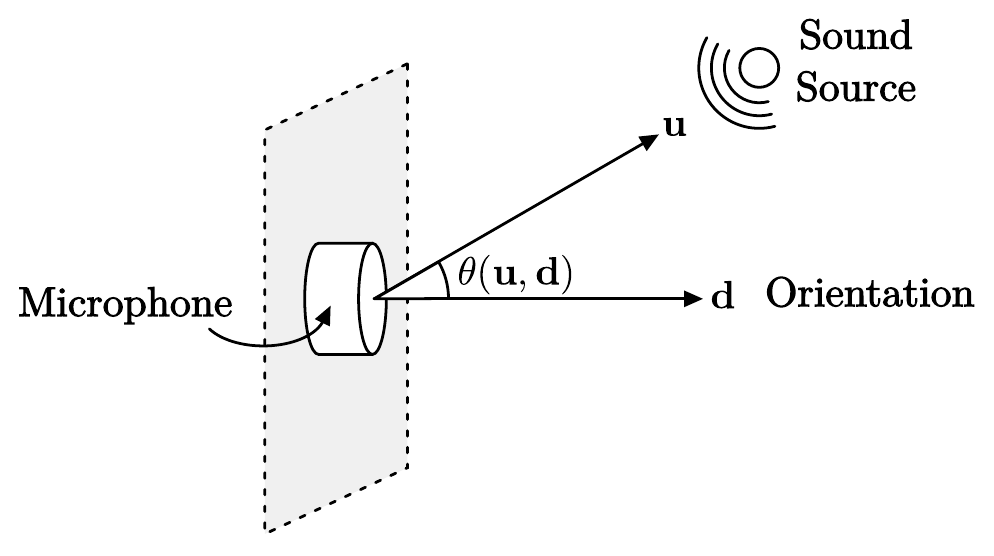}
    \caption{Microphone directivity angle $\theta$ as a function of microphone orientation and source direction}
    \label{fig:ssl_micdir_theta}
  \end{figure}

In a microphone array, microphones are usually assumed to be omnidirectional, i.e., acquiring signals with equal gain from all directions.
In practice however, microphones on a robot platform are often mounted on a rigid body, which may block the direct propagation path between a sound source and a microphone.
The attenuation is mostly due to diffraction, and changes as a function of frequency.
Since the exact diffraction model is not available, the proposed model relies on simpler assumptions: 1) there is a unit gain for sound sources with a direct propagation path, and 2) the gain is null when the path is blocked by the robot body.
As the signal to noise ratio is generally unknown for the blocked microphones, it is safer to assume a low SNR, and setting the gain to zero prevents noise to be injected in the observations.
Moreover, the gain is set constant for all frequencies, and a smooth transition band connects the unit and null gain regions.
This transition band prevents abrupt changes in gains when the sound source position varies.
Figure \ref{fig:ssl_micdir_theta} introduces $\theta(\mathbf{u},\mathbf{d})$, as defined by (\ref{eq:ssl_micdir_theta}), the angle between a sound source located at $\mathbf{u}$, and the orientation of the microphone modeled by the unit vector $\mathbf{d}$.
  \begin{equation}
    \theta(\mathbf{u},\mathbf{d}) = \arccos\left[\frac{\mathbf{u}\cdot \mathbf{d}}{|\mathbf{u}||\mathbf{d}|}\right]
    \label{eq:ssl_micdir_theta}
  \end{equation}

\begin{figure}[!ht]
    \centering
    \includegraphics[width=\columnwidth]{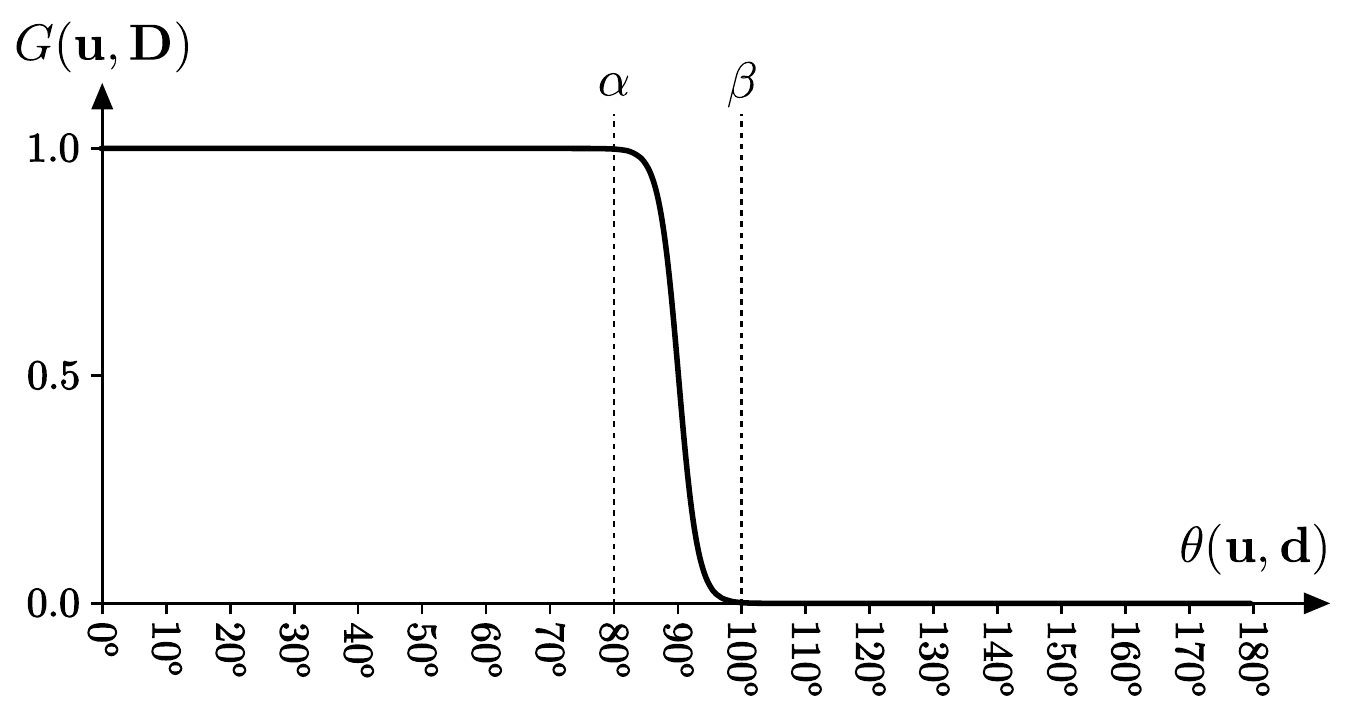}
	\caption{Microphone gain response}    
    \label{fig:ssl_micdir_logistic}
  \end{figure}

Figure \ref{fig:ssl_micdir_logistic} illustrates the logistic function that models the gain $G(\mathbf{u},\mathbf{D})$ as a function of the angle $\theta(\mathbf{u},\mathbf{d})$, given in (\ref{eq:ssl_micdir_G}).
The expression $\mathbf{D}$ is a set that contains the parameters $\{\mathbf{d},\alpha,\beta\}$, where $\alpha$ stands for the angle where the gain is one while $\beta$ corresponds to the angle at which the gain is null.
The region between both angles can be viewed as a transition band.
  \begin{equation}
    G(\mathbf{u},\mathbf{D}) = \frac{1}{1+\exp\left(\left(\frac{20}{\beta-\alpha}\right)\left(\theta(\mathbf{u},\mathbf{d})-\frac{\alpha+\beta}{2}\right)\right)}
    \label{eq:ssl_micdir_G}
  \end{equation}

To make SSL more robust to reverberation, the scan space is restricted to a specific direction.
For instance, the scan space is limited to the hemisphere that points to the ceiling to ignore reflections from the floor.
The unit vector $\mathbf{d}_0$ stands for the orientation of the scan space.

Since microphone directivity introduces some constraints on the scan space, the spatial gains $G(\mathbf{u}_k,\mathbf{D}_p)$ and $G(\mathbf{u}_k,\mathbf{D}_q)$ need to be large enough for a source located in the direction $\mathbf{u}_k$ to excite both microphones $p$ and $q$.
The gain $G(\mathbf{u}_k,\mathbf{D}_0)$ also needs to be large enough for this direction to be part of the scan space.
The mask $\zeta_{pq}(\mathbf{u}_k)$ models this condition, where the constant $G_{min}$ stands for the minimal gain value:
  \begin{equation}
    \zeta_{pq}(\mathbf{u}_k) = \begin{cases}
    1 & G(\mathbf{u}_k,\mathbf{D}_0)G(\mathbf{u}_k,\mathbf{D}_p)G(\mathbf{u}_k,\mathbf{D}_q) \geq G_{min} \\
    0 & \textrm{otherwise} \\
    \end{cases}
    \label{eq:ssl_micdir_zetapqu}
  \end{equation}
  
When the mask $\zeta_{pq}(\mathbf{u}_k)$ is zero, the value of the corresponding sample in the  GCC-PHAT frame is negligible and can be ignored.
When all pairs of microphones are uncorrelated ($\zeta_{pq}(\mathbf{u}_k) = 0$ for all values of $p$ and $q$), the direction $\mathbf{u}_k$ can simply be ignored ($\zeta(\mathbf{u}_k) = 0$):
  \begin{equation}
    \zeta(\mathbf{u}_k) = \begin{cases}
    1 & \sum_{p=1}^{M}{\sum_{q=p+1}^{M}{\zeta_{pq}(\mathbf{u}_k)}} > 0 \\
    0 & \textrm{otherwise} \\
    \end{cases}
    \label{eq:ssl_masking_zetau}
  \end{equation}

Similarly, the GCC-PHAT between microphones $p$ and $q$ needs to be computed only when $\zeta_{pq} = 1$, that is when these microphones are excited simultaneously at least once for a given direction $\mathbf{u}_k$:
  \begin{equation}
    \zeta_{pq} = \begin{cases}
    1 & \sum_{k=1}^{K}{\zeta_{pq}(\mathbf{u}_k)} > 0 \\
    0 & \textrm{otherwise} \\
    \end{cases}
    \label{eq:ssl_masking_zetapq}
  \end{equation}

\subsection{MSW Automatic Calibration}
\label{ssl:subsec:method_tdoaunc}

The TDOA between two microphones $\mathbf{m}_p$ and $\mathbf{m}_q$ is given by the expression $\tau_{pq}(\mathbf{u})$.
Under the far field assumption, the TDOA is set according to (\ref{eq:ssl_tdoaunc_tau1}), where $\mathbf{u}$ represents the normalized direction of the sound source, $f_S$ stands for the sample rate (in samples/sec) and $c$ for the speed of sound (in m/s).
Note that the TDOA is usually given in sec, but is provided in samples here since all processing is performed on discrete-time signals.
  \begin{equation}
    \tau_{pq}(\mathbf{u}) = \frac{f_S}{c}\left(\mathbf{m}_p-\mathbf{m}_q\right)\cdot \mathbf{u}
    \label{eq:ssl_tdoaunc_tau1}
  \end{equation}

The speed of sound varies according to air temperature, humidity and pressure.
These parameters usually lie within a known range in a room (and even outside), but it remains difficult to calculate the exact speed of sound.
In SRP-PHAT-HSDA, the speed of sound is modeled using a random variable $c \sim \mathcal{N}(\mu_c,\sigma_c)$, where $\mu_c$ is the mean and $\sigma_c$ the standard deviation of the normal distribution.
The exact position of each microphone is also modeled by a trivariate normal distribution $\mathbf{m}_p \sim \mathcal{N}(\bm{\mu}_p,\bm{\Sigma}_p)$, where $\bm{\mu}_p$ stands for the $1 \times 3$ mean vector and $\bm{\Sigma}_p$ for the $3 \times 3$ covariance matrix.

The first step consists in solving for the expression $a = f_S / c$ (in samples/m).
To make calculations easier, a normally distributed random variable $\eta \sim \mathcal{N}(0,1)$ is introduced:
  \begin{equation}
	a = \frac{f_S}{\mu_c + \sigma_c \eta}
  \end{equation}

The previous equation can be linearized given that $\mu_c \gg \sigma_c$.
Expanding this function as a Taylor series, the following approximation holds:
  \begin{equation}
    a \approx \frac{f_S}{\mu_c}\left(1 - \frac{\sigma_c}{\mu_c}\eta \right)
  \end{equation}

This results in $a$ being a normally distributed random variable, with mean $\mu_a$ and standard deviation $\sigma_a$.

The second step consists in solving the projection of the distance between both microphones represented by random variables $\mathbf{m}_p$ and $\mathbf{m}_q$, on the deterministic unit vector $\mathbf{u}$, represented below as $b_{pq}(\mathbf{u})$:
  \begin{equation}
    b_{pq}(\mathbf{u}) = \left(\mathbf{m}_p-\mathbf{m}_q\right)\cdot \mathbf{u}
  \end{equation}

The intermediate expression $\left(\mathbf{m}_p-\mathbf{m}_q\right)$ is a random variable with a normal distribution $\sim \mathcal{N}(\bm{\mu}_{pq},\bm{\Sigma}_{pq})$, where $\bm{\mu}_{pq} = \bm{\mu}_{p} - \bm{\mu}_{q}$ and $\bm{\Sigma}_{pq} = \bm{\Sigma}_{p} + \bm{\Sigma}_{q}$.
The position uncertainty is usually significantly smaller than the distance between both microphones, such that $\|\bm{\mu}_{pq}\|^2 \gg \|\bm{\Sigma}_{pq}\|$, where the expression $\|\dots\|$ stands for the vector and matrix norms.
The random variable $b_{pq}(\mathbf{u})$ has a normal distribution:
  \begin{equation}
    b_{pq}(\mathbf{u}) = \mu_{b,pq}(\mathbf{u}) + \sigma_{b,pq}(\mathbf{u}) \eta =  \bm{\mu}_{pq} \cdot \mathbf{u} + \eta\sqrt{\mathbf{u}^T\bm{\Sigma}_{pq}\mathbf{u}}
  \end{equation}

The random variable $\tau_{pq}(\mathbf{u})$ is the product of the normal random variables $a$ and $b_{pq}(\mathbf{u})$, which gives the following expression:
  \begin{equation}
    \tau_{pq}(\mathbf{u}) = (\mu_a + \sigma_a \eta_a)(\mu_{b,pq}(\mathbf{u}) + \sigma_{b,pq}(\mathbf{u}) \eta_b)
  \end{equation}

\noindent where $\eta_a$ and $\eta_b$ are two independent random variables with standard normal distribution.
Since $\mu_a \gg \sigma_a$, and $\mu_{b,pq}(\mathbf{u}) \gg \sigma_{b,pq}(\mathbf{u})$ (for all $p$ and $q$), the following approximation holds:
  \begin{equation}
    \tau_{pq}(\mathbf{u}) \approx \mu_a \mu_{b,pq}(\mathbf{u}) + \mu_a \sigma_{b,pq}(\mathbf{u}) \eta_b + \mu_{b,pq}(\mathbf{u}) \sigma_a \eta_a
  \end{equation}

The random variable $\tau_{pq}(\mathbf{u})$ therefore exhibits a normal distribution $\sim \mathcal{N}(\mu_{\tau,pq}(\mathbf{u}),\sigma_{\tau,pq}(\mathbf{u}))$ where:
  \begin{equation}
    \mu_{\tau,pq}(\mathbf{u}) = \left(\frac{f_S}{\mu_c}\right)(\bm{\mu}_p-\bm{\mu}_q)\cdot \mathbf{u}
  \end{equation}
  \begin{equation}
    \sigma_{\tau,pq}(\mathbf{u}) = \frac{f_S}{\mu_c}\sqrt{\mathbf{u}^{T}(\bm{\Sigma}_p+\bm{\Sigma}_q)\mathbf{u} + [(\bm{\mu}_p-\bm{\mu}_q)\cdot\mathbf{u}]^2\frac{\sigma_c^2}{\mu_c^2}}
  \end{equation}

This models the TDOA estimation uncertainty, and is used to configure MSW size.
In practice, GCC-PHAT based on FFT generates frames with discrete indexes, and therefore the estimated TDOA value (denoted by $\hat{\tau}_{pq}(\mathbf{u}_k)$) for each discrete direction $\mathbf{u}_k$ can be rounded to the closest integer if no interpolation is performed:
  \begin{equation}
    \hat{\tau}_{pq}(\mathbf{u}_k) = \left\lfloor \left(\frac{f_S}{\mu_c}\right)(\bm{\mu}_p-\bm{\mu}_q)\cdot \mathbf{u}_k \right\rceil
  \end{equation}

To cope with the disparity between $\hat{\tau}_{pq}(\mathbf{u}_k)$ and the observation of the random variable $\tau_{pq}(\mathbf{u})$, a MSW filters each GCC-PHAT frame for all pairs of microphones, where $\hat{r}^l_{pq}[n]$ stands for the filtered frame.
The MSW has a size of $2\Delta_{pq} + 1$ samples (the frame index $l$ is omitted here for clarity):
  \begin{equation}
    \hat{r}_{pq}[n] = \max\left\{r_{pq}[n-\Delta\tau_{pq}],\dots,r_{pq}[n+\Delta\tau_{pq}]\right\}
  \end{equation}

\begin{figure}[ht!]
    \centering
    \includegraphics[width=\columnwidth]{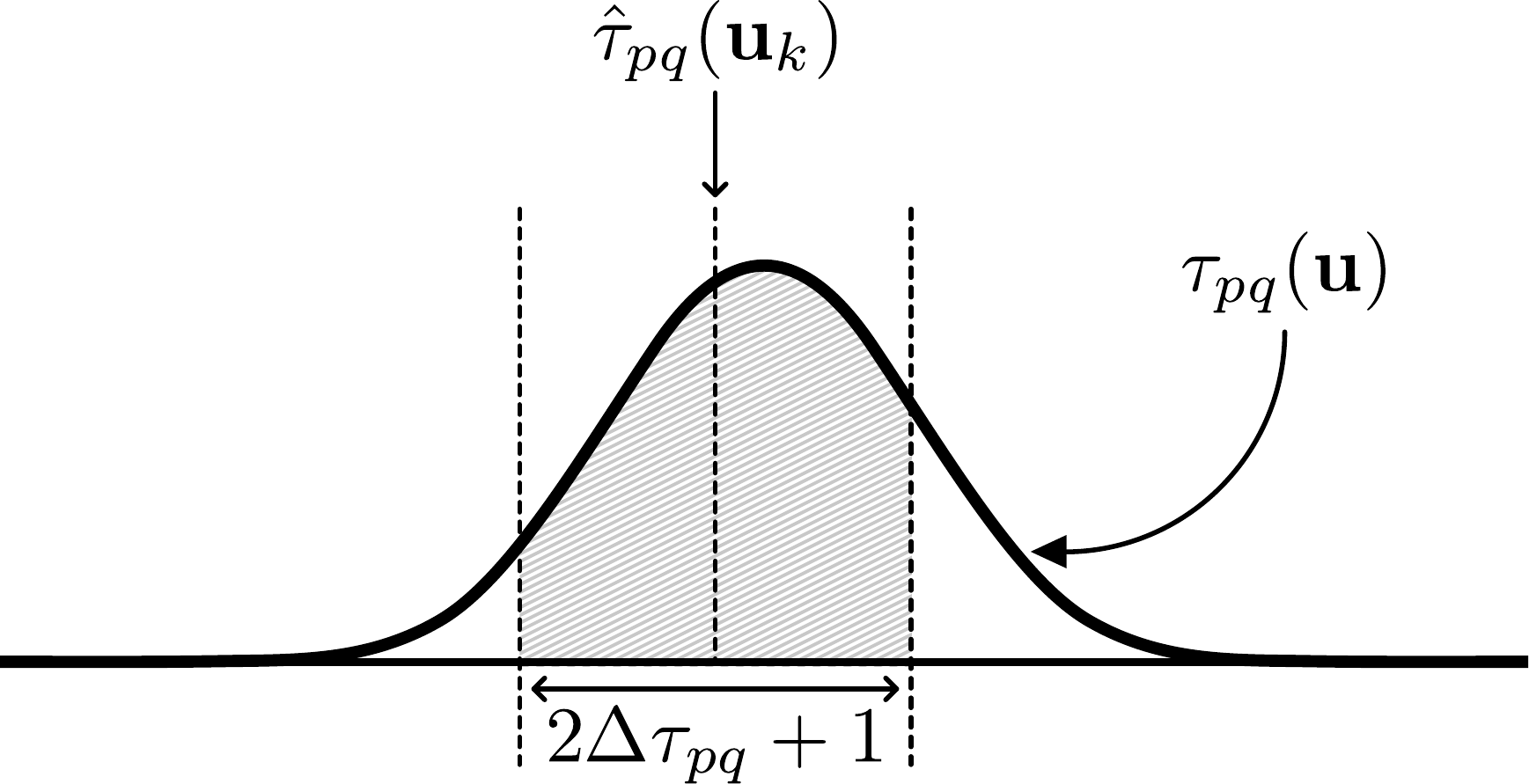}
    \caption{MSW and PDF of the TDOA random variable}    
    \label{fig:ssl_msw_pdf}
  \end{figure}

Figure \ref{fig:ssl_msw_pdf} illustrates how the partial area under the probability density function (PDF) of $\tau_{pq}(\mathbf{u})$ stands for the probability the MSW captures the TDOA value.

The area under the curve corresponds to the integral of the PDF that lies within the interval of the MSW, given by $\mathcal{I}(\mathbf{u}_k) = [\alpha(\mathbf{u}_k),\beta(\mathbf{u}_k)]$:
\begin{equation}
\alpha(\mathbf{u}_k) = \hat{\tau}_{pq}(\mathbf{u}_k)-\Delta\tau_{pq}-0.5
\end{equation}
\begin{equation}
\beta(\mathbf{u}_k) = \hat{\tau}_{pq}(\mathbf{u}_k)+\Delta\tau_{pq}+0.5
\end{equation}
\begin{equation}
P(\tau_{pq}(\mathbf{u}) \in \mathcal{I}(\mathbf{u}_k)) = \displaystyle\int_{\alpha(\mathbf{u}_k)}^{\beta(\mathbf{u}_k)}{\frac{1}{\sqrt{2\pi\sigma^2}}\exp{\left[-\frac{(\tau-\mu)^2}{2\sigma^2}\right]}d\tau}
\end{equation}
where $\mu = \mu_{\tau,pq}(\mathbf{u})$ and $\sigma = \sigma_{\tau,pq}(\mathbf{u})$.

A discrete integration over space $D_k$ is used to estimate the probability $P_{pq}$ that the MSW captures a source in a neighboring direction to $\mathbf{u}$:
  \begin{equation}
    P_{pq}(\mathbf{u} \in D_k) \approx \sum_{e=1}^{E}{\frac{P(\tau_{pq}(\mathbf{v}_{k,e}) \in \mathcal{I}(\mathbf{u}_k))}{E}}
  \end{equation}

An octagon made of $E = 4(2^{\mathcal{D}} + 4^{\mathcal{D}})+1$ points estimates the discretized surface, where $\mathcal{D}$ stands for the number of recursive iterations.
The radius of the octagon corresponds to the distance between $\mathbf{u}_k$ and its closest neighbor.
Figure \ref{fig:ssl_octagons} shows octagons for $\mathcal{D} = 0$, $1$ and $2$ iterations:
\begin{figure}[!ht]
\centering
\subfloat[$\mathcal{D} = 0$]{\includegraphics[width=0.27\columnwidth]{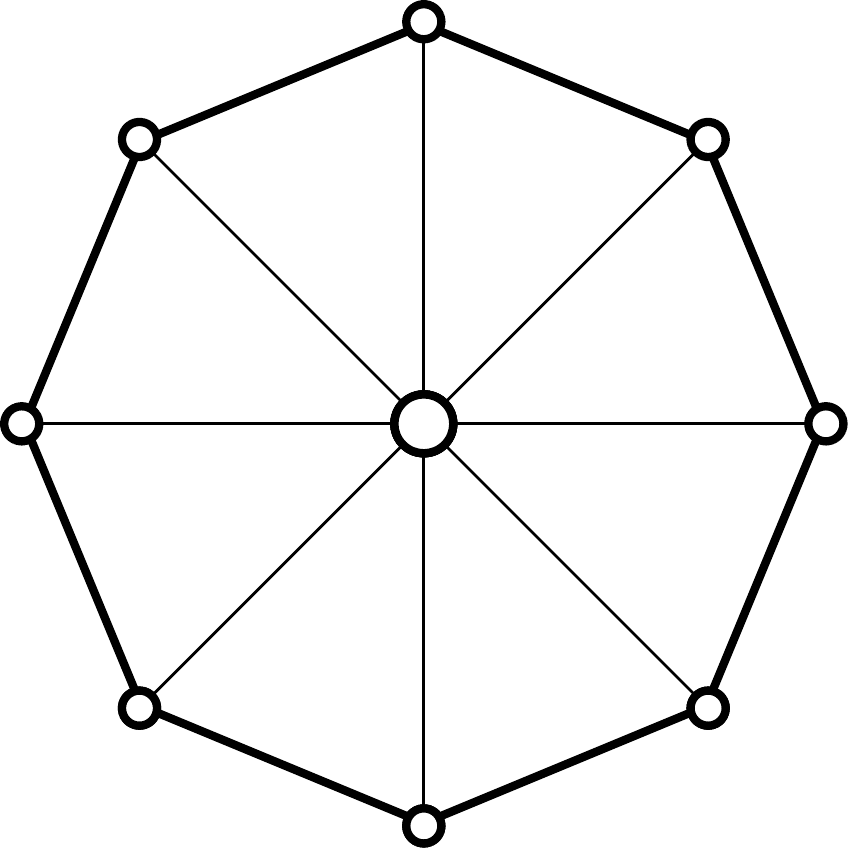}}
\hfil
\subfloat[$\mathcal{D} = 1$]{\includegraphics[width=0.27\columnwidth]{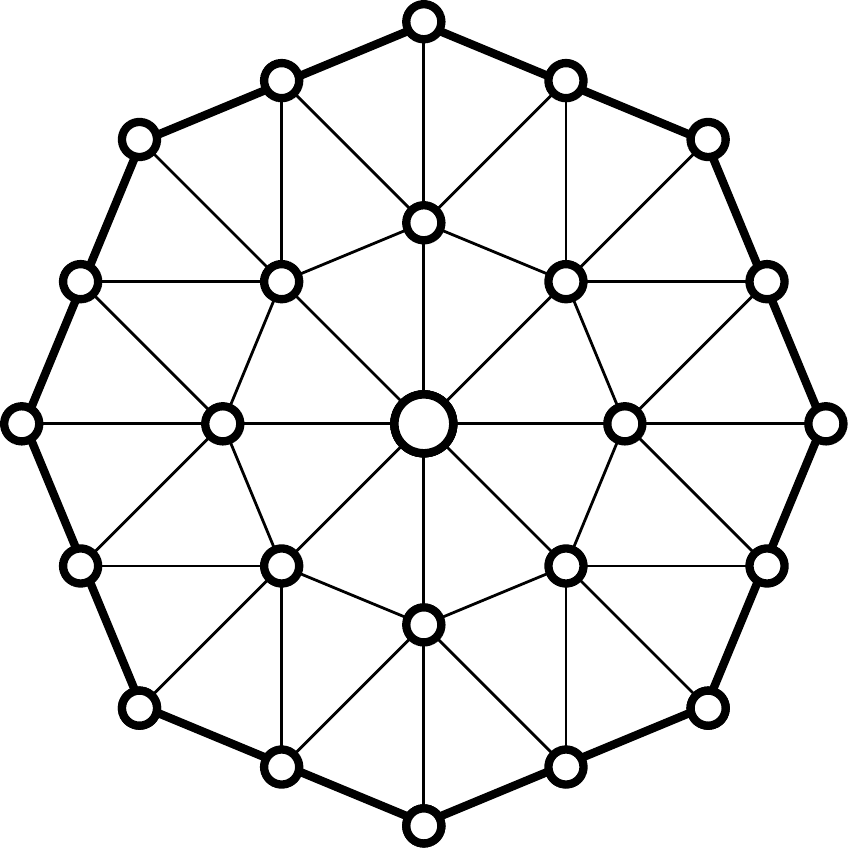}}
\hfil
\subfloat[$\mathcal{D} = 2$]{\includegraphics[width=0.27\columnwidth]{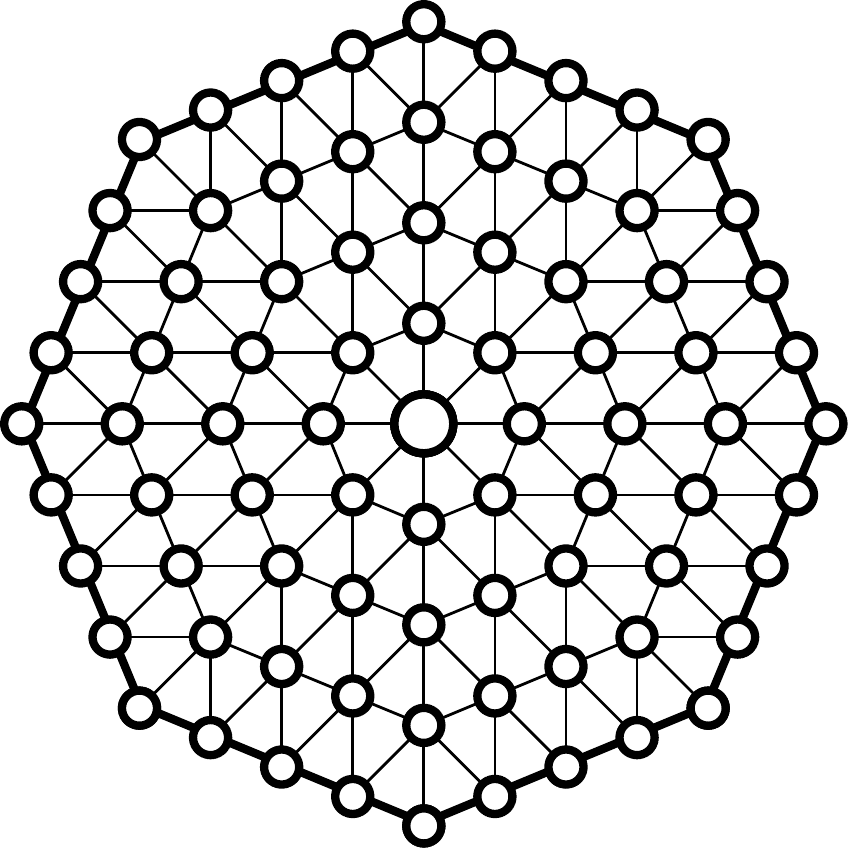}}
\caption{Discrete octogons}
\label{fig:ssl_octagons}
\end{figure}

For all directions in the set $S = \{\mathbf{u}_1, \mathbf{u}_2, \dots, \mathbf{u}_{K}\}$, the probability that the MSW captures sources in neighboring directions is estimated as follows:
  \begin{equation}
    P_{pq}(\mathbf{u} \in D_k, 1 \leq k \leq K) \approx \sum_{k=1}^{K}\sum_{e=1}^{E}{\frac{P(\tau_{pq}(\mathbf{v}_{k,e}) \in \mathcal{I}(\mathbf{u}_k))}{K E}}
  \label{eq:P_pq_u_Dk}
  \end{equation}

For a given discrete direction $\mathbf{u}_k$, the probability that the discrete point $\mathbf{v}_{k,e}$ is captured for all pairs of microphones is estimated with the following expression:
  \begin{equation}
    P(\mathbf{v}_{k,e},\mathbf{u}_k) \approx \sum_{p=1}^{M}{\sum_{q=p+1}^{M}\frac{P(\tau_{pq}(\mathbf{v}_{k,e}) \in \mathcal{I}(\mathbf{u}_k))}{M(M-1)}}
    \label{eq:P_v_ke}
  \end{equation}

The objective is to maximize $P(\mathbf{v}_{k,e},\mathbf{u}_k)$ for all directions $\mathbf{u}_k$ and discrete points $\mathbf{v}_{k,e}$, while keeping the MSW window size as small as possible to preserve the localization accuracy.
To achieve this, Algorithm \ref{alg:ssl_msw} increments the parameters $\Delta\tau_{pq}$ progressively until the threshold $C_{min}$ is reached.
This calibration is performed once at initialization.
  \begin{algorithm}[!ht]
  \caption{MSW Automatic Calibration -- Offline}
  \label{alg:ssl_msw}
  \begin{algorithmic}[1]
  \For{all pairs $pq$}
  \State $\Delta\tau_{pq} \gets 0$
  \State Compute $P_{pq}(\mathbf{u} \in D_k, 1 \leq k \leq K)$ with (\ref{eq:P_pq_u_Dk})
  \EndFor
  \State Compute $P(\mathbf{v}_{k,e},\mathbf{u}_k)$ for all $k$ and $e$ with (\ref{eq:P_v_ke})
  \While{$\min_{k,e}{\{P(\mathbf{v}_{k,e},\mathbf{u}_k)\}} < C_{min}$}
  \State $(pq)^* \gets \argmin_{pq}{\{P_{pq}(\mathbf{u} \in D_k, 1 \leq k \leq K)\}}$
  \State $\Delta\tau_{(pq)^*} \gets \Delta\tau_{(pq)^*} + 1$
  \State Update $P_{(pq)^*}(\mathbf{u} \in D_k, 1 \leq k \leq K)$
  \State Update $P(\mathbf{v}_{k,e},\mathbf{u}_k)$ for all $k$ and $e$
  \EndWhile
  \end{algorithmic}
  \end{algorithm}

\subsection{Hierarchical Search}
\label{subsec:ssl_hiersearch}

Hierarchical search involves two discrete grids: one with a coarse resolution and the other with a fine resolution.
A matching matrix $\mathcal{M}$ provides a mean to connect at initialization the coarse and fine resolution grids, which are then used to perform the hierarchical search.

Algorithm \ref{alg:ssl_scan} first performs a scan using the coarse resolution grid, and then a second scan over a region of the fine resolution grid to improve accuracy.
The expressions $\hat{r}'_{pq}$ and $\hat{r}''_{pq}$ stand for the GCC-PHAT frames at pair $pq$ filtered by the MSW for the coarse and fine resolutions grids, respectively.
To consider the microphone directivity in the scanning process, the GCC-PHAT result for each pair $pq$ and directions $\mathbf{u}'_{c}$ or $\mathbf{u}''_{f}$ is summed only when the binary masks $\zeta_{pq}(\mathbf{u}'_c)$ or $\zeta_{pq}(\mathbf{u}''_f)$ are set to $1$.
The energy levels (defined by the expressions $\mathcal{E}'$ and $\mathcal{E}''$) are normalized with the number of active pairs for each direction (expressed by $\mathcal{T}$).
The variable $\epsilon$ is set to a small value to avoid division by zero.

The coarse scan returns the maximum index $c^*$ on the coarse grid, and then the fine scan searches all points $f$ where $\mathcal{M}(c^*,f) = 1$.
The point $f^*$ then corresponds to the index of the point on the fine grid with the maximum value.
Scanning for the $v$ potential source returns the DoA $\bm{\lambda}_v = \mathbf{u}''_{f^*}$ and the corresponding energy level $\Lambda_v = \mathcal{E}''(f^*)$.

The proposed Hierarchical Search involves $K' + K''U/K'$ directions to scan in average, compared with $K''$ directions for a fixed grid with the same resolution.
For instance, with $\mathcal{L'}=2$, $\mathcal{L''}=4$ and $U=10$ for SRP-PHAT-HSDA, and $\mathcal{L}=4$ for SRP-PHAT, there are in average $320$ directions to scan, instead of $2562$.

\begin{algorithm}[!ht]
\caption{Hierarchical Search Scanning -- Online}
\label{alg:ssl_scan}
\begin{algorithmic}[1]
\For{$c = 1$ \textbf{to} $K'$}
\State $\mathcal{E}'(c) \gets 0$, $\mathcal{T} \gets 0$
\For{all pairs $pq$}
\If{$\zeta_{pq}(\mathbf{u}'_c) = 1$}
\State $\mathcal{E}'(c) \gets \mathcal{E}'(c) + \hat{r}'_{pq}[\hat{\tau}_{pq}(\mathbf{u}'_c)]$
\State $\mathcal{T} \gets \mathcal{T} + 1$
\EndIf
\EndFor
\State $\mathcal{E}'(c) \gets \mathcal{E}'(c) / (\mathcal{T} + \epsilon)$
\EndFor
\State $c^* \gets \argmax_{c}{\mathcal{E}'(c)}$
\For{$f = 1$ \textbf{to} $K''$}
\State $\mathcal{E}''(f) \gets 0$, $\mathcal{T} \gets 0$
\If{$\mathcal{M}(c^*,f) = 1$}
\For{all pairs $pq$}
\If{$\zeta_{pq}(\mathbf{u}''_f) = 1$}
\State $\mathcal{E}''(f) \gets \mathcal{E}''(f) + \hat{r}''_{pq}[\hat{\tau}_{pq}(\mathbf{u}''_f)]$
\State $\mathcal{T} \gets \mathcal{T} + 1$
\EndIf
\EndFor
\State $\mathcal{E}''(f) \gets \mathcal{E}''(f) / (\mathcal{T} + \epsilon)$
\EndIf
\EndFor
\State $f^* \gets \argmax_{f}{\mathcal{E}''(f)}$
\State \Return $\{\mathbf{u}''_{f^*},\mathcal{E}''(f^*)\}$
\end{algorithmic}
\end{algorithm}

The matching matrix $\mathcal{M}$ that connects the coarse and fine resolution spaces (denoted by $S'$ and $S''$, respectively) needs to be generated offline prior to online hierarchical search.
This $K' \times K''$ matrix, denoted by the variable $\mathcal{M}$, connects each direction from the fine resolution grid (composed of $K''$ directions) to many directions in the coarse resolution grid (made of $K'$ directions).

The similitude between a direction $\mathbf{u}'_c$ in $S'$ and a direction $\mathbf{u}''_f$ in $S''$ is given by $\delta_{pq}(c,f)$, which is the length of the intersection between subsets $I'_{pq}(\mathbf{u}'_c)$ and $I''_{pq}(\mathbf{u}''_f)$:
\begin{equation}
\delta_{pq}(c,f) = \left| I'_{pq}(\mathbf{u}'_c) \cap I''_{pq}(\mathbf{u}''_f) \right|
\end{equation}
where:
\begin{equation}
I'_{pq}(\mathbf{u}'_c) = [\hat{\tau}_{pq}(\mathbf{u}'_c)-\Delta\tau'_{pq},\hat{\tau}_{pq}(\mathbf{u}'_c)+\Delta\tau'_{pq}]
\end{equation}
\begin{equation}
I''_{pq}(\mathbf{u}''_f) = [\hat{\tau}_{pq}(\mathbf{u}''_f)-\Delta\tau''_{pq},\hat{\tau}_{pq}(\mathbf{u}''_f)+\Delta\tau''_{pq}]
\end{equation}

The expressions $\Delta\tau'_{pq}$ and $\Delta\tau''_{pq}$ depend on the window size of the MSWs, computed for the coarse and fine resolutions grids, for each pair of microphones $pq$.
Each direction $\mathbf{u}''_{f}$ in the fine resolution grid is mapped to the $U$ most similar directions in the coarse resolution grid, derived using Algorithm \ref{alg:ssl_mapping}.
\begin{algorithm}[!ht]
\caption{Hierarchical Search Matching -- Offline}
\label{alg:ssl_mapping}
\begin{algorithmic}[1]
\For{$f = 1$ \textbf{to} $K''$}
\For{$c = 1$ \textbf{to} $K'$}
\State $\mathcal{M}(c,f) \gets 0$
\State $\mathcal{V}(c) \gets 0$
\For{all pairs $pq$}
\If{$\zeta_{pq}(\mathbf{u}'_c) = 1$ \textbf{and} $\zeta_{pq}(\mathbf{u}''_f) = 1$}
\State $\mathcal{V}(c) \gets \mathcal{V}(c) + \delta_{pq}(c,f)$
\EndIf
\EndFor
\EndFor
\For{$u = 1$ \textbf{to} $U$}
\State $c^* \gets \argmax_{c}{\mathcal{V}(c)}$
\State $\mathcal{M}(c^*,f) \gets 1$
\State $\mathcal{V}(c^*) \gets 0$       
\EndFor
\EndFor
\end{algorithmic}
\end{algorithm}

\section{Modified 3D Kalman Filters for SST}
\label{sec:sst}

\begin{figure*}[!ht]
  \centering
  \includegraphics[width=\textwidth]{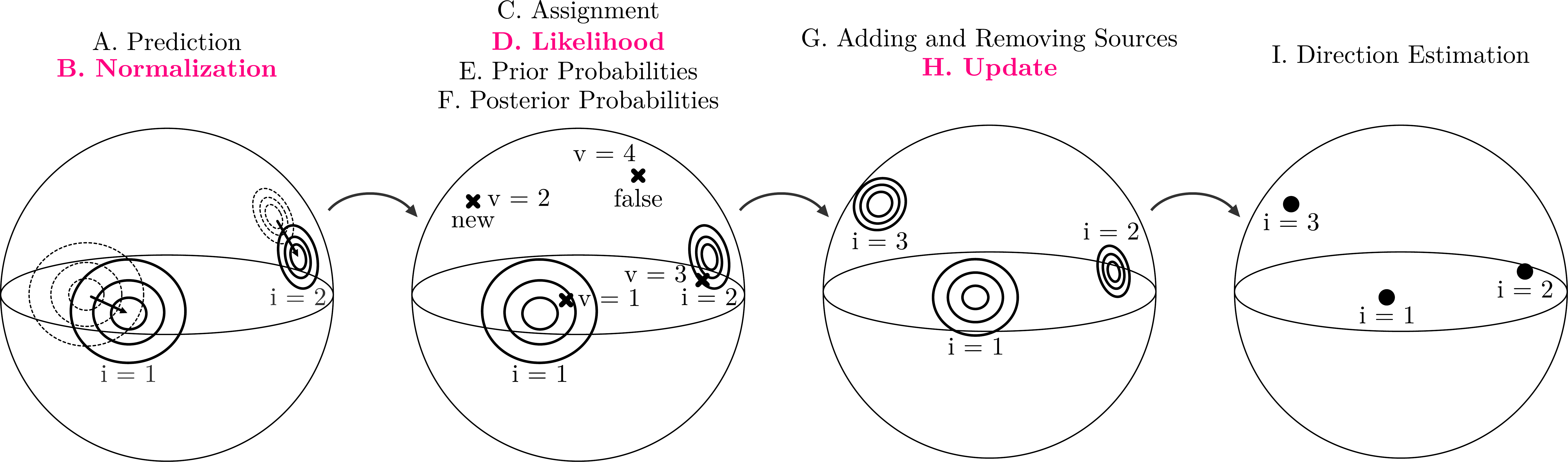}
  \caption{Tracking simultaneous sound sources using M3K. Tracked sources are labeled $i=1,2,3$ and potential sources are labeled $v=1,2,3,4$.}
  \label{fig:sst_overview}
  \end{figure*}

Figure \ref{fig:sst_overview} illustrates the nine steps of the M3K method, where  Normalization (Step B), Likelihood (Step D) and Update (Step H) are introduced to include Kalman filtering and replace particle filtering in the multiple sources tracking method presented by Valin et al. \cite{valin2007robust}.
First, the new states of each tracked source are predicted (Step A) and normalized (Step B) in relation to the search space.
Second, the potential sources are then assigned (Step C) to either a source currently tracked, a new source of a false detection (Steps D, E, F).
Third, the method then adds (Step G) new sources to be tracked if needed, and removes inactive sources previously tracked.
Fourth, the states of each tracked source are updated (Step H) with the relevant observations, and the direction of each tracked source is finally estimated (Step I) from the Gaussian distributions. 

Before presenting in more details these nine steps in the following subsections, let us first define the Kalman filter model used in M3K.
A Kalman filter estimates recursively the state of each source and provides the estimated source direction.
Normally distributed random variables model the 3D-direction ($(d_x)^l_i$, $(d_y)^l_i$ and $(d_z)^l_i$) and 3D-velocity ($(s_x)^l_i$, $(s_y)^l_i$ and $(s_z)^l_i$), where $\{\dots\}^T$ stands the transpose operator:
  \begin{equation}
  \mathbf{d}^l_i = \left[
  \begin{array}{ccc}
  (d_x)^l_i & (d_y)^l_i & (d_z)^l_i
  \end{array}
  \right]^T
  \label{eq:sst_d}
  \end{equation}
  \begin{equation}
  \mathbf{s}^l_i = \left[
  \begin{array}{ccc}
  (s_x)^l_i & (s_y)^l_i & (s_z)^l_i
  \end{array}
  \right]^T
  \label{eq:sst_ddot}
  \end{equation}

The $6 \times 1$ random vector $\mathbf{x}^l_i$ concatenates these positions and velocities:
  \begin{equation}
  \mathbf{x}^l_i = \left[
  \def\arraystretch{1.1}
  \begin{array}{c}
  \mathbf{d}^l_i \\ 
  \mathbf{s}^l_i \\
  \end{array}
  \right]
  \label{eq:sst_x1}
  \end{equation}
The Kalman model assumes the state evolves over time according to the following linear model: 
  \begin{equation}
  \mathbf{x}^l_i = \mathbf{F}\mathbf{x}^{l-1}_i + \mathbf{B}\mathbf{u}^l_i + \mathbf{w}_i
  \label{eq:sst_x2}
  \end{equation}
where the matrix $\mathbf{F}$ stands for the state transition model, $\mathbf{B}$ represents the control-input matrix, $\mathbf{u}^i_l$ is the control vector and $\mathbf{w}_i$ models the process noise.
In the $6 \times 6$ matrix $\mathbf{F}$, the expression $\Delta T = \Delta N/f_S$ denotes the time interval (in second) between two successive frames, with $\Delta N$ being the hop size in samples between two frames, and $f_S$ the sample rate in samples per second:
  \begin{equation}
  \mathbf{F} = \left[ 
  \def\arraystretch{1.1}
  \begin{array}{cccccc}
  \ \,1\ \,&\ \,0\ \,&\ \,0\ \,& \Delta T &\ \,0\ \,&\ \,0\ \,\\
  0 & 1 & 0 &\ \,0\ \,& \Delta T & 0 \\
  0 & 0 & 1 & 0 & 0 & \Delta T \\
  0 & 0 & 0 & 1 & 0 & 0 \\
  0 & 0 & 0 & 0 & 1 & 0 \\
  0 & 0 & 0 & 0 & 0 & 1 \\
  \end{array}
  \right]
  \label{eq:sst_F}
  \end{equation}

With M3K, there is no control input, and therefore the expression ($\mathbf{B}\mathbf{u}^i_l$) in (\ref{eq:sst_x2}) is ignored.
The process noise $\mathbf{w}_i$ exhibits a multivariate normal distribution, where $\mathbf{w}_i \sim \mathcal{N}(\mathbf{0},\mathbf{Q})$.
In M3K, the process noise lies in the velocity state and is parametrized with the variance $\sigma_Q^2$.
In this method, the parameter $\sigma_Q^2$ is set to a constant value, but it would be possible to have it depend on $\Delta T$ as uncertainty increases for larger values of $\Delta T$.
  \begin{equation}
  \mathbf{Q} = \left[ 
  \def\arraystretch{1.1}
  \begin{array}{cccccc}
  \ \,0\ \,&\ \,0\ \,&\ \,0\ \,&\ \,0\ \,&\ \,0\ \,&\ \,0\ \,\\
  0 & 0 & 0 & 0 & 0 & 0 \\
  0 & 0 & 0 & 0 & 0 & 0 \\
  0 & 0 & 0 & \sigma_Q^2 & 0 & 0 \\
  0 & 0 & 0 & 0 & \sigma_Q^2 & 0 \\
  0 & 0 & 0 & 0 & 0 & \sigma_Q^2 \\
  \end{array}
  \right]
  \label{eq:sst_Q}
  \end{equation}

The observations $\mathbf{z}^i_l$ are represented by random variables in the $x$-, $y$- and $z$-directions, obtained from the states:
  \begin{equation}
  \mathbf{z}^i_l = \mathbf{H}\mathbf{x}^i_l + \mathbf{v}
  \label{eq:sst_z1}
  \end{equation}
where
  \begin{equation}
  \mathbf{z}^i_l = \left[
  \def\arraystretch{1.1}
  \begin{array}{ccc}
  (z_x)^l_i & (z_y)^l_i & (z_z)^l_i \\
  \end{array}
  \right]^T
  \label{eq:sst_z2}
  \end{equation}

The $3 \times 6$ matrix $\mathbf{H}$ stands for the observation model:
  \begin{equation}
  \mathbf{H} = \left[ 
  \def\arraystretch{1.1}
  \begin{array}{cccccc}
  \ \,1\ \,&\ \,0\ \,&\ \,0\ \,&\ \,0\ \,&\ \,0\ \,&\ \,0\ \,\\
  0 & 1 & 0 & 0 & 0 & 0 \\
  0 & 0 & 1 & 0 & 0 & 0 \\
  \end{array}
  \right]
  \label{eq:sst_H}
  \end{equation}

Expression $\mathbf{v} \sim \mathcal{N}(\mathbf{0},\mathbf{R})$ models the observation noise, where the $3 \times 3$ diagonal covariance matrix $\mathbf{R}$ is defined as:
  \begin{equation}
  \mathbf{R} = \left[ 
  \def\arraystretch{1.1}
  \begin{array}{ccc}
  \sigma_R^2&\ \,0\ \,&\ \,0\ \,\\
  \ \,0\ \,& \sigma_R^2 & 0 \\
  0 & 0 & \sigma_R^2 \\
  \end{array}
  \right]
  \label{eq:sst_R}
  \end{equation}

M3K therefore requires only two manually-tuned parameters, $\sigma^2_Q$ and $\sigma^2_R$, which influence the tracking sensitivity and inertia of each tracked source.

\subsection{Prediction (Step A)}
\label{subsec:sst_prediction}

The vector $\hat{\mathbf{x}}_i^{l|l}$ is the posterior mean of the state, where the $6 \times 6$ matrix $\mathbf{P}_i^{l|l}$ stands for the state error posterior covariance matrix of each tracked source $i$.
The tracking method predicts new states (also referred to as prior states) for each sound source $i$.
Predicted mean vector and covariance matrix are obtained as follows:
  \begin{equation}
  \hat{\mathbf{x}}_i^{l|l-1} = \mathbf{F}\hat{\mathbf{x}}_i^{l-1|l-1}
  \label{eq:sst_prediction_xhat}
  \end{equation}
  \begin{equation}
  \mathbf{P}_i^{l|l-1} = \mathbf{F}\mathbf{P}_i^{l-1|l-1}\mathbf{F}^T + \mathbf{Q}
  \label{eq:sst_prediction_P}
  \end{equation}

Each prediction step increases the states uncertainty, which is then reduced in the Update step (Step H) when the tracked source is associated to a relevant observation.

\subsection{Normalization (Step B)}
\label{subsec:sst_normalization}

The observations $\bm{\lambda}^l_v$ stands for sound direction of the potential source $v$, which constitutes a unit vector.
A normalization constraint is therefore introduced and generates a new state mean vector $(\hat{\mathbf{x}}')_i^{l|l-1}$, for which the predicted direction $(\hat{\mathbf{d}}')_i^{l|l-1}$ lies on a unitary sphere and velocity $(\hat{\mathbf{s}}')_i^{l|l-1}$ is tangential to the sphere surface:
  \begin{equation}
  (\hat{\mathbf{x}}')_i^{l|l-1} = \left[
    \def\arraystretch{1.1}
    \begin{array}{c}
    (\hat{\mathbf{d}}')_i^{l|l-1} \\ 
    (\hat{\mathbf{s}}')_i^{l|l-1} \\
    \end{array}
    \right]
  \end{equation}
  where
  \begin{equation}
  (\hat{\mathbf{d}}')_i^{l|l-1} = \frac{\hat{\mathbf{d}}_i^{l|l-1}}{\|\hat{\mathbf{d}}_i^{l|l-1}\|}
  \end{equation}
  and
  \begin{equation}
  (\hat{\mathbf{s}}')_i^{l|l-1} = \hat{\mathbf{s}}_i^{l|l-1} - \hat{\mathbf{d}}_i^{l|l-1}\left(\frac{\hat{\mathbf{s}}_i^{l|l-1} \cdot \hat{\mathbf{d}}_i^{l|l-1}}{\|\hat{\mathbf{d}}_i^{l|l-1}\|^2}\right) 
  \end{equation}

This manipulation violates the Kalman filter assumptions, which state that all processes are Gaussian and that the system is linear \cite{julier1997new}.
In practice however, Kalman filtering remains efficient as this normalization only involves a slight perturbation in direction and velocity, which makes the nonlinearity negligible.

During the normalization process, the covariance matrix remains unchanged.
The Update step (Step H) ensures that the radial component of the matrix $\mathbf{P}_i^{l|l-1}$ stays as small as possible such that the PDF (Probability Density Function) lies mostly on the unit sphere surface.

\subsection{Assignment (Step C)}
\label{subsec:sst_assignment}

Assuming that $I$ sources are tracked, the function $f_g(v)$ assigns the potential source at index $v$ to either a false detection ($-2$), a new source ($-1$), or a previously tracked source (from $1$ to $I$):
  \begin{equation}
  f_g(v) \in \{ -2, -1, 1, 2, \dots, I\}
  \label{eq:sst_assignment_fgs}
  \end{equation}

There are $V$ potential sources that can be assigned to $I+2$ values, which leads to $G = {(I+2)}^V$ possible assignments.
The vector for the assignation $g$ concatenates the $V$ assignment functions for the potential sources:
  \begin{equation}
  \mathbf{f}_g = \left[
  \begin{array}{cccc}
  f_g(1) & f_g(2) & \dots & f_g(V)
  \end{array}
  \right]
  \label{eq:sst_assignment_fg}
  \end{equation}

\subsection{Likelihood (Step D)}
\label{subsec:sst_likelihood}

The energy level $\Lambda^l_v$ gives significant information regarding sound source activity.
When a sound source is active ($\mathcal{A}$), which means the source emits a sound, a Gaussian distribution models the energy level and the PDF is given by:
  \begin{equation}
  p(\Lambda^l_v|\mathcal{A}) = \mathcal{N}(\Lambda^l_v|\mu_\mathcal{A}, \sigma_\mathcal{A})
  \label{eq:sst_likelihood_pLambdaA}
  \end{equation}
where $\mu_{\mathcal{A}}$ and $\sigma_{\mathcal{A}}$ stand for the mean and standard deviation of the normal distribution, respectively (for simplicity, we omit $\mu_{\mathcal{A}}$ and $\sigma_{\mathcal{A}}$ from the left-hand side of the equation).

When the source is inactive, the energy level is modeled with the same distribution but different parameters:
  \begin{equation}
  p(\Lambda^l_v|\mathcal{I}) = \mathcal{N}(\Lambda^l_v|\mu_\mathcal{I}, \sigma_\mathcal{I})
  \label{eq:sst_likelihood_pLambdaI}
  \end{equation}
where $\mu_{\mathcal{I}}$ and $\sigma_{\mathcal{I}}$ also represent the mean and standard deviation of the normal distribution, respectively.
Typically the standard deviation $\sigma_{\mathcal{I}}$ is similar to $\sigma_{\mathcal{A}}$, but the mean $\mu_{\mathcal{I}}$ is smaller than $\mu_{\mathcal{A}}$.
Moreover, the probability that the potential source $\bm{\lambda}^l_v$ is generated by the tracked source $i$ is obtained with the following volume integral:
  \begin{equation}
  P(\bm{\lambda}^l_v|\mathcal{C}_i) = \iiint{p(\bm{\lambda}^l_v|(\mathbf{d}')^{l|l-1}_i)p((\mathbf{d}')^{l|l-1}_i)}\,dx\,dy\,dz
  \label{eq:sst_likelihood_plambdaC}
  \end{equation}

The symbol $\mathcal{C}_i$ stands for a coherent source, which describes a sound source located at a specific direction in space.
As modeled by the Kalman filter, the probability $p((\mathbf{d}')_i^{l|l-1})$ follows this normal distribution:
  \begin{equation}
  (\mathbf{d}')_i^{l|l-1} \sim \mathcal{N}\left(\bm{\mu}^{l}_i,\bm{\Sigma}_i^{l}\right)
  \label{eq:sst_likelihood_di}
  \end{equation}
where
  \begin{equation}
  \bm{\mu}_i^{l} = \mathbf{H}(\hat{\mathbf{x}}')^{l|l-1}_i
  \label{eq:sst_likelihood_mui}
  \end{equation}
and
  \begin{equation}
  \bm{\Sigma}_i^{l} = \mathbf{H}\mathbf{P}_i^{l|l-1}\mathbf{H}^T
  \label{eq:sst_likelihood_sigmai}
  \end{equation}

Given the normalized tracked source direction $(\mathbf{d}')_i^{l|l-1}$, the following expression represents the PDF when the potential source $\bm{\lambda}_v^l$ is observed :
  \begin{equation}
  p(\bm{\lambda}^l_v|(\mathbf{d}')_i^{l|l-1}) = \mathcal{N}\left(\bm{\lambda}^l_v|(\mathbf{d}')_i^{l|l-1},\mathbf{R}\right)
  \label{eq:sst_likelihood_lambdad}
  \end{equation}

Note that swapping the mean and the random variable leads to the same PDF (the expression $|\dots|$ stands for the matrix determinant):
  \begin{equation}
  p(\bm{\lambda}^l_v|(\mathbf{d}')_i^{l|l-1}) = \frac{(2\pi)^{\frac{-3}{2}}}{|\mathbf{R}|^{1/2}}e^{-\frac{1}{2}(\bm{\lambda}^l_v-(\mathbf{d}')_i^{l|l-1})^T \mathbf{R}^{-1}(\bm{\lambda}^l_v-(\mathbf{d}')_i^{l|l-1})}
  \label{eq:sst_likelihood_plambdad}
  \end{equation}
  \begin{equation}
  p((\mathbf{d}')_i^{l|l-1}|\bm{\lambda}^l_v) = \frac{(2\pi)^{\frac{-3}{2}}}{|\mathbf{R}|^{1/2}}e^{-\frac{1}{2}((\mathbf{d}')_i^{l|l-1}-\bm{\lambda}^l_v)^T \mathbf{R}^{-1}((\mathbf{d}')_i^{l|l-1}-\bm{\lambda}^l_i)}
  \label{eq:sst_likelihood_pdlambda}
  \end{equation}

The following PDF is therefore defined:
  \begin{equation}
  p((\mathbf{d}')_{l|l-1}^i|\bm{\lambda}^l_v) = \mathcal{N}\left((\mathbf{d}')_{l|l-1}^i|\bm{\mu}_v^{l},\bm{\Sigma}_v^{l}\right)
  \label{eq:sst_likelihood_dz}
  \end{equation}
where
  \begin{equation}
  \bm{\mu}_v^{l} = \bm{\lambda}^l_v
  \label{eq:sst_likelihood_muq}
  \end{equation}
and
  \begin{equation}
  \bm{\Sigma}_v^{l} = \mathbf{R}
  \label{eq:sst_likelihood_sigmaq}
  \end{equation}

The expression $p(\bm{\lambda}^l_v|(\mathbf{d}')_i^{l|l-1})$ is equivalent to $p((\mathbf{d}')_i^{l|l-1}|\bm{\lambda}^l_v)$, which results in the product of two Gaussian distributions. 
According to \cite{bromiley2003products}, this results in a new Gaussian distribution, scaled by the factor $\omega^l_{iv}$:
  \begin{equation}
  p(\bm{\lambda}^l_v|(\mathbf{d}')_i^{l|l-1})p((\mathbf{d}')_i^{l|l-1}) = \omega_{iv}^{l} \mathcal{N}((\mathbf{d}')_i^{l|l-1}|\bm{\mu}_{iv}^{l},\bm{\Sigma}_{iv}^{l})
  \label{eq:sst_likelihood_pzdpd}
  \end{equation}

As derived in \cite{bromiley2003products}, the mean vector $\bm{\mu}_{iv}^{l}$ and covariance matrix $\bm{\Sigma}_{iv}^{l}$ of the resulting distribution are equal to:
\begin{equation}
\bm{\mu}_{iv}^{l} = \bm{\Sigma}_{iv}^{l}((\bm{\Sigma}_i^{l})^{-1}\bm{\mu}_i^{l}+(\bm{\Sigma}_v^{l})^{-1}\bm{\mu}_v^{l})
\label{eq:sst_likelihood_muiq}
\end{equation}
\begin{equation}
\bm{\Sigma}_{iv}^{l} = ((\bm{\Sigma}_i^{l})^{-1} + (\bm{\Sigma}_v^{l})^{-1})^{-1}
\label{eq:sst_likelihood_sigmaiq}
\end{equation}

The scaling factor equals to:
  \begin{equation}
  \omega_{iv}^{l} = e^{\left(\frac{1}{2}\left[(C_1)_{iv}^{l} + (C_2)_{iv}^{l} - (C_3)_{iv}^{l} - (C_4)_{iv}^{l}\right]\right)}
  \label{eq:sst_likelihood_omega}
  \end{equation}
where
  \begin{equation}
  (C_1)_{iv}^{l} = \log|\bm{\Sigma}_{iv}^{l}| - \log\left(8\pi^3|\bm{\Sigma}_{i}^{l}||\bm{\Sigma}_{v}^{l}|\right)
  \label{eq:sst_likelihood_c1}
  \end{equation}
  \begin{equation}
  (C_2)_{iv}^{l} = (\bm{\mu}^{l}_{iv})^T(\bm{\Sigma}_{iv}^{l})^{-1}\bm{\mu}^{l}_{iv}
  \label{eq:sst_likelihood_c2}
  \end{equation}
  \begin{equation}
  (C_3)_{iv}^{l} = (\bm{\mu}^{l}_{i})^T(\bm{\Sigma}_{i}^{l})^{-1}\bm{\mu}^{l}_{i}
  \label{eq:sst_likelihood_c3}
  \end{equation}
  \begin{equation}
  (C_4)_{iv}^{l} = (\bm{\mu}^{l}_{v})^T(\bm{\Sigma}_{v}^{l})^{-1}\bm{\mu}^{l}_{v}
  \label{eq:sst_likelihood_c4}
  \end{equation}

\begin{figure}[!ht]
\centering
\includegraphics[width=\columnwidth]{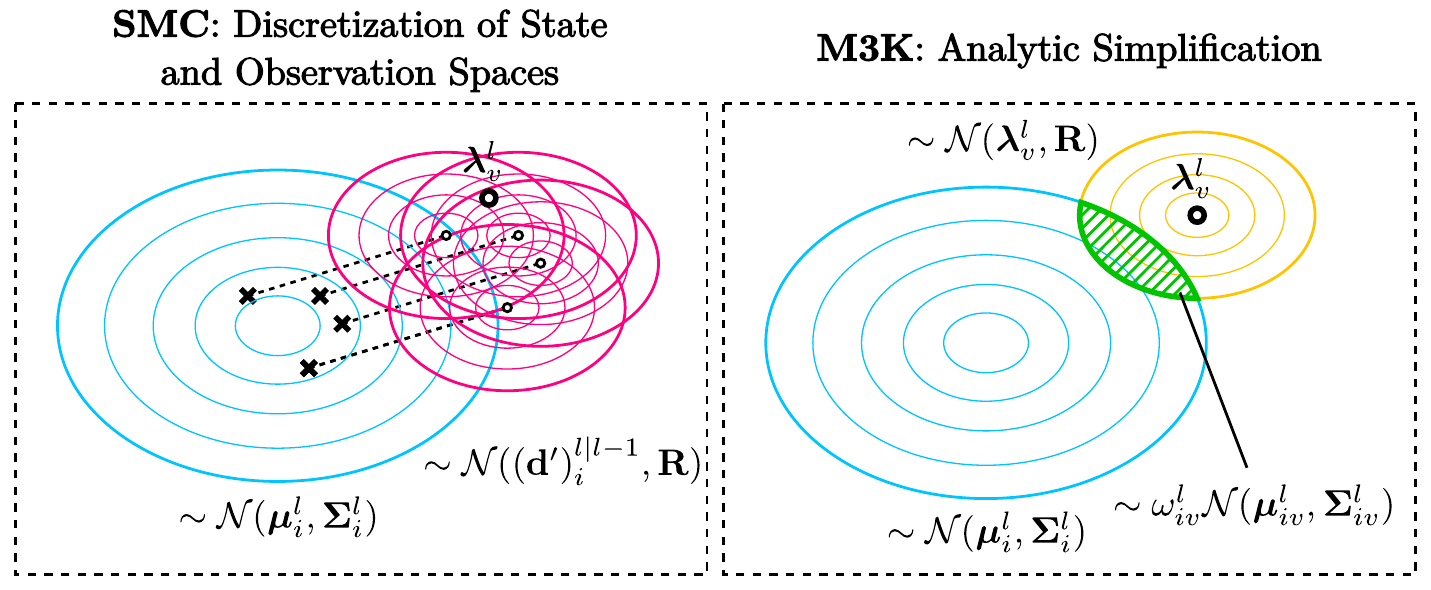}
\caption{Analytic simplification of M3K compared to SMC. The probability that the observation $\bm{\lambda}^l_v$ occurs corresponds to the sum of the product of the probability of each state (in blue) with the probability this state generates the observation $\bm{\lambda}^l_v$ (in pink).}
\label{fig:sst_likelihood_coherence}
\end{figure}

The new Gaussian distribution is substituted in (\ref{eq:sst_likelihood_plambdaC}), and since the volume integral over a trivariate normal PDF is equal to $1$, the probability is simply equal to the scaling factor computed in (\ref{eq:sst_likelihood_omega}):
  \begin{equation}
  P(\bm{\lambda}_v^l|\mathcal{C}_i) = \omega_{iv}^{l}\iiint{\mathcal{N}((\mathbf{d}')_i^{l|l-1}|\bm{\mu}_{iv}^{l},\bm{\Sigma}_{iv}^{l})}\,dx\,dy\,dz = \omega_{iv}^{l}
  \label{eq:sst_likelihood_pzC1}
  \end{equation}
This provides a direct way to compute the probability $p(\bm{\lambda}_v^l|\mathcal{C}_i)$, which is far more efficient than SMC where the probability is estimated by sampling the distribution.
Figure \ref{fig:sst_likelihood_coherence} illustrates the analytic simplification of how M3K simplifies the computation of the triple integral introduced in (\ref{eq:sst_likelihood_plambdaC}).

Each state probability (in the blue area) is multiplied by the probability that $\bm{\lambda}^l_v$ is observed (in pink) by this state.
This involves a significant amount of computations to sample the state space: this is in fact what the particle filter does, and why the computational load is important.
It is therefore more efficient to compute the closed expression that represents the overlap (in green) between state PDF (in blue) and the PDF obtained in (\ref{eq:sst_likelihood_dz}) from swapping variables (in yellow).

When a new source appears or a false detection occurs, the observation lies anywhere on the scanned space.
This is denoted by the symbol $\mathcal{D}$ for diffused signal. 
The SSL module generates DoAs from the scanned space around the microphone array.
This space is modeled by a unit sphere around the array, but the search often partially covers the complete area due to blind spots introduced by the microphone array geometry and other constraints. 
A uniform distribution therefore models the PDF, where $\hat{K}$ denotes the number of points scanned, and $K$ the total number of points needed to discretize the entire sphere: 
\begin{equation}
p(\bm{\lambda}_v^l|\mathcal{D}) = \frac{\hat{K}}{K}\left(\frac{1}{4\pi}\right) = \frac{\hat{K}}{4\pi K}
\label{eq:sst_likelihood_pLambdaD}
\end{equation}
where $1/4\pi$ stands for the uniform distribution over a complete sphere.

\begin{figure}[!ht]
\centering
\includegraphics[width=0.95\columnwidth]{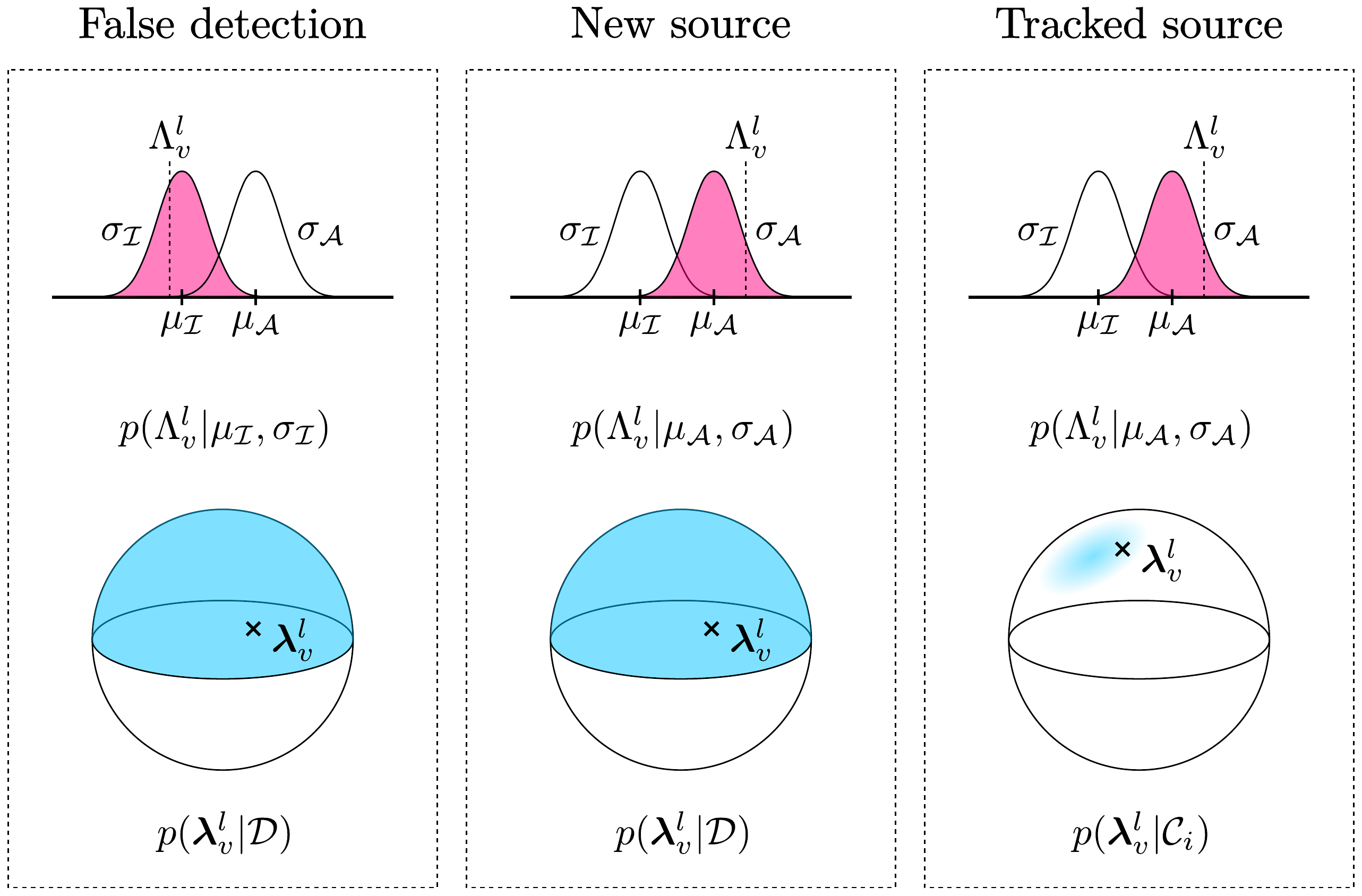}
\caption{Types of assignments for each potential source}
\label{fig:sst_likelihood_assignments}
\end{figure}

The overall likelihood for each possible potential-tracked source assignation results in the combination of the energy level $\Lambda_v^l$ and potential source positions $\bm{\lambda}_v^l$ observations, concatenated in the vector $\bm{\psi}_v^l$.
Figure \ref{fig:sst_likelihood_assignments} illustrates the three types of assignment: 
\begin{enumerate}
\item False detection: the perceived signal is diffused ($\mathcal{D}$) and the source is inactive ($\mathcal{I}$). 
\item New source: the perceived signal is diffused ($\mathcal{D}$) and the source is active ($\mathcal{A}$). 
\item Tracked source $i$: the potential source direction is coherent with the tracked source $i$ ($\mathcal{C}_i$) and the source is active ($\mathcal{A}$).
\end{enumerate}
In Fig. \ref{fig:sst_likelihood_assignments}, it is assumed that the potential sources are generated only in the top hemisphere, which motivates the use of a uniform distribution in this region only for new sources and false detection assignments.

The probability $P(\bm{\psi}^q_l|f_g(q))$ is therefore computed as follows:
\begin{equation}
P(\bm{\psi}^l_v|f_g(v)) = \begin{cases}
P(\Lambda_v^l|\mathcal{I})P(\bm{\lambda}_v^l|\mathcal{D}) & f_g(v)=-2 \\
P(\Lambda_v^l|\mathcal{A})P(\bm{\lambda}_v^l|\mathcal{D}) & f_g(v)=-1 \\
P(\Lambda_v^l|\mathcal{A})P(\bm{\lambda}_v^l|\mathcal{C}_{f_g(v)}) & f_g(v)\geq 1\\
\end{cases}
\label{eq:sst_likelihood_ppsifg}
\end{equation}

Assuming conditional independence, the product of the individual probabilities generates the probability for each assignment:
\begin{equation}
P(\bm{\Psi}^l|\mathbf{f}_g) = \prod_{v=1}^{V}{P(\bm{\psi}_v^l|f_g(v))}
\label{eq:sst_likelihood_pPsifg}
\end{equation}

\subsection{Prior Probabilities (Step E)}
\label{sst:subsec:method_apriori}

The prior probabilities that a false detection, a new source or a tracked source occur are simply defined with the constant parameters $P_{false}$, $P_{new}$, and $P_{track}$, respectively:

\begin{equation}
P(f_g(v)) = \begin{cases}
P_{false} & f_g(v) = -2 \\
P_{new} & f_g(v) = -1 \\
P_{track} & f_g(v) \geq 1 \\
\end{cases}
\label{eq:sst_apriori_pfg}
\end{equation}

These parameters are set empirically but it is observed they have little impact on the performance of tracking.
The prior probability for a given permutation corresponds to the product of each individual assignment:

\begin{equation}
P(\mathbf{f}_g) = \prod_{v=1}^{V}{P(f_g(v))}
\label{eq:sst_apriori_pfgs}
\end{equation}

\subsection{Posterior Probabilities (Step F)} 
\label{subsec:sst_posteriori}

Bayes' theorem provides a method to obtain the posterior probability for each permutation $\mathbf{f}_g$:
\begin{equation}
P(\mathbf{f}_g|\bm{\Psi}^l) = \frac{P(\bm{\Psi}^l|\mathbf{f}_g)P(\mathbf{f}_g)}{\sum_{g=1}^{G}{P(\bm{\Psi}^l|\mathbf{f}_g)P(\mathbf{f}_g)}}
\label{eq:sst_posteriori_fc}
\end{equation}

To calculate the probability that a specific assignment is observed, the discrete Kronecker delta $\delta[n]$ is introduced:
\begin{equation}
\delta[n] = \begin{cases}
0 & n \neq 0 \\
1 & n = 0 \\
\end{cases}
\label{eq:sst_posteriori_delta}
\end{equation}

The probability that the tracked source $i$ generates the potential source $v$ is therefore computed as follows:
\begin{equation}
P(i|\bm{\psi}_v^l) = \sum_{g=1}^{G}{P(\mathbf{f}_g|\bm{\Psi}^l)\delta[f_g(v)-i]}
\label{eq:sst_posteriori_pi}
\end{equation}

The probability that a new source is observed is computed similarly:
\begin{equation}
P(\text{new}|\bm{\psi}_v^l) = \sum_{g=1}^{G}{P(\mathbf{f}_g|\bm{\Psi}^l)\delta[f_g(v)+1]}
\label{eq:sst_posteriori_pnew}
\end{equation}

Finally, the probability that a tracked source is observed by any potential sources is computed using the combinations where there is an assignment between at least one potential source and the tracked source:
\begin{equation}
P(i|\bm{\Psi}^l) = \sum_{g=1}^{G}{P(\mathbf{f}_g|\bm{\Psi}^l)\left(1 - \prod_{v=1}^{V}{\left(1 - \delta[f_g(v)-i]\right)}\right)}
\label{eq:sst_posteriori_pis}
\end{equation}

\subsection{Adding and Removing Sources (Step G)}
\label{sst:subsec:method_addrem}

Sound sources may appear and disappear dynamically as a new sound source starts or a tracked source stops being active. 
When a new source is detected ($p(\text{new}|\bm{\psi}_v^l) > \theta_{new}$), this step waits $N_{prob}$ frames to confirm this is really a valid source and not just a sporadic detection.
During this probation interval, the observation noise variance is set to the parameter $(\sigma^2_R)_{prob}$ to take a small value, as it is assumed the observations should lie close to each others during this time interval.
The average of the probability $p(i|\bm{\psi}_v^l)$ of the newly tracked sound source is evaluated, and the source is kept only if the average exceeds the threshold $\theta_{prob}$.

Once the existence of a source is confirmed, it is tracked until the source becomes inactive ($p(i|\bm{\Psi}^l) < \theta_{dead}$) for at least $N_{dead}$ frames.
During this active state, the observation noise variance is increased to the value of $(\sigma^2_R)_{active}$, to deal with noisy observations and possible motion of the sources. 
When the source no longer exists, it is deleted and tracking of this source stops. 

\subsection{Update (Step H)}
\label{eq:sst_update}

For each tracked source, the Kalman gain is computed as follows:
\begin{equation}
\mathbf{K}_i^{l|l-1} = \mathbf{P}_i^{l-1|l}\mathbf{H}^T(\mathbf{H}\mathbf{P}_i^{l|l-1}\mathbf{H}^T+\mathbf{R})^{-1}
\label{eq:sst_update_K}
\end{equation}

The expression $\hat{v}(i)$ stands for the index of the potential source that maximizes the probability $p(i|\bm{\Psi}^l)$):

\begin{equation}
\hat{v}(i) = \argmax_v{\left\{p(i|\bm{\Psi}^l)\right\}}
\label{eq:sst_update_q}
\end{equation}

This process is similar to gating, excepts that the probability $p(i|\bm{\Psi}^l)$ is used instead of the Mahalanobis distance between the observation and the tracked source position \cite{leonard1991mobile}.
The weighting factor $p(i|\bm{\Psi}^l)$ modulates the update rate of the mean vector and covariance matrix:
\begin{equation}
\hat{\mathbf{x}}_i^{l|l} = (\hat{\mathbf{x}}')_i^{l|l-1} + p(i|\bm{\Psi}^l)\mathbf{K}_i^{l|l-1}(\bm{\lambda}_{\hat{v}(i)}^l - \mathbf{H}(\hat{\mathbf{x}}')_i^{l|l-1})
\label{eq:sst_update_x}
\end{equation}
\begin{equation}
\mathbf{P}_i^{l|l} = \mathbf{P}_i^{l|l-1} - p(i|\bm{\Psi}^l)\mathbf{K}_i^{l|l-1}\mathbf{H}\mathbf{P}_i^{l|l-1}
\label{eq:sst_update_P}
\end{equation}

When no potential source is clearly associated to the tracked source $i$, the probability $P(i|\bm{\Psi}^l)$ gets close to zero.
The mean of the updated state $\hat{\mathbf{x}}^{l|l}_i$ is then similar to the mean of the predicted state $\hat{\mathbf{x}}_i^{l|l-1}$.
Similarly, the updated covariance matrix $\mathbf{P}_i^{l|l}$ is similar to the predicted matrix $\mathbf{P}_i^{l|l-1}$, which grows after each prediction steps.
In other words, when the observations do not provide useful information, the tracked source moves according to its inertia while the exact position uncertainty grows.

\subsection{Direction Estimation (Step I)}
\label{subsec:sst_direction}

The updated states provide an estimation for each sound source direction.
This estimated direction $\bm{\phi}_i^l$ corresponds to the first moment of the posterior random variable $(\mathbf{d}')_i^{l|l}$:
\begin{equation}
\bm{\phi}_i^l = \iiint{\mathcal{N}((\mathbf{d}')_i^{l|l}|\mathbf{H}\hat{\mathbf{x}}_i^{l|l},\mathbf{H}\mathbf{P}_i^{l|l}\mathbf{H}^T)(\mathbf{d}')_i^{l|l}\,dx\,dy\,dz}
\end{equation}
which simplifies to
\begin{equation}
\bm{\phi}_i^l = \mathbf{H}\hat{\mathbf{x}}_i^{l|l}
\end{equation}

\section{Experimental Setup}
\label{sec:setup}

\begin{figure}[!ht]
  \centering
  \begin{tabular}{cc}
  \subfloat[OMA]{%
  \includegraphics[width=0.45\columnwidth]{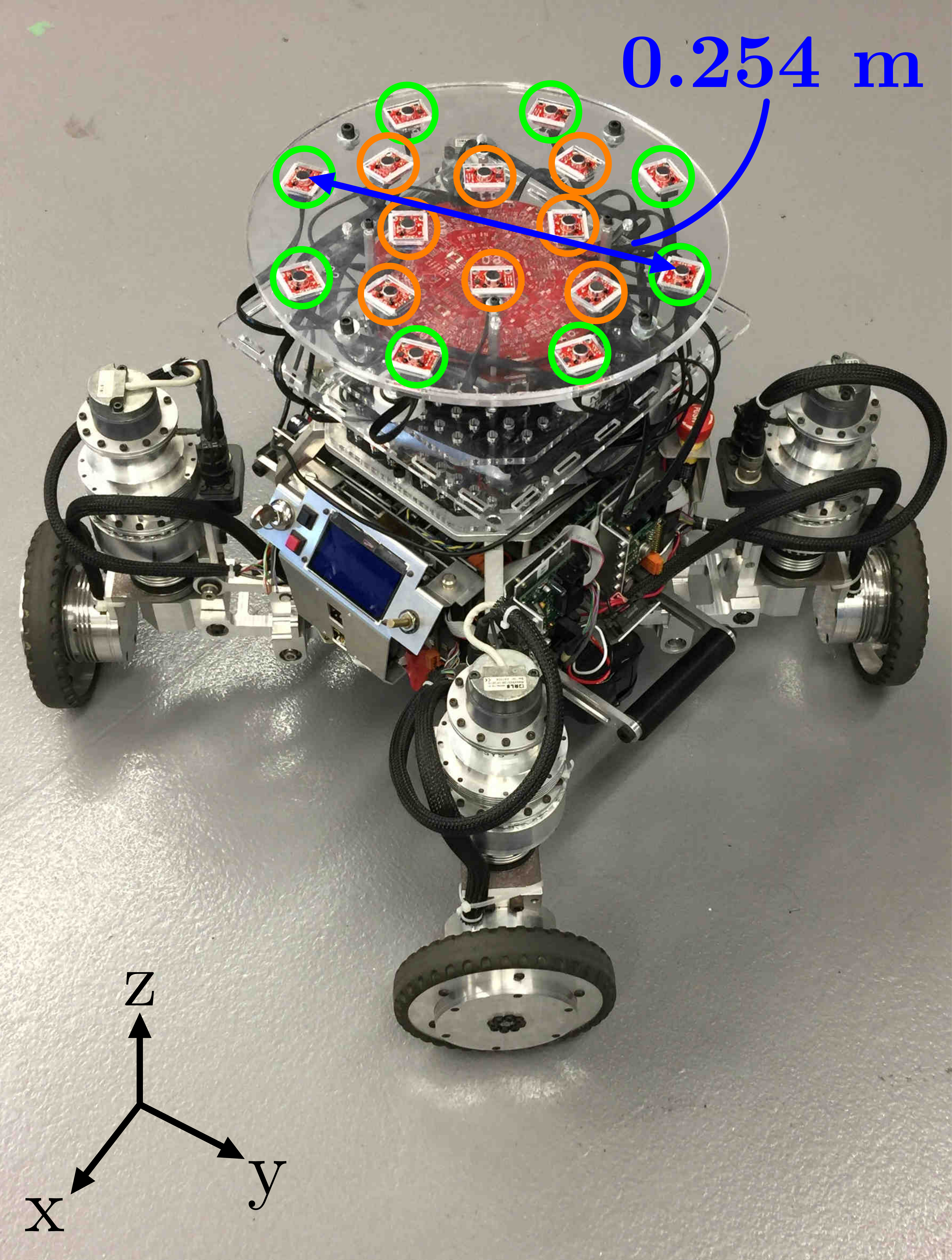}} 
  & 
  \subfloat[CMA]{%
  \includegraphics[width=0.45\columnwidth]{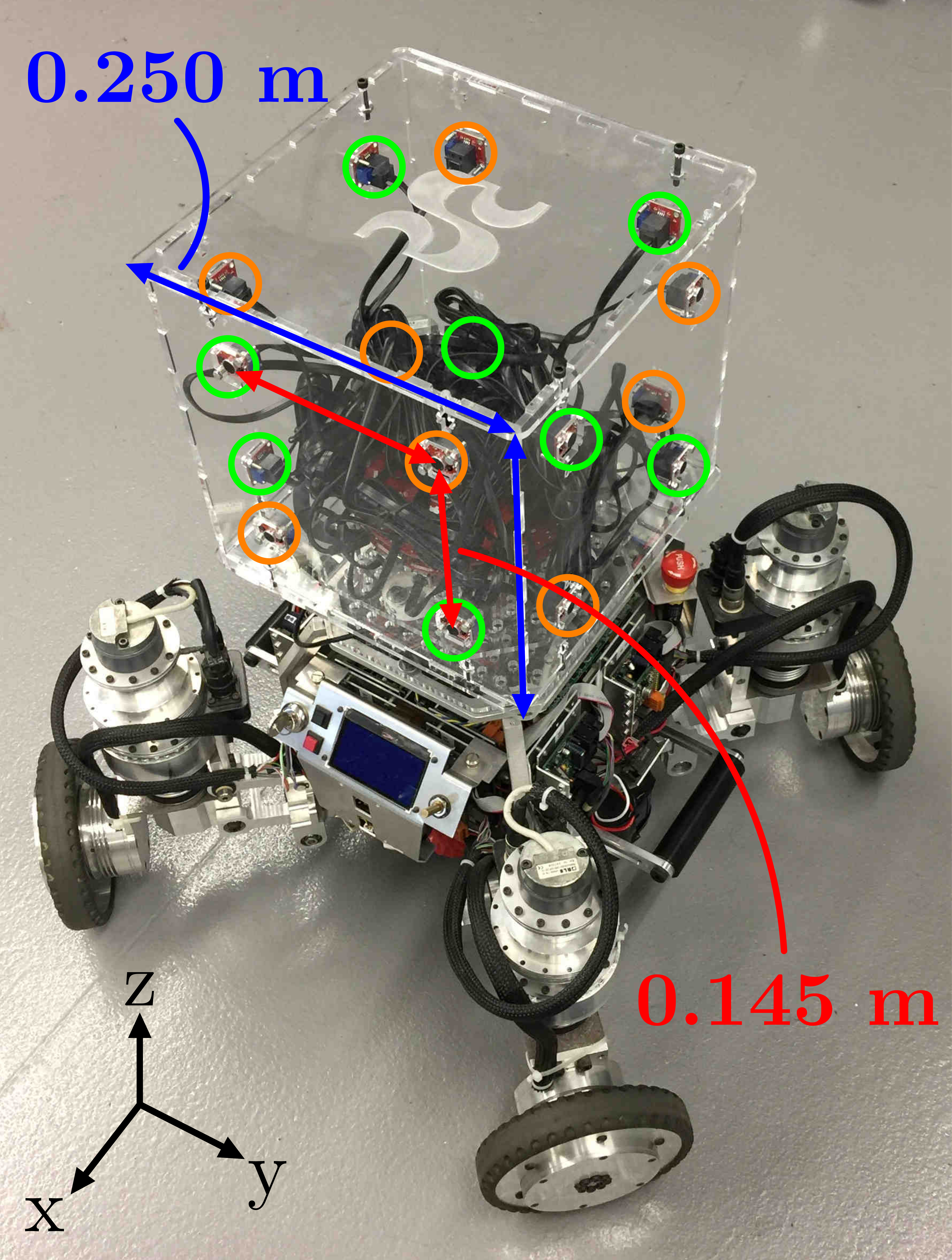}}  
  \end{tabular}
  \caption{8- and 16-microphone array configurations}
  \label{fig:setup_robots}
  \end{figure}

Experiments involve two 16-microphone array configurations installed on a mobile robot: 1) an opened microphone array (OMA) with microphones placed on a circular plane, and 2) a close microphone array (CMA) with microphones placed on a cubic structure.
Figure \ref{fig:setup_robots} shows these two configurations.
The microphones in green are used for experiments that involve only 8 microphones, while both green and orange microphones are considered for experiments with 16 microphones.
The OMA configuration consists of a circular surface with a diameter of $0.254$ m, while the CMA configuration involves a cubic structure with $0.250$ m edges, where microphones form many squares with $0.145$ m edges.
With the OMA configuration, all microphones have a direct path with the sound sources around the robot, while with the CMA sounds may be blocked by the cubic structure.

A diagonal covariance matrix models the uncertainty of microphone positions:
\begin{equation}
\bm{\Sigma}_p = \left[ 
\begin{array}{ccc}
(\sigma_{xx})_p & 0 & 0 \\
0 & (\sigma_{yy})_p & 0 \\
0 & 0 & (\sigma_{zz})_p \\
\end{array}
\right]
\end{equation}
and is set according to the microphone array configuration:
\begin{itemize}
\item For OMA, the variance $(\sigma_{zz})_p$ in the $z$-direction is set to zero, while the variances $(\sigma_{xx})_p$ and $(\sigma_{yy})_p$ in dimensions $x$ and $y$ are equal to $\sigma_{mic}^2$.
All microphones point to the ceiling, and therefore the direction unit vector $\mathbf{d}_p$ is oriented in the positive $z$-axis for all microphones.
\item For CMA, all microphones point outwards the cube, and therefore the direction unit vector $\mathbf{d}_p$ is oriented in the positive or negative $x$ and $y$-axes.
The variances $(\sigma_{xx})_p$, $(\sigma_{yy})_p$ and $(\sigma_{zz})_p$ are set respectively to $\sigma_{mic}^2$ if the microphone lies on a face in the plane that spans the corresponding $x$-, $y$- or $z$-axis, or are set to $0$ otherwise.
%% FG: Reviewer 1: Ajout d'une justification pour cette géométrie
This shape fits in the robot perimeter and leaves room for additional sensors on the top of the cube.
\end{itemize}

%% FG: Reviewer 1: Ajout des dimensions de la pièce
All experiments are performed in a 10\thinspace m x 10\thinspace m x 5\thinspace m room.
The reverberation level in the room (measured between 0\thinspace Hz and 24000\thinspace Hz) reaches $RT_{60} = 600$\thinspace msec, and there is no background noise.
The robot is positioned at the center of the room.

Table \ref{tab:setup_ssl_parameters} lists SRP-PHAT-HSDA parameters used for the experiments.
The frame sizes $N$ correspond to a duration of $16$\thinspace msec ($N/f_S$) to match speech stationarity. 
The hop size $\Delta N$ is set to provide a $50\%$ overlap between frames.
The refining level is set to $\mathcal{D} = 1$, which ensures a reliable integration and maintains the memory usage and execution time as low as possible during initialization.
We set $V = 4$ to detect up to four simultaneously active sound sources.
The parameter $\sigma^2_{mic}$ is chosen to model the uncertainty introduced by the membrane area of all microphones.
The minimum gain $G_{min}$ gets a value close to zero to generate the appropriate masks to limit the search space.
The mean and standard deviation $\mu_c$ and $\sigma_c$ are set to model the speed of sound in typical indoor and outdoor conditions.
The minimum probability threshold $C_{min}$ is chosen to ensure a good coverage of the random distribution of the TDOAs, while keeping the resolution high.
The scan space directivity $\mathbf{d}_0$ points to the ceiling to remove reflections from the floor, which corresponding values of $\alpha_0$ and $\beta_0$ to keep only the top hemisphere.
The number of links $U$ between the directions in the fine and coarse resolution grids is chosen to ensure an effective mapping while minimizing the number of scanned directions.
The resolution levels of the coarse ($\mathcal{L}'$) and fine ($\mathcal{L}''$) grids are set to minimize the number of lookups while maintaining a good resolution.
The parameters $\alpha_p$ and $\beta_p$ are chosen to ensure a smooth directivity response for all microphones that neglects signals coming from behind the microphones.

\begin{table}[!ht]
\centering
\caption{SRP-PHAT-HSDA parameters}
\vspace{0.1cm}
\renewcommand{\arraystretch}{1.2}
\begin{tabular}{|c|c|c|c|c|}
\cline{1-2}\cline{4-5}
Parameter & Value & & Parameter & Value \\
\cline{1-2}\cline{4-5}
$f_S$ (samples/sec) & $16000$ & & $\mu_c$ (m/s) & $343.0$ \\
$N$ (samples) & $256$ & & $\sigma_c$ (m/s) & $5.0$ \\
$\Delta N$ (samples) & $128$ & & $C_{min}$ & $0.3$ \\
$\mathcal{D}$ & $1$ & & $\mathbf{d}_0$ & $[\ 0\ 0\ 1\ ]$ \\
$U$ & $10$ & & $\mathcal{L}'$ & $2$ \\
$V$ & $4$ & & $\mathcal{L}''$ & $4$ \\
$\alpha_p\ (p > 0)$ & $80^{\circ}$ & & $\alpha_0$ & $80^{\circ}$ \\
$\beta_p\ (p > 0)$ & $100^{\circ}$ & & $\beta_0$ & $90^{\circ}$\\
$G_{min}$ & $0.1$ & & $\sigma^2_{mic}$ & $1\textrm{E}{-6}$ \\
\cline{1-2}\cline{4-5}
\end{tabular}
\label{tab:setup_ssl_parameters}
\end{table}

Table \ref{tab:setup_sst_parameters} lists the M3K parameters used for the experiments.
In these experiments, M3K can track up to $I_{max} = 10$ sources simultaneously.
The number of tracked sources can exceed the number of potential sources provided by SSL because the sources are active at different frames.
The energy level for active and inactive sound sources follows a Gaussian distribution with means $\mu_{\mathcal{A}}$ and $\mu_{\mathcal{I}}$, and variances $\sigma_{\mathcal{A}}^2$ and $\sigma_{\mathcal{I}}^2$.
The Bayesian Extension method has been used to automatically tune these parameters \cite{otsuka2011bayesian}.
However, for these experiments, setting these parameters empirically leads to good detection rates and tracking accuracy.
The parameters $(\sigma_R^2)_{prob}$ and $(\sigma_R^2)_{active}$ match standard deviations of approximately $3^{\circ}$ and $8^{\circ}$ on the grid, respectively.
The expression $(\sigma_R^2)_{prob}$ is smaller than $(\sigma_R^2)_{active}$ since the observations lie close to each other during the short probation interval.
The variance of the process noise $\sigma_Q^2$ is set to a value high enough to follow a source that changes direction, but low enough to preserve the source inertia.
Parameters $P_{false}$, $P_{new}$ and $P_{track}$ are chosen empirically: they have little impact on 
%% FM corrigea trakcing %% FG: OK
tracking performance as long as $P_{track}$ is greater than $P_{false}$ and $P_{new}$.
New source detection requires $\theta_{new}$ to be close to a probability of $1$, but small enough to detect new sources, and is therefore set empirically to $0.7$.
Probation corresponds to the time interval while a source is tracked but not displayed.
It is defined as $N_{prob}\Delta N/f_S$ sec, which lies within the duration range of a single phoneme (i.e., $40$ msec) \cite{anastasakos1995duration}. 
The number of inactive frames $N_{dead}\Delta N/f_S$ is set to match a duration of 1.2 sec, which is a reasonable silence period to consider a source is no longer active.
\begin{table}[!ht]
    \centering
    \caption{SST Module Parameters}
    \vspace{0.1cm}
    \renewcommand{\arraystretch}{1.2}
    \begin{tabular}{|c|c|c|c|c|}
    \cline{1-2}\cline{4-5}
    Parameters & Values & & Parameters & Values \\
    \cline{1-2}\cline{4-5}
    $I_{max}$ & $10$ & & $P_{false}$ & $0.1$ \\
    $\mu_{\mathcal{A}}$ & $0.20$ & & $P_{new}$ & $0.1$ \\
    $\sigma_{\mathcal{A}}^2$ & $0.0025$ & & $P_{track}$ & $0.8$ \\
    $\mu_{\mathcal{I}}$ & $0.10$ & & $\theta_{new}$ & 0.7 \\
    $\sigma_{\mathcal{I}}^2$ & $0.0025$ & & $N_{prob}$ & 5 \\
    $(\sigma^2_R)_{prob}$ & $0.0015$ & & $\theta_{prob}$ & 0.8 \\
    $(\sigma^2_R)_{active}$ & $0.0030$ & & $N_{dead}$ & $150$ \\
    $\sigma^2_Q$ & $0.000009$ & & $\theta_{dead}$ & $0.9$ \\
    \cline{1-2}\cline{4-5}
    \end{tabular}
    \label{tab:setup_sst_parameters}
  \end{table}

\section{Results}
\label{sec:results}

%% FM: Modifia %% FG: OK
In this section results are presented for SSL by comparing SRP-PHAT-HSDA with SRP-PHAT, and for SST by comparing M3K with SMC and using SRP-PHAT-HSDA. 
%SRP-PHAT-HSDA returns noisy potential sources direction of arrivals, which are then filtered with M3K to provide a smooth sources tracking over time.
%Results demonstrate the improved accuracy and reduced computational load of the proposed SRP-PHAT-HSDA and M3K methods when compared to the SRP-PHAT and SMC methods.

\subsection{Sound Source Localization}
\label{subsec:results_ssl}

Localization accuracy is computed with the proposed method to validate the MSW Automatic Calibration method.
The CPU usage for a single core between SRP-PHAT and the proposed SRP-PHAT-HSDA is measured to compare computational load on a Raspberry Pi 3 (equipped with a ARM Cortex-A53 Quad-Core processor clocked at 1.2 GHz).
Performance with multiple speech sources around the robot is also evaluated.
Both methods are implemented with the ODAS framework (Open embeddeD Audition System) in C language (without Neon/SSE optimization), which is available online as open source\footnote{http:/odas.io}.

\subsubsection{Localization Accuracy}
\label{subsubsec:results_ssl_accuracy}

\begin{figure}[!ht]
\includegraphics[width=\columnwidth]{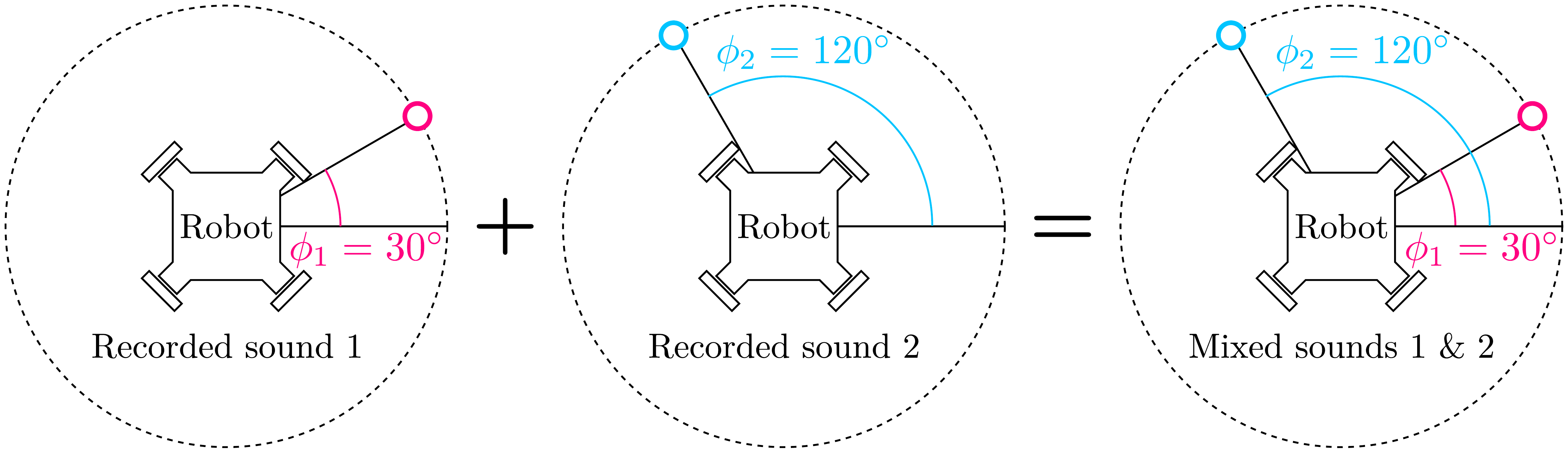}
\caption{Setup for SSL}
\label{fig:results_ssl_accuracy_setup}
\end{figure}

To evaluate localization accuracy, the robot is installed in the middle of a large room and a loudspeaker is positioned $r = 3$ m away at an height of $h = 1.15$ m referenced to origin of the microphone array, at azimuths of $\phi = 0^{\circ}$, $10^{\circ}$, $\dots$, $350^{\circ}$, for a total of $36$ positions.
For each position, a white noise signal plays in the loudspeaker for $2$ sec.
The recorded signals are then mixed to generate two active sources at different azimuths $\phi_1$ and $\phi_2$, as shown in Fig. \ref{fig:results_ssl_accuracy_setup}.
%% FM: Pourquoi avoir enlevé la figure 8? 
%% FG: Figure ajoutée
All permutations of $\{\phi_1,\phi_2\}$, where $\phi_1 \neq \phi_2$, ($\{0^{\circ},10^{\circ}\}$, $\{0^{\circ},20^{\circ}\}$, $\dots$, $\{350^{\circ},340^{\circ}\}$) are investigated, for a total of $1260$ permutations.
The expression $\bm{\gamma}(\phi)$ stands for the DoA in Cartesian coordinates that corresponds to azimuth $\phi$:
\begin{equation}
\bm{\gamma}(\phi) = \frac{1}{\sqrt{r^2+h^2}}
\left[
\begin{array}{ccc}
r\cos(\phi) & r\sin(\phi) & h
\end{array}
\right]
\end{equation}

The Root Mean Square Error (RMSE) corresponds to the smallest distance between potential source $v$ and both theoretical DoAs at angles $\phi_1$ and $\phi_2$:
\begin{equation}
\textrm{RMSE}_v = \min\{\|\bm{\lambda}_v - \bm{\gamma}(\phi_1)\|,\|\bm{\lambda}_v - \bm{\gamma}(\phi_2)\|\}
\end{equation}

The average RMSE for both potential sources $v=1$ and $v=2$ provide insight regarding  localization accuracy:
\begin{equation}
\textrm{RMSE} = \frac{\textrm{RMSE}_1+\textrm{RMSE}_2}{2}
\end{equation}

Table \ref{tab:results_ssl_rmse_16} present the RMSE results with 16 microphones, respectively.
The SRP-PHAT method for a single grid with refining level $\mathcal{L} = 1,2,3,4$ and omni-directional microphone model performs localization with fixed values for the size of MSW ($\Delta\tau_{pq} = 0, 1, 2, 3$), and is compared with the SRP-PHAT-HSDA method which automatically calibrates $\Delta\tau_{pq}$ (MSWAC) and uses the microphone directivity model.
Moreover, Hierarchical Search using two grids with different refining levels $\mathcal{L} = \{2,4\}$ (which means $\mathcal{L'} = 2$ and $\mathcal{L''} = 4$) is also examined, for fixed and automatically selected values of $\Delta\tau_{pq}$.
The RMSE values in bold stand for the smallest values across $\Delta\tau_{pq} = 0,1,2,3$ and MSWAC, for a given refining level.
\begin{table*}[!ht]
\centering
\caption{SSL RMSE with two active sound sources and 16 microphones}
\vspace{0.1cm}
\renewcommand{\arraystretch}{1.2}
\begin{tabular}{|c|c|c||c|c|c|c|c|}
\hline
\multirow{2}{*}{Configuration} & \multirow{2}{*}{$\Delta\tau_{pq}$} & \multirow{2}{*}{Microphone Directivity} & \multicolumn{5}{c|}{$\mathcal{L}$} \\
\cline{4-8}
 &  & & $1$ & $2$ & $3$ & $4$ & $\{2,4\}$\\
\hline
\hline
\multirow{5}{*}{OMA} &  $0$ & Omni-directional & $0.279$ & $\mathbf{0.140}$ & $\mathbf{0.081}$ & $\mathbf{0.064}$ & $\mathbf{0.064}$ \\
                     &  $1$ & Omni-directional & $\mathbf{0.231}$ & $0.186$ & $0.208$ & $0.222$ & $0.222$ \\
                     &  $2$ & Omni-directional & $0.311$ & $0.349$ & $0.375$ & $0.384$ & $0.385$ \\
                     &  $3$ & Omni-directional & $0.501$ & $0.559$ & $0.584$ & $0.596$ & $0.596$ \\  
\cline{2-8}                     
                     & MSWAC & Directive & $\mathbf{0.232}$ & $\mathbf{0.148}$ & $\mathbf{0.080}$ & $\mathbf{0.063}$ & $\mathbf{0.064}$ \\
\hline
\hline
\multirow{5}{*}{CMA} &  $0$ & Omni-directional & $0.528$ & $\mathbf{0.282}$ & $\mathbf{0.153}$ & $\mathbf{0.128}$ & $\mathbf{0.165}$ \\
                     &  $1$ & Omni-directional & $\mathbf{0.414}$ & $0.286$ & $0.272$ & $0.273$ & $0.285$ \\
                     &  $2$ & Omni-directional & $0.477$ & $0.448$ & $0.454$ & $0.456$ & $0.454$ \\
                     &  $3$ & Omni-directional & $0.599$ & $0.624$ & $0.649$ & $0.657$ & $0.649$ \\
\cline{2-8}                     
                     & MSWAC & Directive & $\mathbf{0.300}$ & $\mathbf{0.195}$ & $\mathbf{0.117}$ & $\mathbf{0.094}$ & $\mathbf{0.103}$ \\
\hline
\end{tabular}
\label{tab:results_ssl_rmse_16}
\end{table*}

Results suggest that when set to a constant, the ideal value of $\Delta\tau_{pq}$ changes according to the grid resolution level $\mathcal{L}$.
For the OMA configuration, setting automatically the value of $\Delta\tau_{pq}$ leads to the same accuracy as when the best constant value of $\Delta\tau_{pq}$ is chosen except for the case when $\mathcal{L}=1$ and $\mathcal{L}=2$, where the RMSE is slightly higher, yet almost equal.
This occurs when the size of the MSW is overestimated.
For the CMA configuration, the automatic calibration combined with the directivity model leads to better accuracy compared to constant values chosen empirically and the omni-directional model.
Finally, Hierarchical Search ($\mathcal{L} = \{2,4\}$) provides the same accuracy as the high resolution ($\mathcal{L} = 4$) grid with OMA, and increases the RMSE marginally with CMA.
It is possible to improve the accuracy of Hierarchical Search such that it matches the fixed grid with $\mathcal{L} = 4$ by increasing the parameters $U$, at the cost of increasing slightly the computational load.

\subsubsection{Computational Load}
\label{subsubsec:results_ssl_load}

\begin{figure}[!ht]
\centering
\includegraphics[width=\columnwidth]{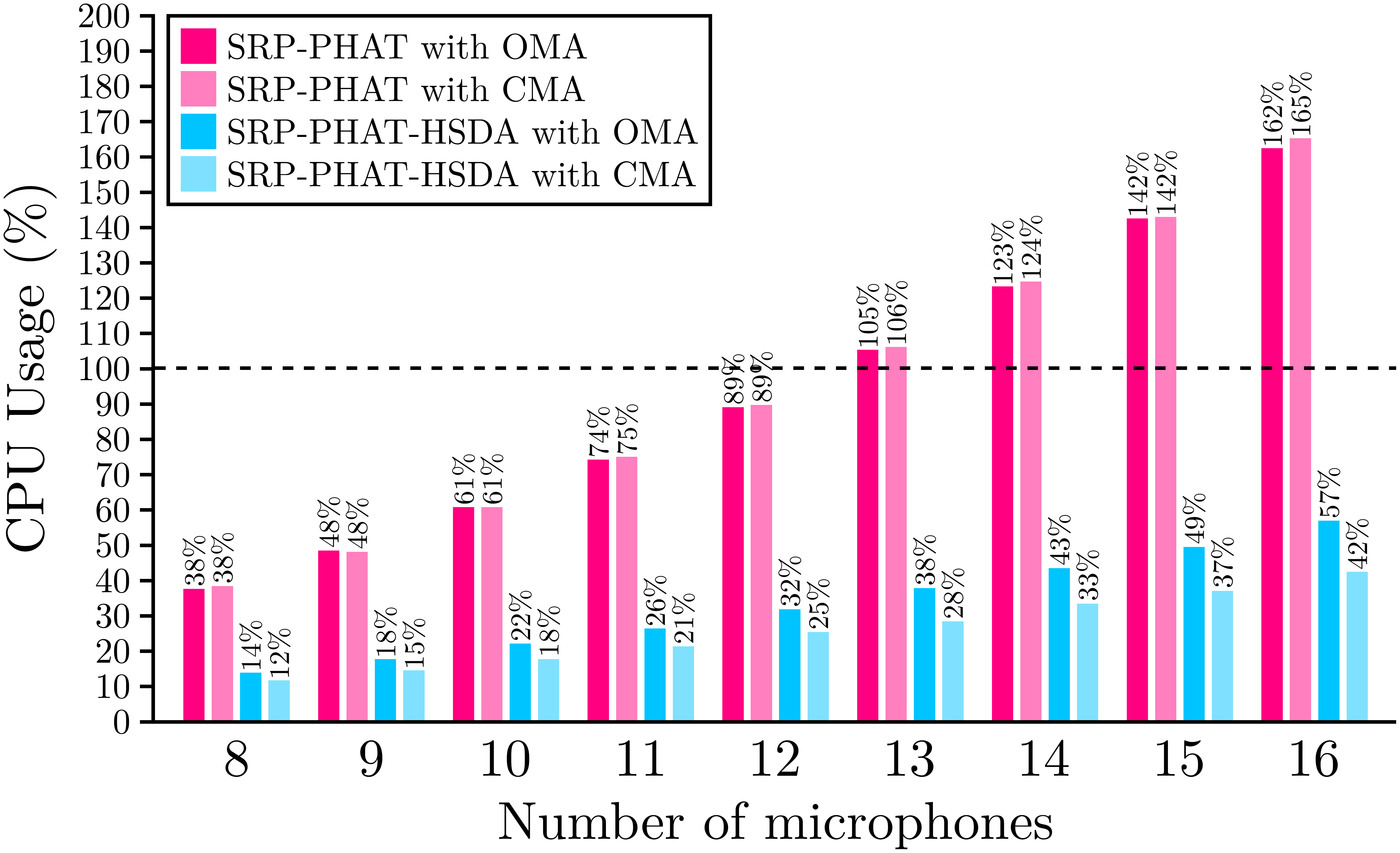}
\caption{CPU Usage on a single core on a Raspberry Pi 3}
\label{fig:results_ssl_cpu}
\end{figure}

Figure \ref{fig:results_ssl_cpu} shows the CPU usage for a single core on the Raspberry Pi 3.
Results demonstrate that SRP-PHAT-HSDA reduces considerably  computational load.
CPU usage reduces by a factor of four for the CMA configuration with $16$ microphones.
The SRP-PHAT-HSDA uses less computations with CMA than OMA, which is due to  the microphone directivity model that disregards the non significant pairs of microphones.
SRP-PHAT-HSDA is capable of online performance with $16$ microphones, while 
%% FM: Enleva "the previous ... method" et modifia légèrement la phrase
%% FG: OK
SRP-PHAT can not be processed online past $12$ microphones, as the CPU usage exceeds 100\% (106\% and 105\% with $13$ microphones for CMA and OMA, respectively).

\subsubsection{Localization of Multiple Speech Sources}
\label{subsubsec:results_ssl_speech}

\begin{figure}[!ht]
\centering
\subfloat[SRP-PHAT]{\includegraphics[width=\columnwidth]{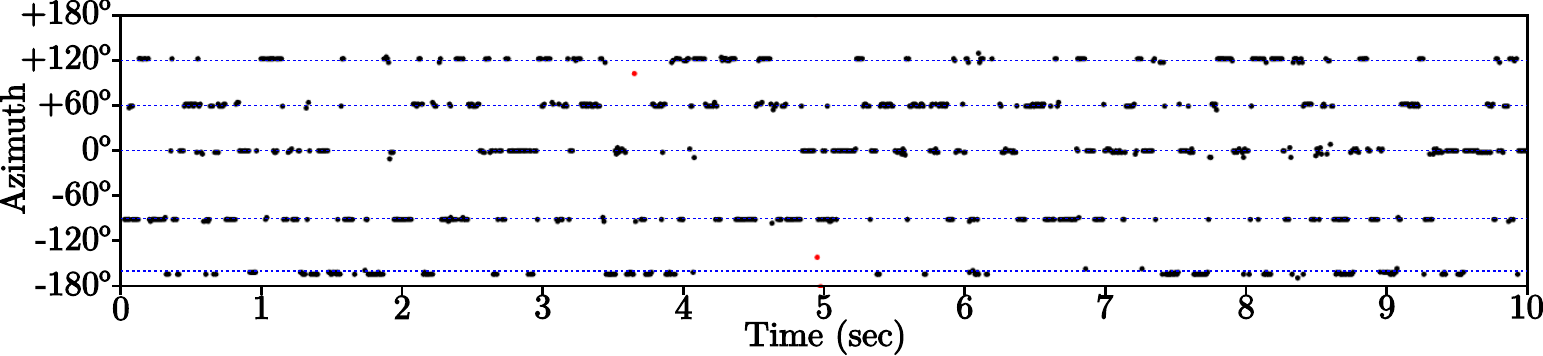}}\\
\subfloat[SRP-PHAT-HSDA]{\includegraphics[width=\columnwidth]{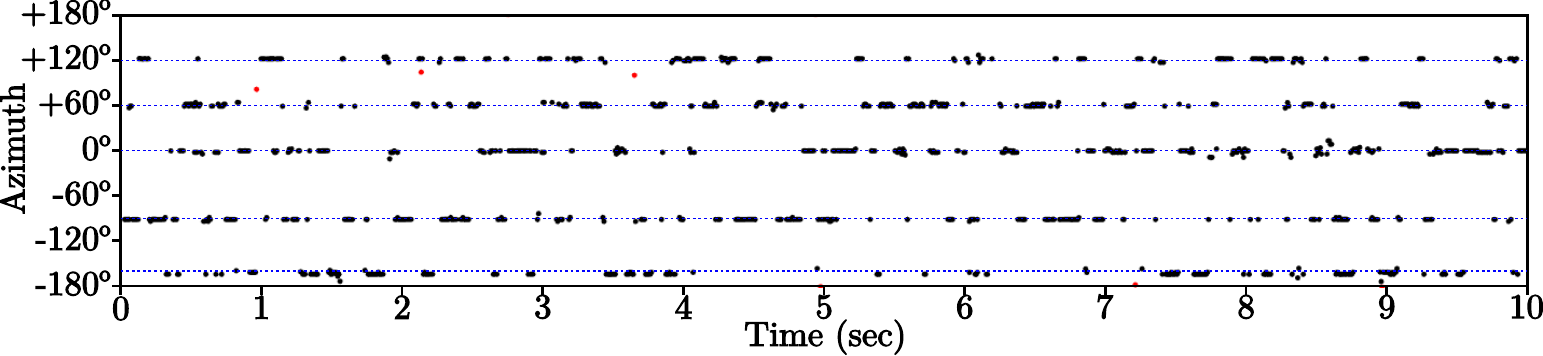}}\\
\caption{Azimuths obtained with the OMA configuration for five speech sources and 8 microphones (true azimuths are plotted in blue, and false detections are in red)}
\label{fig:results_ssl_multiple_oma_08}
\end{figure}

\begin{figure}[!ht]
\centering
\subfloat[SRP-PHAT]{\includegraphics[width=\columnwidth]{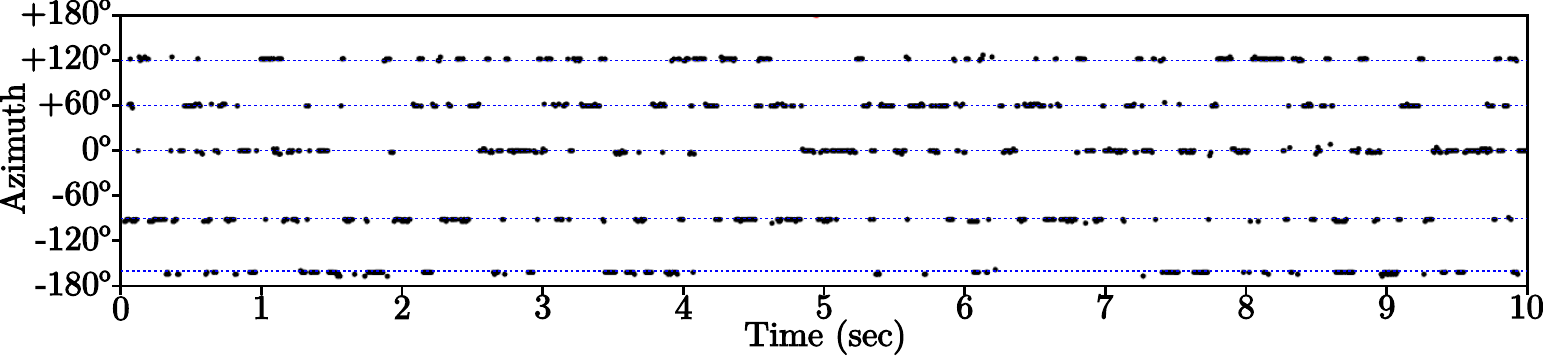}}\\
\subfloat[SRP-PHAT-HSDA]{\includegraphics[width=\columnwidth]{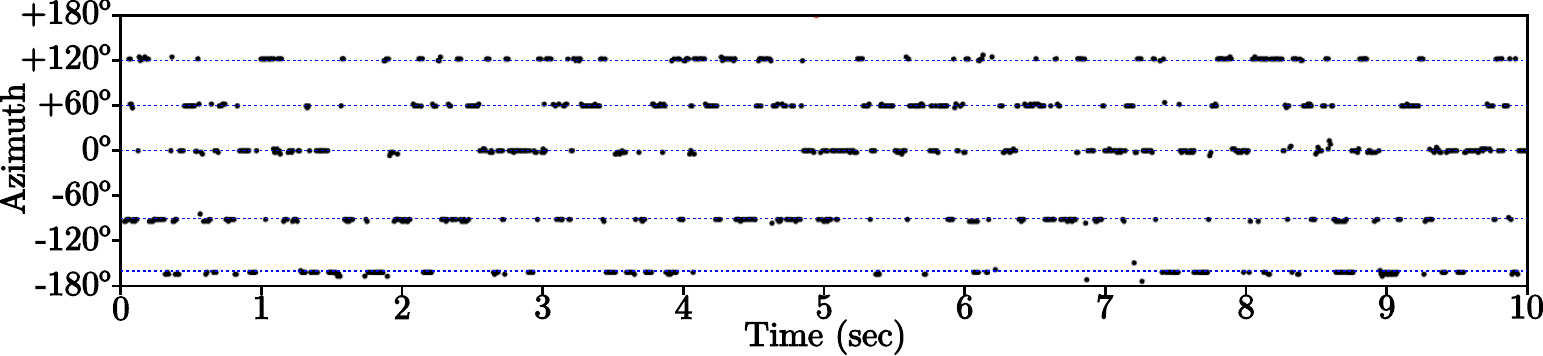}}\\
\caption{Azimuths obtained with the OMA configuration for five speech sources and 16 microphones (true azimuths are plotted in blue, and false detections are in red)}
\label{fig:results_ssl_multiple_oma_16}
\end{figure}

\begin{figure}[!ht]
\centering
\subfloat[SRP-PHAT]{\includegraphics[width=\columnwidth]{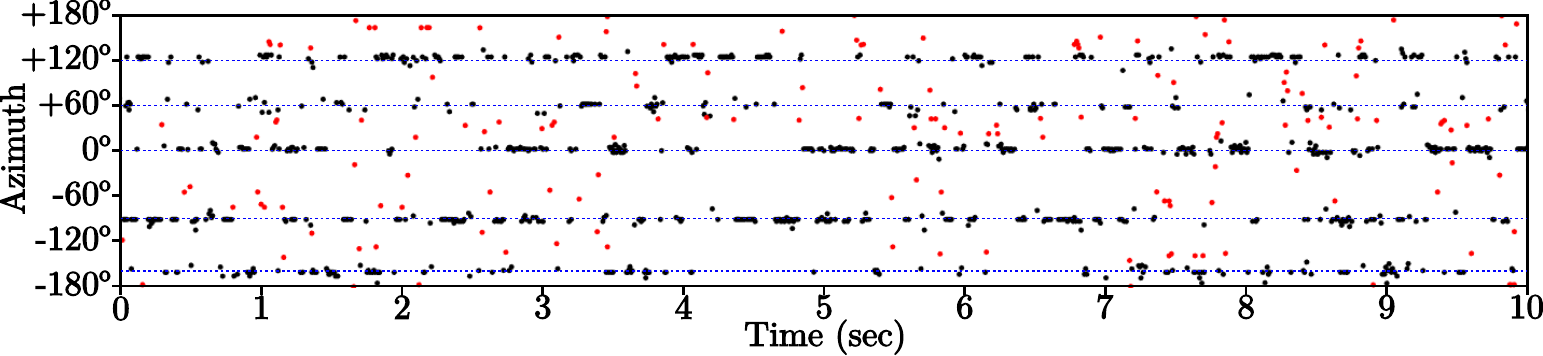}}\\
\subfloat[SRP-PHAT-HSDA]{\includegraphics[width=\columnwidth]{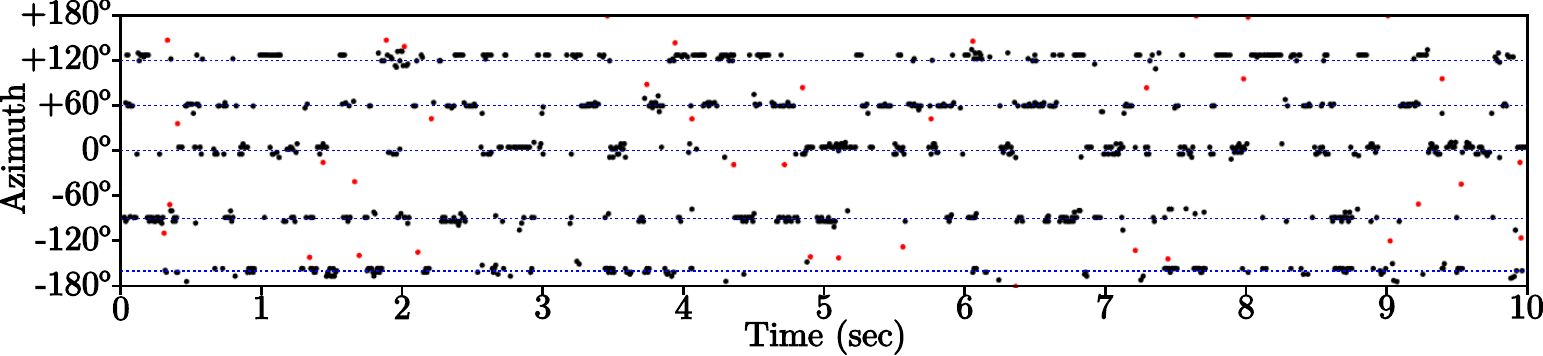}}\\
\caption{Azimuths obtained with the CMA configuration for five speech sources and 8 microphones (true azimuths are plotted in blue, and false detections are in red)}
\label{fig:results_ssl_multiple_cma_08}
\end{figure}

\begin{figure}[!ht]
\centering
\subfloat[SRP-PHAT]{\includegraphics[width=\columnwidth]{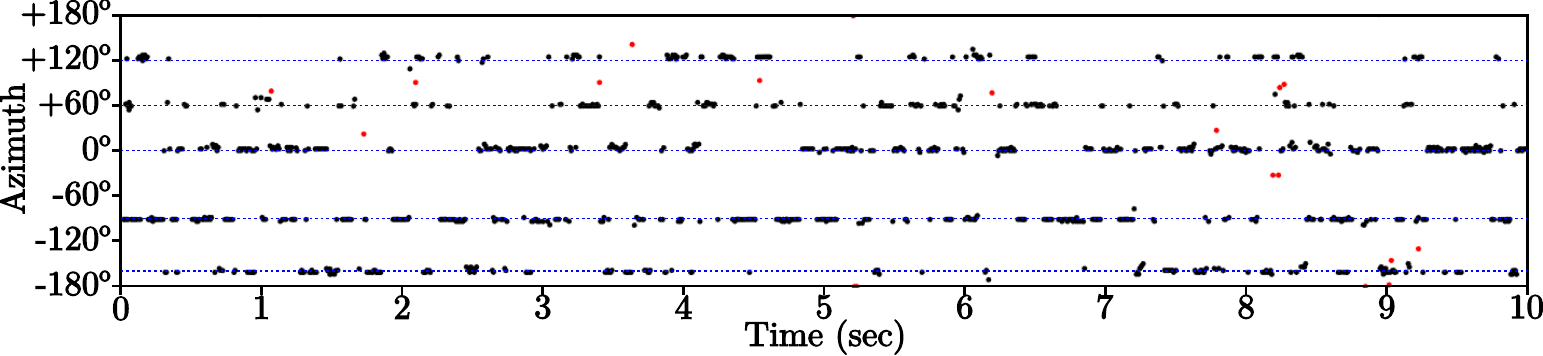}}\\
\subfloat[SRP-PHAT-HSDA]{\includegraphics[width=\columnwidth]{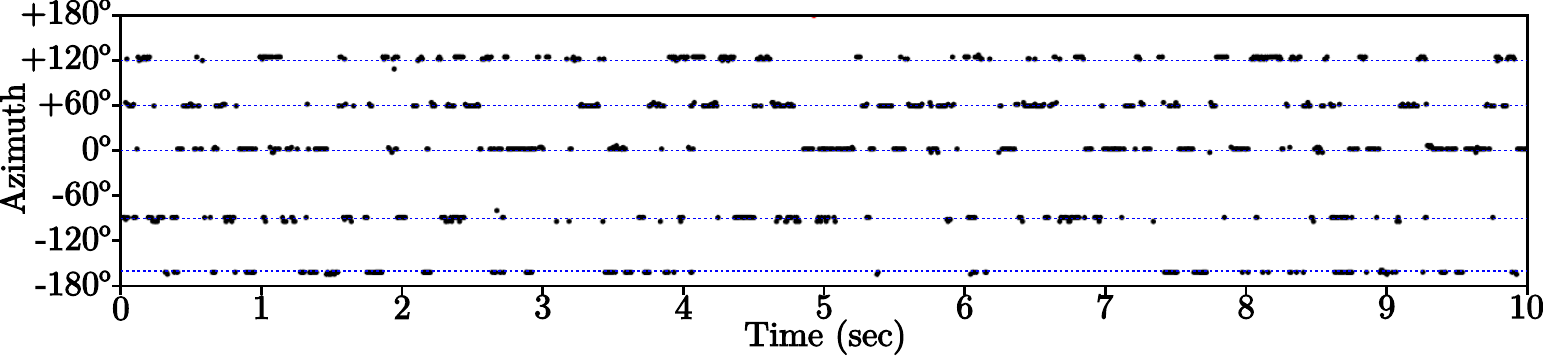}}\\
\caption{Azimuths obtained with the CMA configuration for five speech sources and 16 microphones (true azimuths are plotted in blue, and false detections are in red)}
\label{fig:results_ssl_multiple_cma_16}
\end{figure}

Five speech sources are played in loudspeakers located at azimuths $\phi_1 = 0^{\circ}$, $\phi_2 = 60^{\circ}$, $\phi_3 = 120^{\circ}$, $\phi_4 = 200^{\circ}$ and $\phi_5 = 270^{\circ}$.
Sources $1$, $3$ and $5$ are male speakers while sources $2$ and $4$ are female speakers.
They all speak continuously during $10$ seconds.
Since potential sources are significant when the energy level $\Lambda_v$ is high, only the most energetic potential sources (25\% of all potential sources which have the highest values of $\Lambda_v$) are plotted in Fig. \ref{fig:results_ssl_multiple_oma_08} and Fig. \ref{fig:results_ssl_multiple_oma_16}, for the OMA configuration with 8 and 16 microphones, respectively, and in Fig. \ref{fig:results_ssl_multiple_cma_08} and Fig. \ref{fig:results_ssl_multiple_cma_16}, for the CMA configuration with 8 and 16 microphones, respectively.

Results show that for the OMA configuration, both SRP-PHAT and SRP-PHAT-HSDA perform similarly: with both methods, there are few false detections.
The HSDA method therefore performs similarly with the standard PHAT for OMA configurations, but requires less computations.
On the other hand, the SRP-PHAT-HSDA outperforms the SRP-PHAT method with the CMA configuration: there are numerous false detections with SRP-PHAT for 8 microphones, while there are fewer with SRP-PHAT-HSDA, and there are many false detections with SRP-PHAT with 16 microphones, while there is only one at 5 sec with the proposed SRP-PHAT-HSDA.
This robustness is due to the Microphone Directivity model that exploits the direct path of sound propagation with closed microphone array shapes.
This suggests that the HSDA method should be used with CMA configurations to reduce false detections.

\subsection{Sound Source Tracking}
\label{subsec:results_sst}

The proposed M3K method is tested in a real environment on a mobile robot and compared with the SMC method.
The computational load of M3K is first measured on a low-cost embedded hardware, and compared to the load with SMC.
Tracking experiments with static and moving sound sources are then examined. 
Male and female speakers talking in English are used as sound sources.
Only the tracking of the azimut is presented, because for both methods elevation matched the height of the sound sources for all trials.

\subsubsection{Computational Load}
\label{subsubsec:results_sst_cpu}

\begin{figure}[!ht]
\centering
\includegraphics[width=\columnwidth]{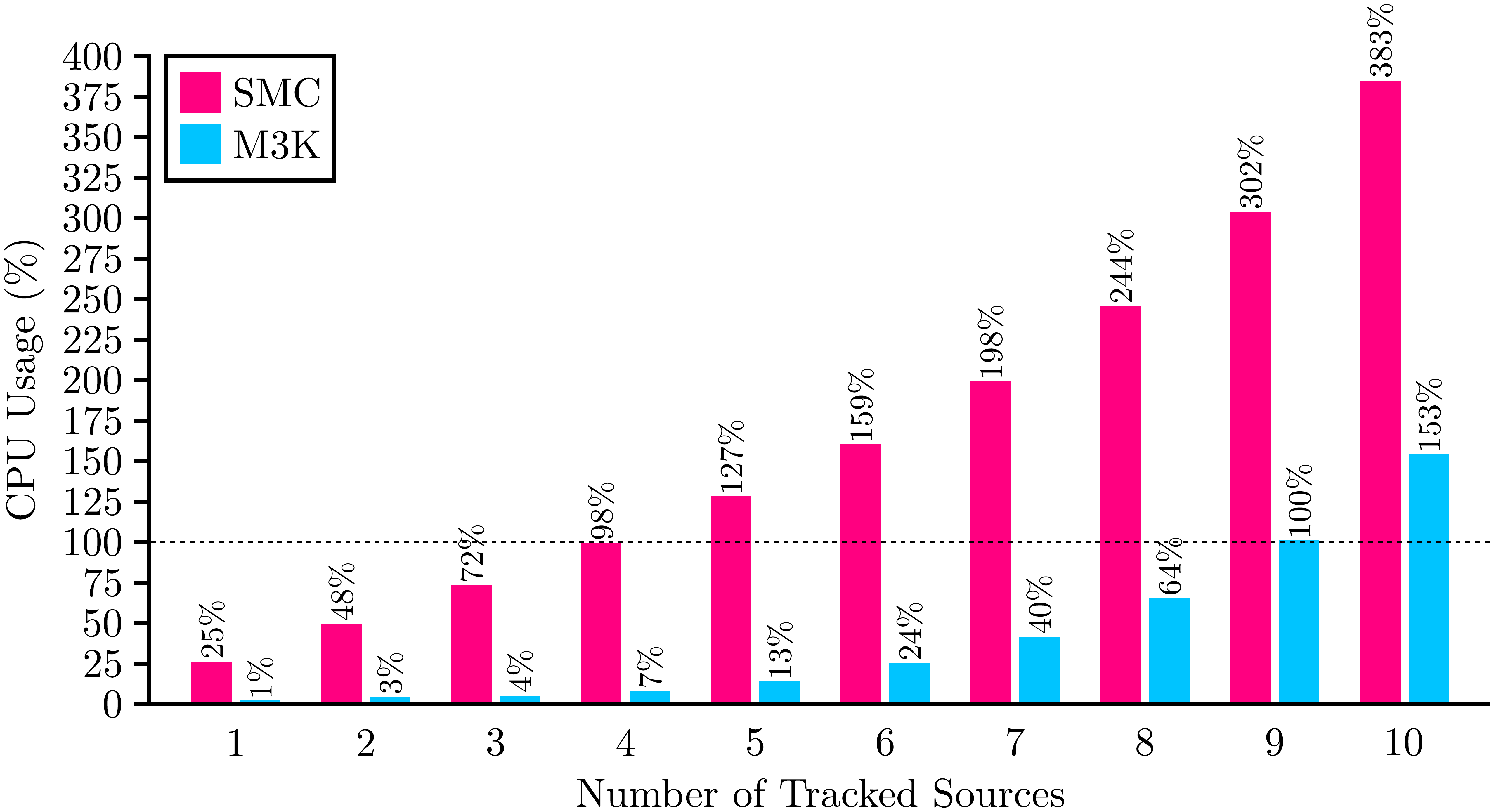}
\caption{CPU Usage of SMC and M3K on a Raspberry Pi 3}
\label{fig:results_sst_cpu}
\end{figure}

A Raspberry Pi 3 is used to compare CPU usage of M3K and SMC for a single core with C code, which is not optimized with Neon/SSE instructions.
To assess the performance in terms of the number of tracked sources, the maximum number of simultaneously tracked sources $I_{max}$ is set from $1$ to $10$, while there are $10$ active sound sources located at the following azimuths: $0^{\circ}$, $40^{\circ}$, $70^{\circ}$, $100^{\circ}$, $140^{\circ}$, $180^{\circ}$, $220^{\circ}$, $260^{\circ}$, $300^{\circ}$ and $330^{\circ}$.
Figure \ref{fig:results_sst_cpu} shows the CPU usage with both methods.
The SMC method allows online processing for up to four tracked sources, and then CPU usage exceeds $100\%$ (the usage increases to $127\%$ with five tracked sources).
The M3K reduces significantly the amount of computations, providing online processing with eight tracked sources (the usage is slightly above $100\%$ 
%% FM: Enleva (and rounded to $100\%$) %% FG: OK
with nine tracked sources).
When a single source is tracked, M3K uses $0.8\%$ of the CPU (rounded to $1\%$ on Fig. \ref{fig:results_sst_cpu}), while SMC reaches a CPU usage of $24\%$.
The M3K method is therefore up to $30$ times more effective in terms of computational load.
As the number of tracked sources increases, the number of assignments $(I+2)^V$ rises exponentially, which explains the high CPU usage for large values of $I$.

\subsubsection{Static Sound Sources}
\label{subsubsec:results_sst_static}

\begin{figure}[!ht]
\includegraphics[width=\columnwidth]{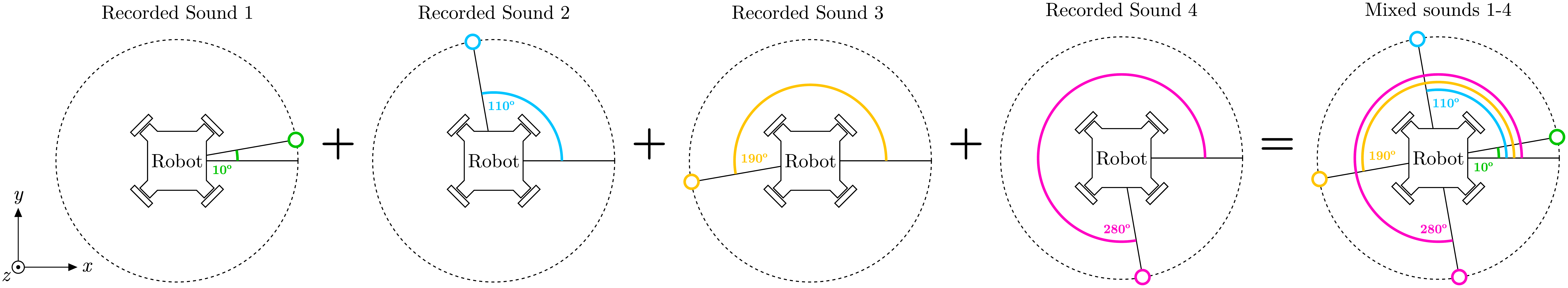}
\caption{Setup for SST}
\label{fig:results_sst_setup}
\end{figure}

\begin{figure*}[!ht]
\centering
\subfloat[Potential sources (OMA)]{\label{fig:results_static_04_oma_pots}\includegraphics[width=0.5\textwidth]{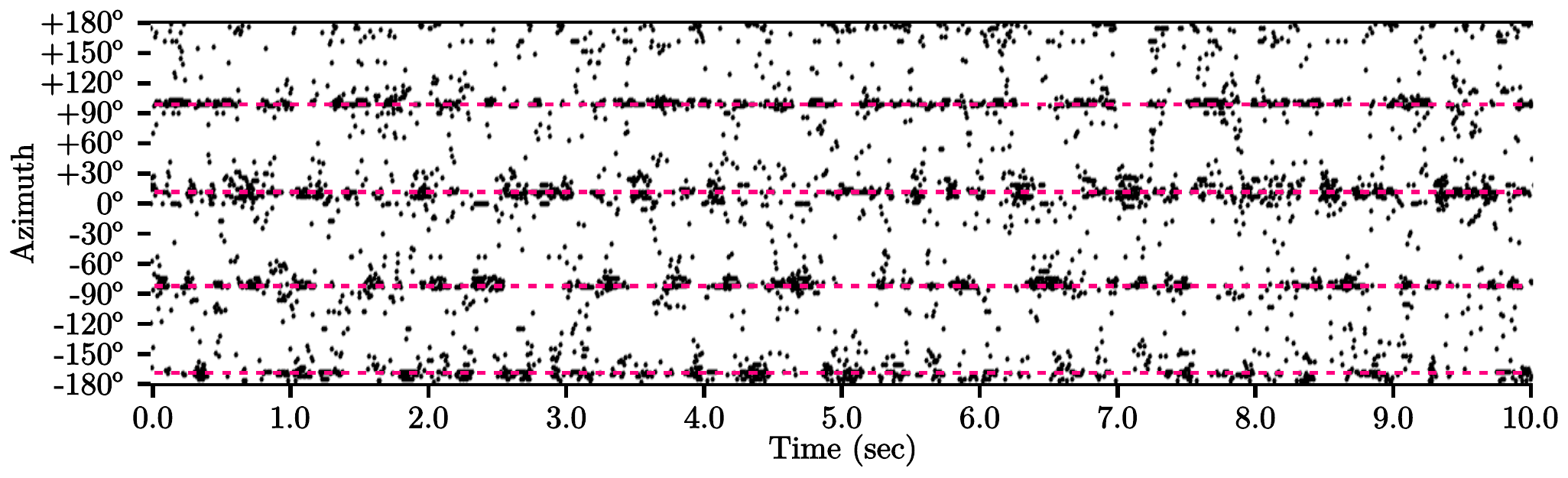}} 
\subfloat[Potential sources (CMA)]{\label{fig:results_sst_static_04_cma_pots}\includegraphics[width=0.5\textwidth]{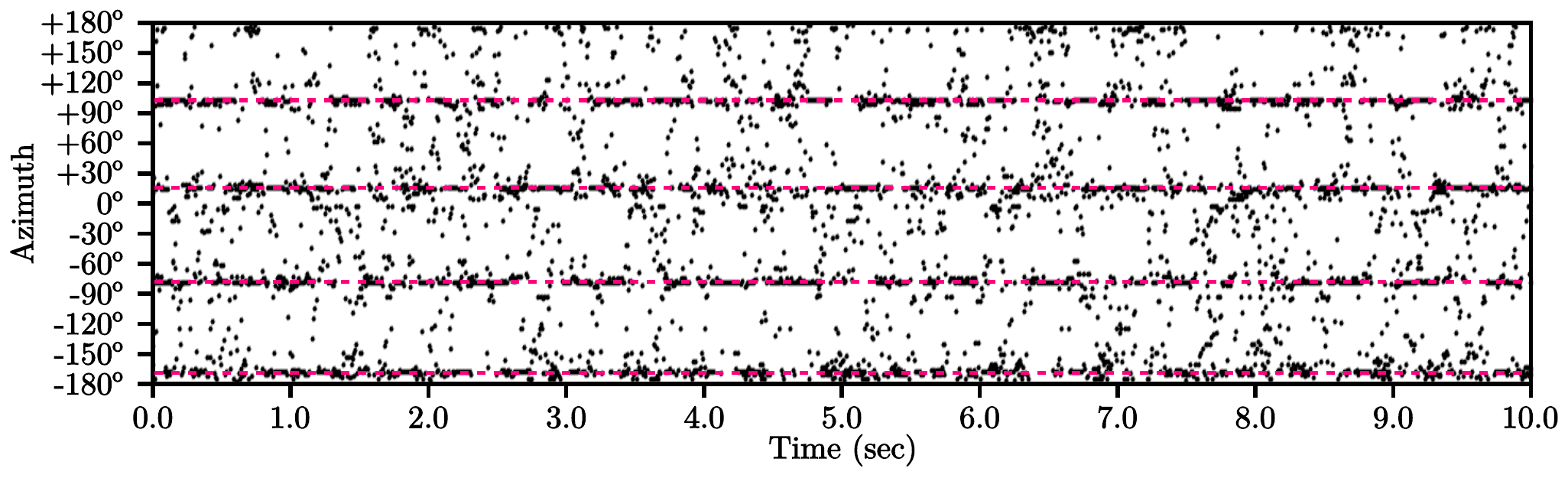}}\\
\subfloat[Tracked sources with SMC (OMA)]{\label{fig:results_sst_static_04_oma_tracked_particle}\includegraphics[width=0.5\textwidth]{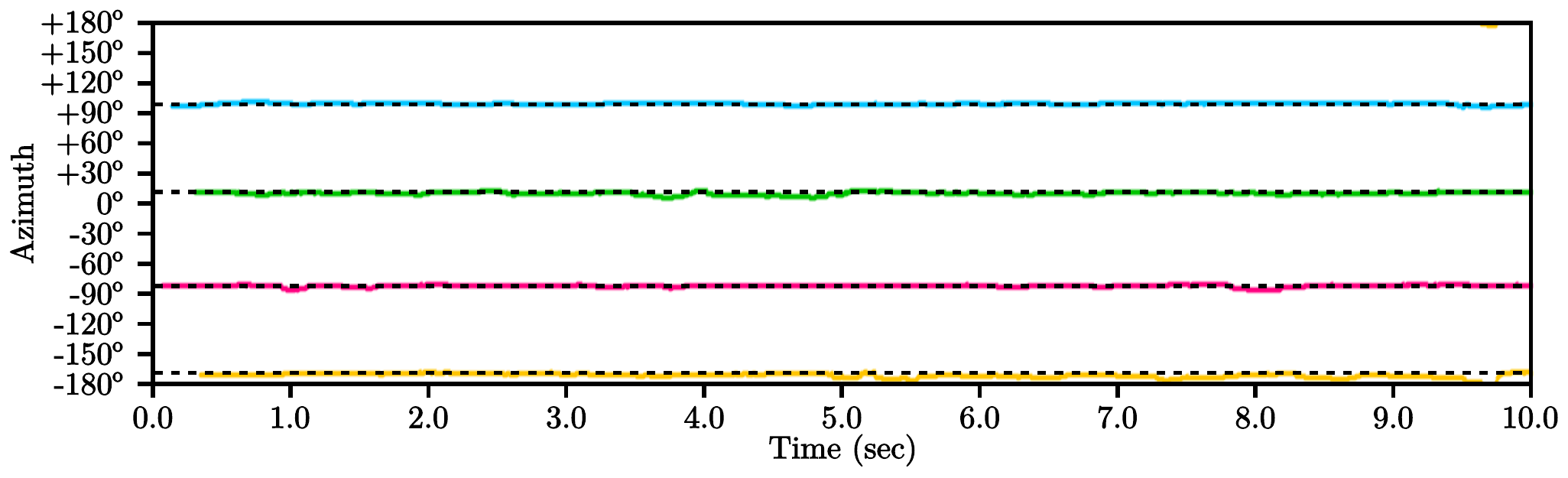}}
\subfloat[Tracked sources with SMC (CMA)]{\label{fig:results_sst_static_04_cma_tracked_particle}\includegraphics[width=0.5\textwidth]{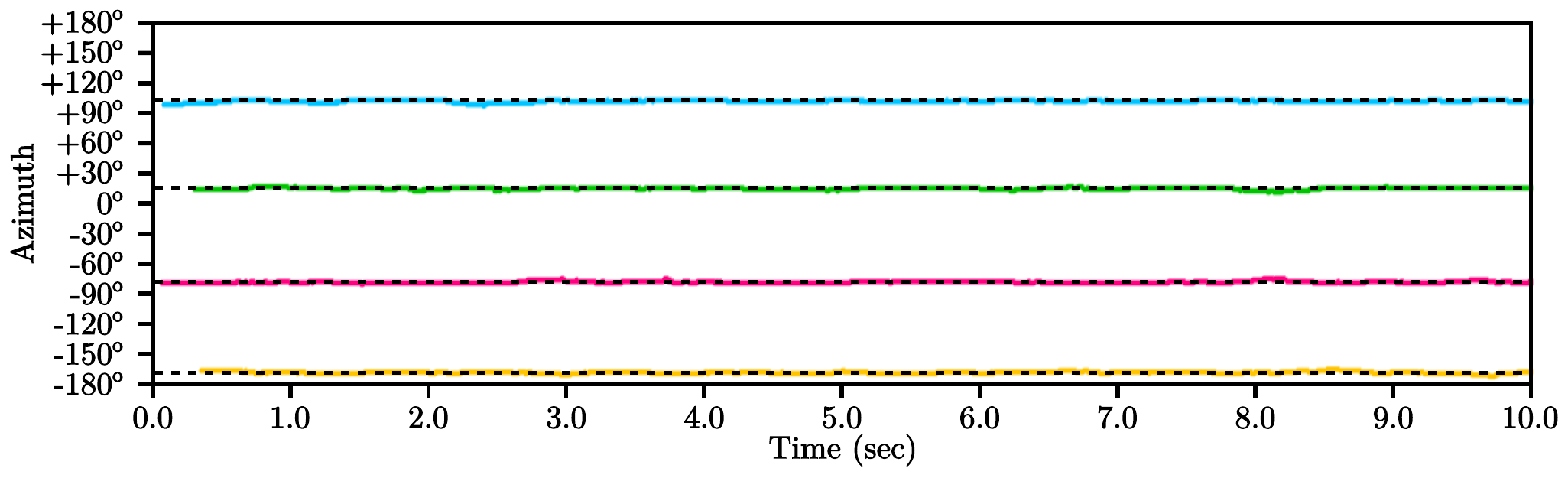}}\\
\subfloat[Tracked sources with M3K (OMA)]{\label{fig:results_sst_static_04_oma_tracked_kalman}\includegraphics[width=0.5\textwidth]{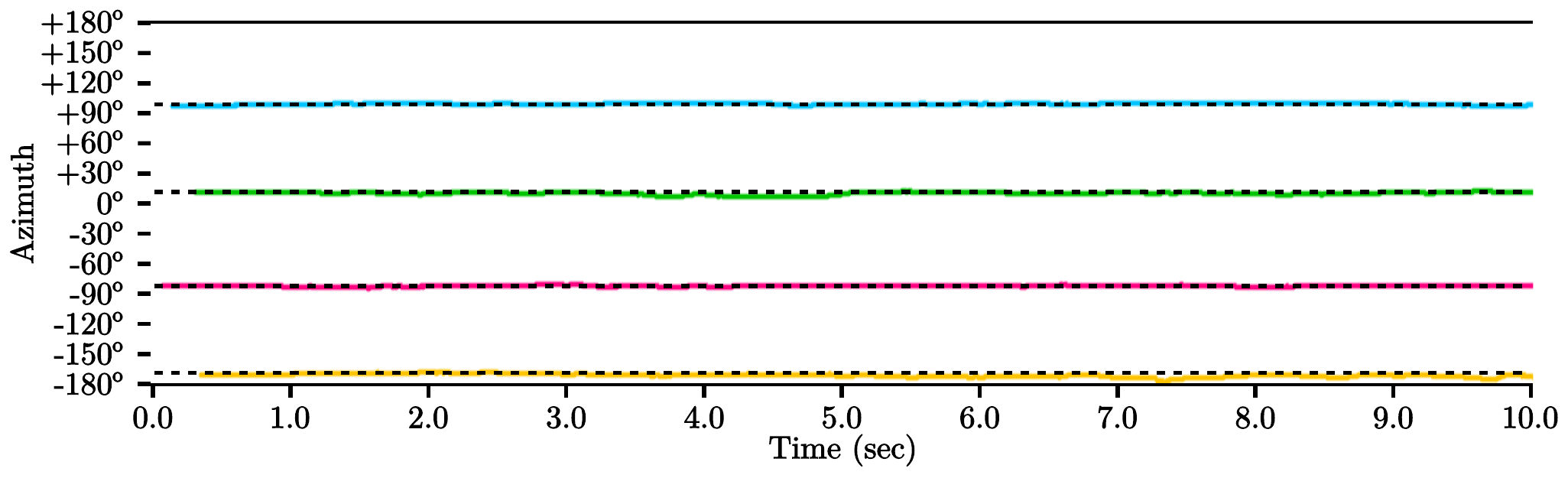}}
\subfloat[Tracked sources with M3K (CMA)]{\label{fig:results_sst_static_04_cma_tracked_kalman}\includegraphics[width=0.5\textwidth]{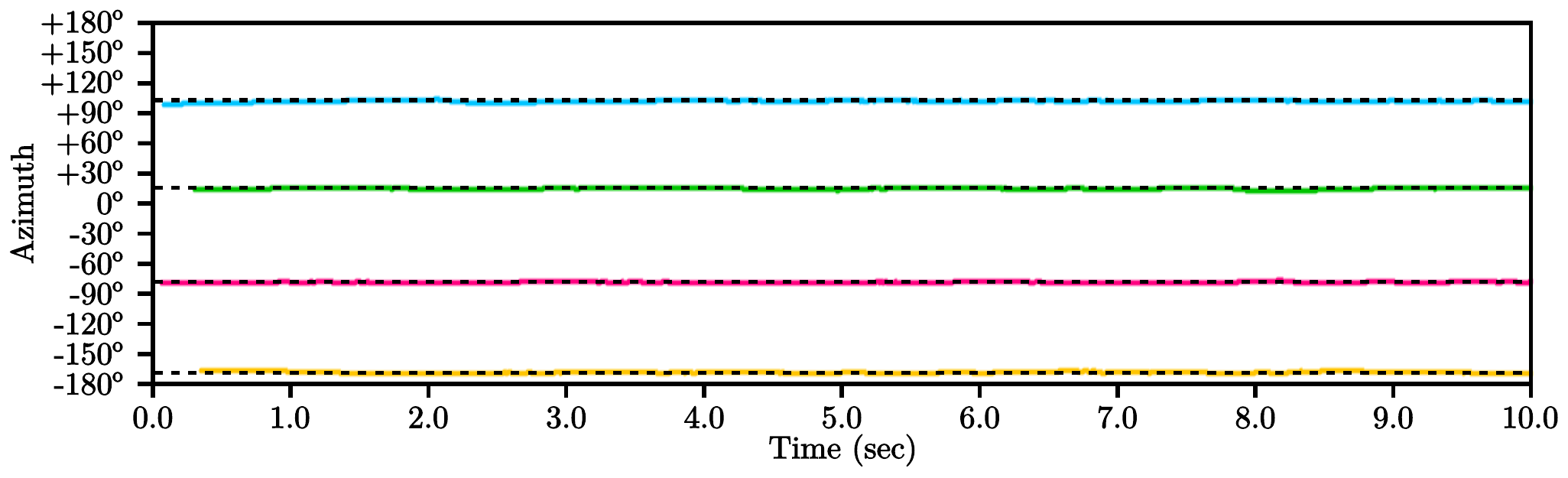}}
\caption{Azimuths of four static speech sources}
\label{fig:results_sst_static_04}
\end{figure*}

The first expriment conducted involves four static sources, to reproduce the test conditions of \cite{valin2007robust}. 
In this experiment, a loudspeaker is positioned $r=3$ m away from the robot, at azimuths of $10^{\circ}$, $100^{\circ}$, $190^{\circ}$, and $280^{\circ}$, and a height of $1.2$ m related to the robot microphone array origin, as shown in Fig. \ref{fig:results_sst_setup}.
%
%% FM: Pourquoi avoir enlevé la figure 7: Figure 7 illustrates how these signals for each position are recorded individually, and then mixed together.
%% FG: Tout comme dans le cas de SSL, je jugeais que c'était suffisant l'explication textuelle pour sauver de l'espace
Figure \ref{fig:results_sst_static_04} shows the potential sources generated by 
%% FM: Précisa la méthode utilisée pour SSL: the SSL module %% FG: OK
SRP-PHAT-HSDA and the corresponding tracked sources using SMC and M3K for the OMA and CMA configurations.
Tracked sources trajectories illustrate that M3K performs as well as SMC for both OMA and CMA configurations.

\begin{figure*}[!ht]
\centering
\subfloat[Potential sources (OMA)]{\label{fig:results_sst_static_09_oma_pots}\includegraphics[width=0.5\textwidth]{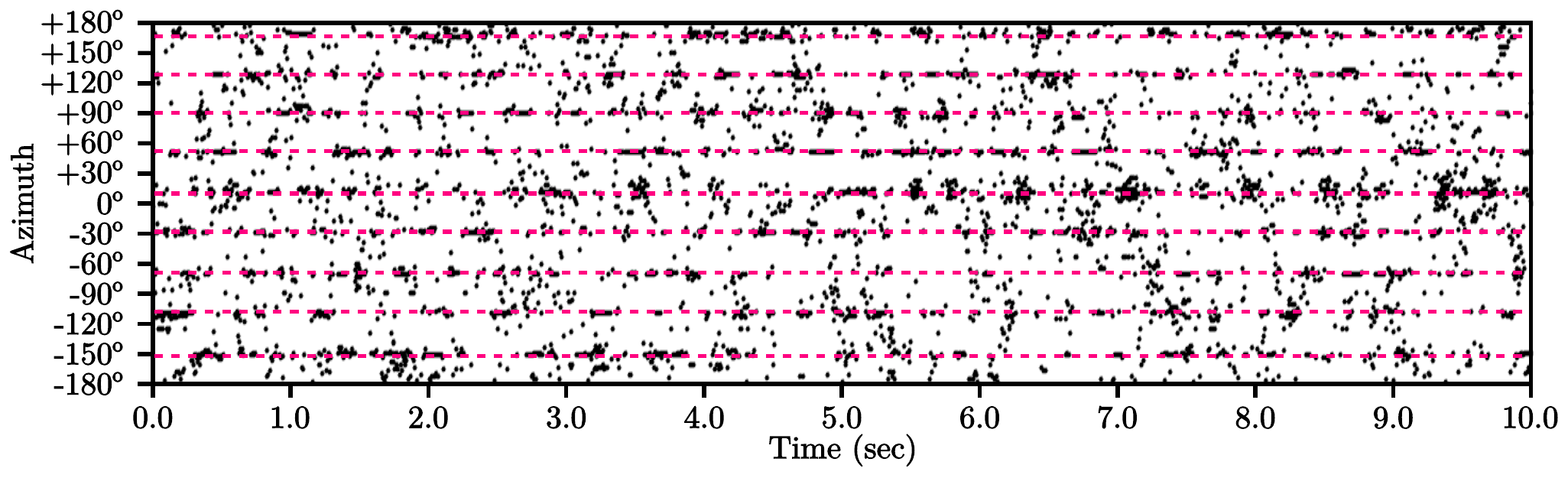}}
\subfloat[Potential sources (CMA)]{\label{fig:results_sst_static_09_cma_pots}\includegraphics[width=0.5\textwidth]{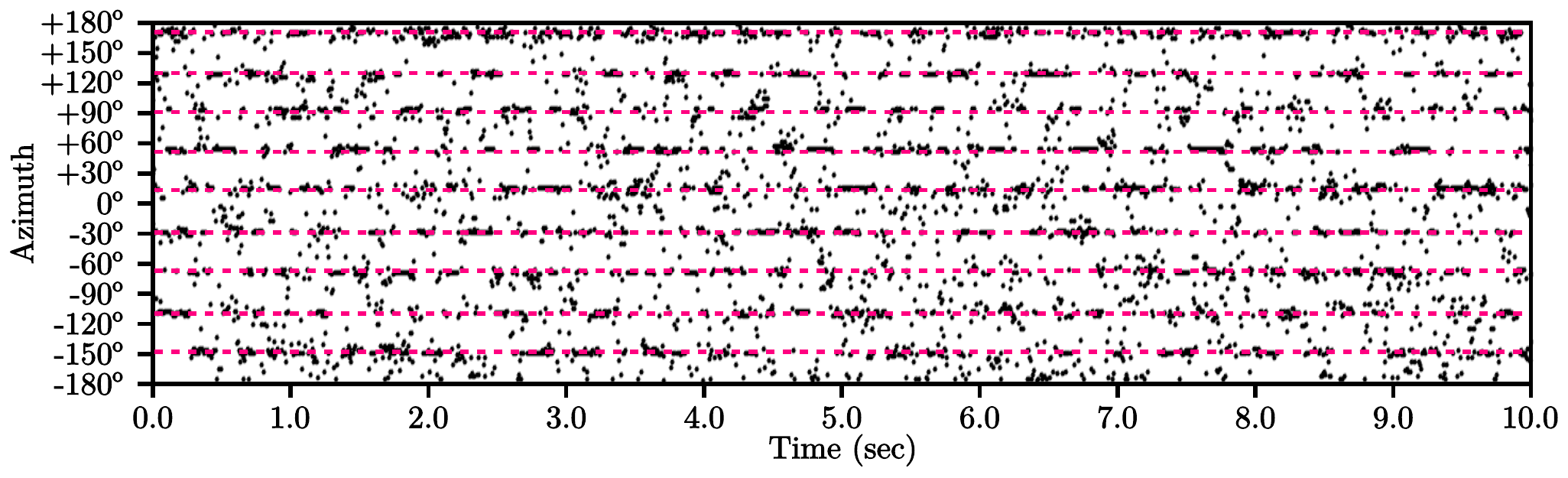}}\\
\subfloat[Tracked sources with SMC (OMA)]{\label{fig:results_sst_static_09_oma_tracked_particle}\includegraphics[width=0.5\textwidth]{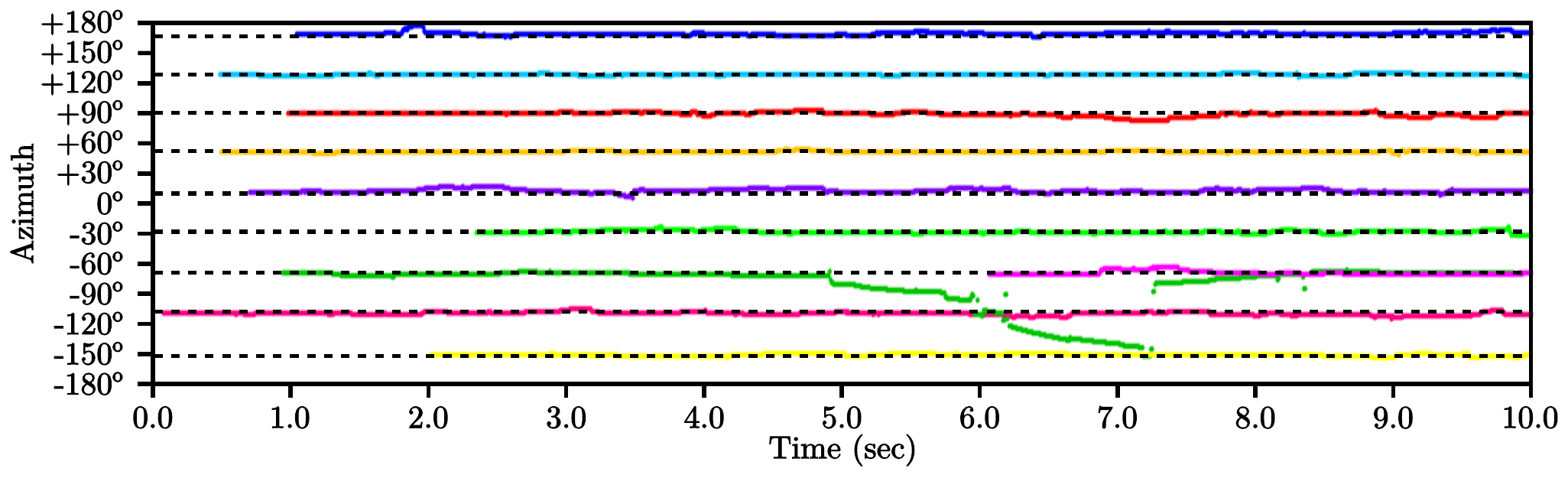}}
\subfloat[Tracked sources with SMC (CMA)]{\label{fig:results_sst_static_09_cma_tracked_particle}\includegraphics[width=0.5\textwidth]{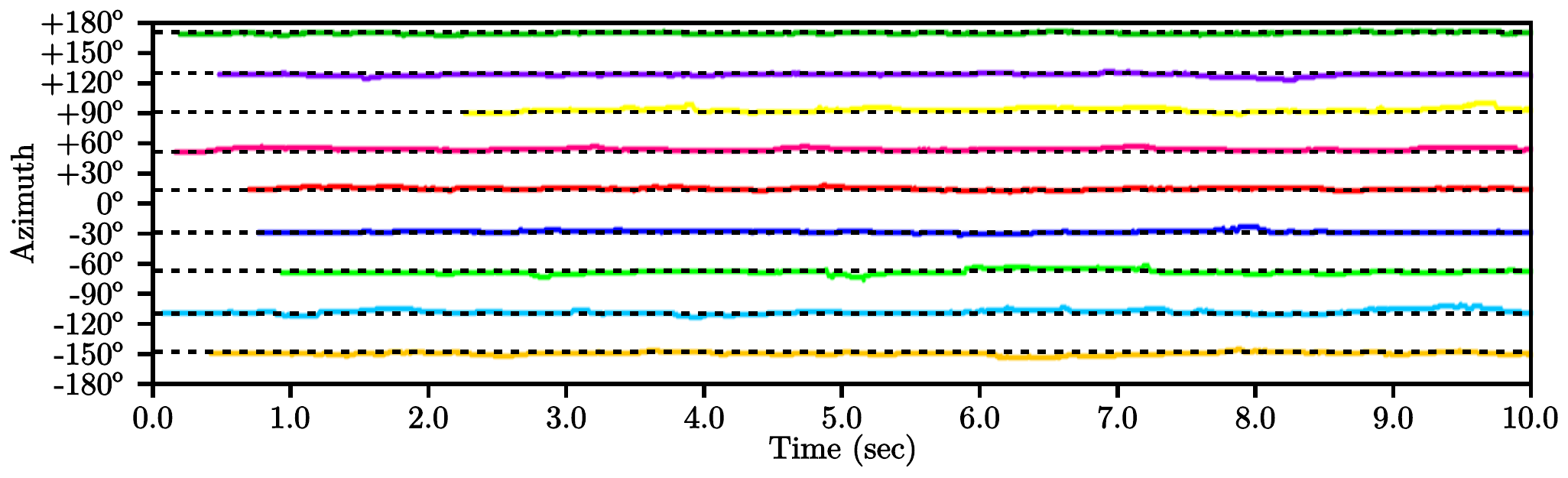}}\\
\subfloat[Tracked sources with M3K (OMA)]{\label{fig:results_sst_static_09_oma_tracked_kalman}\includegraphics[width=0.5\textwidth]{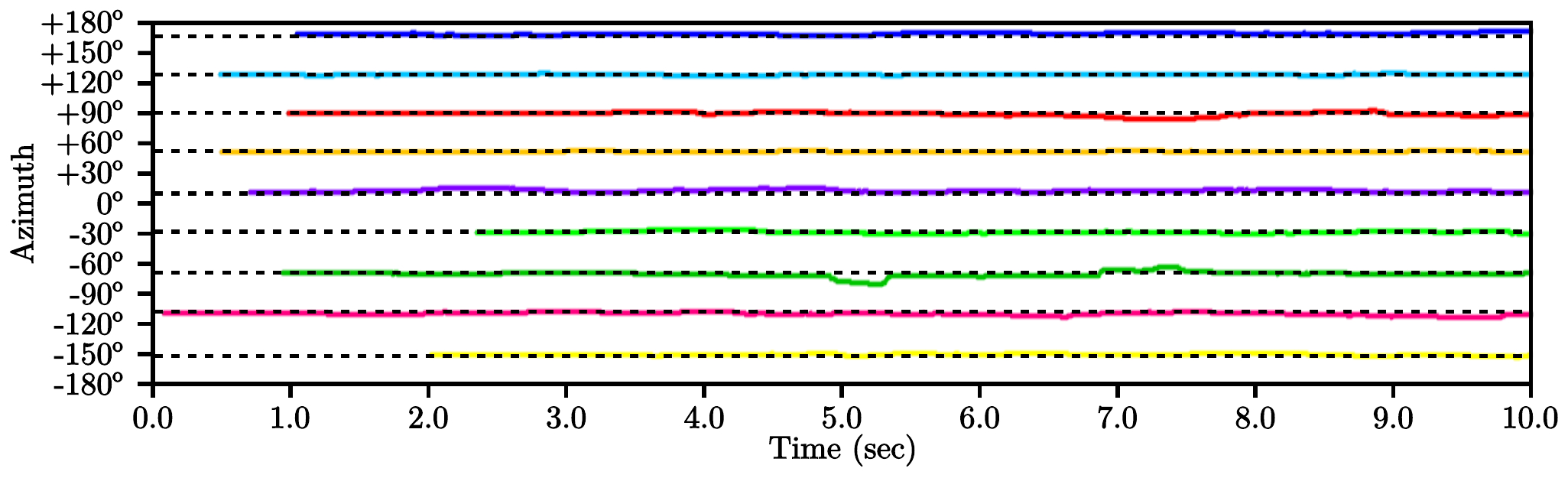}}
\subfloat[Tracked sources with M3K (CMA)]{\label{fig:results_sst_static_09_cma_tracked_kalman}\includegraphics[width=0.5\textwidth]{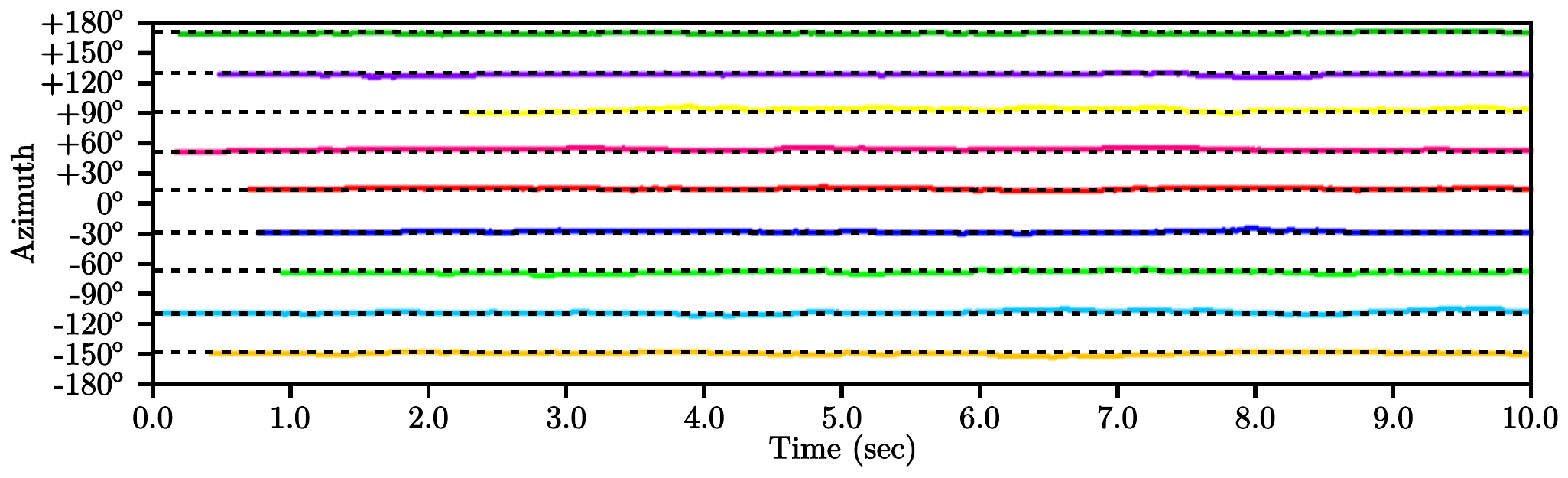}}
\caption{Azimuths of nine static sources}
\label{fig:results_sst_static_09}
\end{figure*}

We then increased the number of tracked sound sources to nine, reaching the limit of acceptable tracking performance. 
These nine static speech sources, at azimuths of $10^{\circ}$, $50^{\circ}$, $90^{\circ}$, $130^{\circ}$, $170^{\circ}$, $210^{\circ}$, $250^{\circ}$, $290^{\circ}$ and $330^{\circ}$.
Figure \ref{fig:results_sst_static_09} shows the potential sources and the tracking results.
The high number of sources makes detection and tracking more challenging for two reasons: 1) sources are closer to each other, which makes differentiation difficult between two static sources; and 2) observations sparsity increases, which means each sound source gets assigned fewer potential sources from the SSL module as they are distributed over all active sources.
SMC performs tracking accurately with the CMA configuration, but there is an error in tracking with the OMA that starts at 5 sec.
The source drifts, and another source is created to track the source at $290^{\circ}$.
M3K, which assumes all sources have a null or constant velocity, models more accurately the static sound source dynamics with both configurations.
With nine sources, both M3K and SMC also take more time to detect all sources (up to 1 sec of latency) due to observations sparsity.

\subsubsection{Moving Sound Sources}
\label{subsubsec:results_sst_moving}

\begin{figure}[!ht]
\centering
\subfloat[Crossing sources]
{\includegraphics[width=0.45\columnwidth]{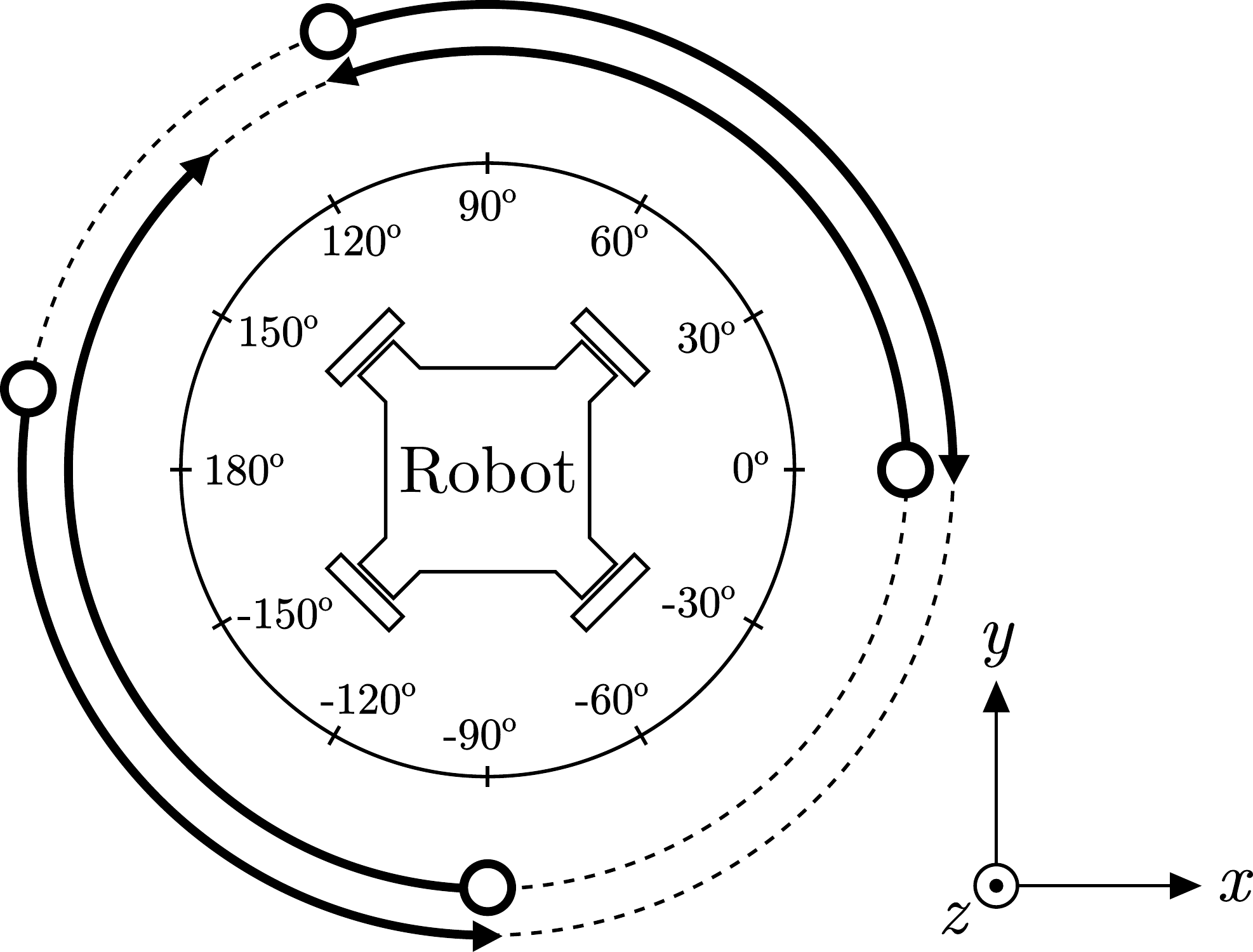}\label{fig:results_sst_moving_trajectories_crossing}}
\subfloat[Parallel sources]
{\includegraphics[width=0.45\columnwidth]{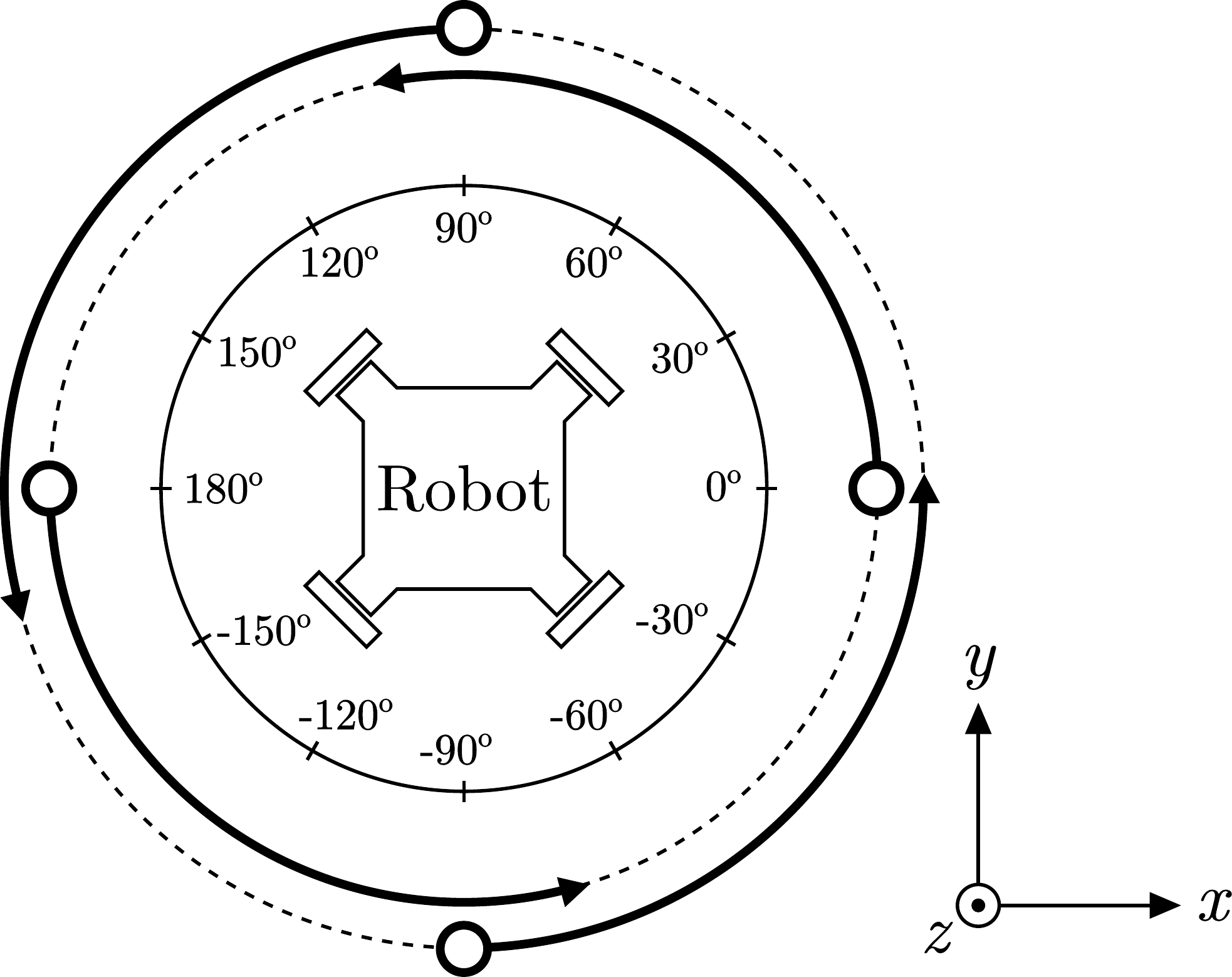}\label{fig:results_sst_moving_trajectories_parallel}}
\caption{Trajectories of four moving sources}
\label{fig:results_sst_moving_trajectories}
\end{figure}

Two tests conditions are examined.
The first involves four moving sources crossing, tested with OMA and CMA.
For this experiment, a male speaker performs four trajectories at $r=3$ m away from the robot, starting from different positions as illustrated by Fig. \ref{fig:results_sst_moving_trajectories_crossing}. 
These recordings are combined to generate the case of four simultaneous moving sources that cross each other.
The second test 
%% FM: Enleva le s à conditions %% FG: OK
condition involves moving sound sources following each other.
A male speaker performed four trajectories at $r=3$ m away from the robot, which are mixed together such that sources are following each other, as shown in Fig. \ref{fig:results_sst_moving_trajectories_parallel}.

\begin{figure*}[!ht]
\centering
\subfloat[Potential sources (OMA)]{\label{fig:results_sst_moving_crossing_oma_pots}\includegraphics[width=0.5\textwidth]{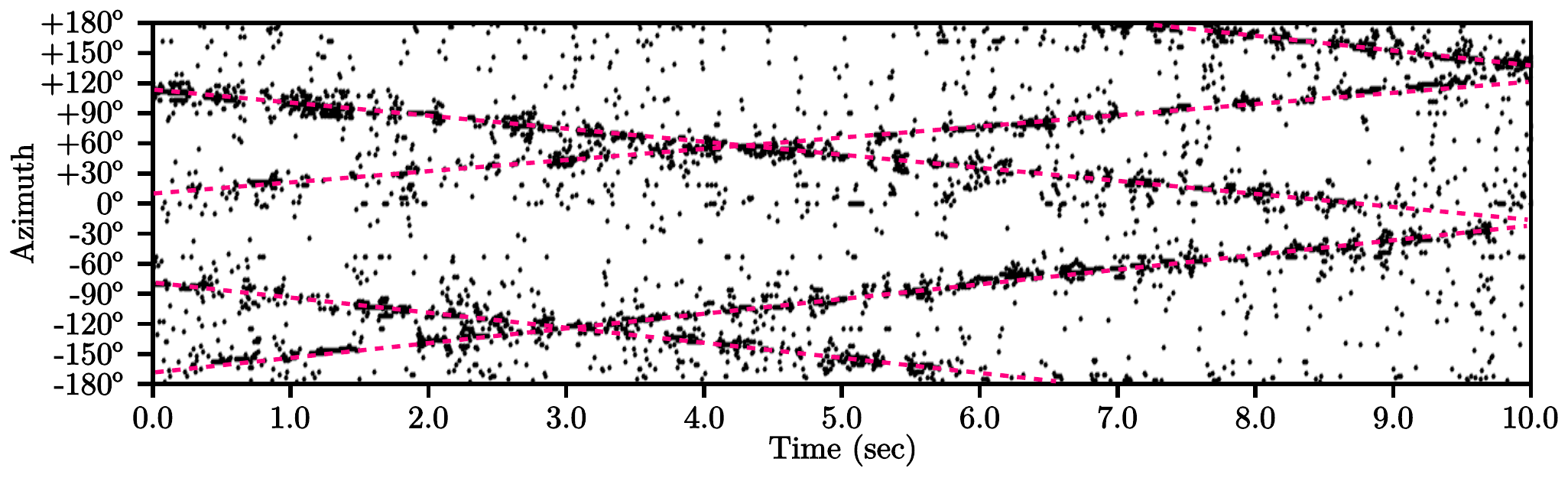}}
\subfloat[Potential sources (CMA)]{\label{fig:results_sst_moving_crossing_cma_pots}\includegraphics[width=0.5\textwidth]{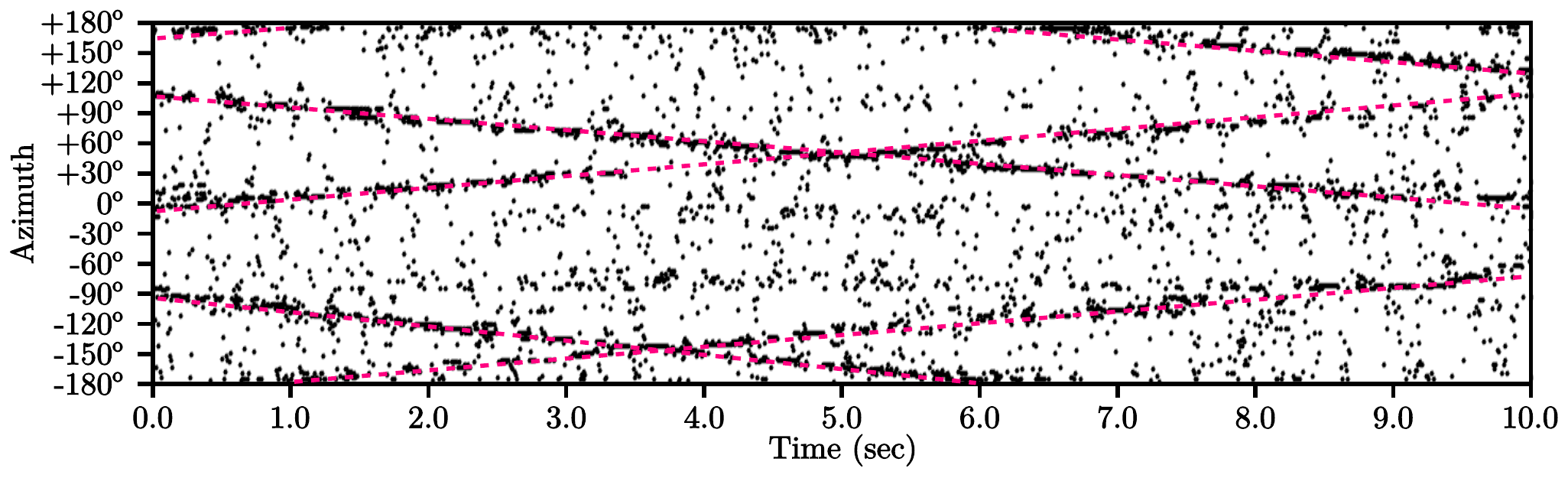}}\\
\subfloat[Tracked sources with SMC (OMA)]{\label{fig:results_sst_moving_crossing_oma_tracked_particle}\includegraphics[width=0.5\textwidth]{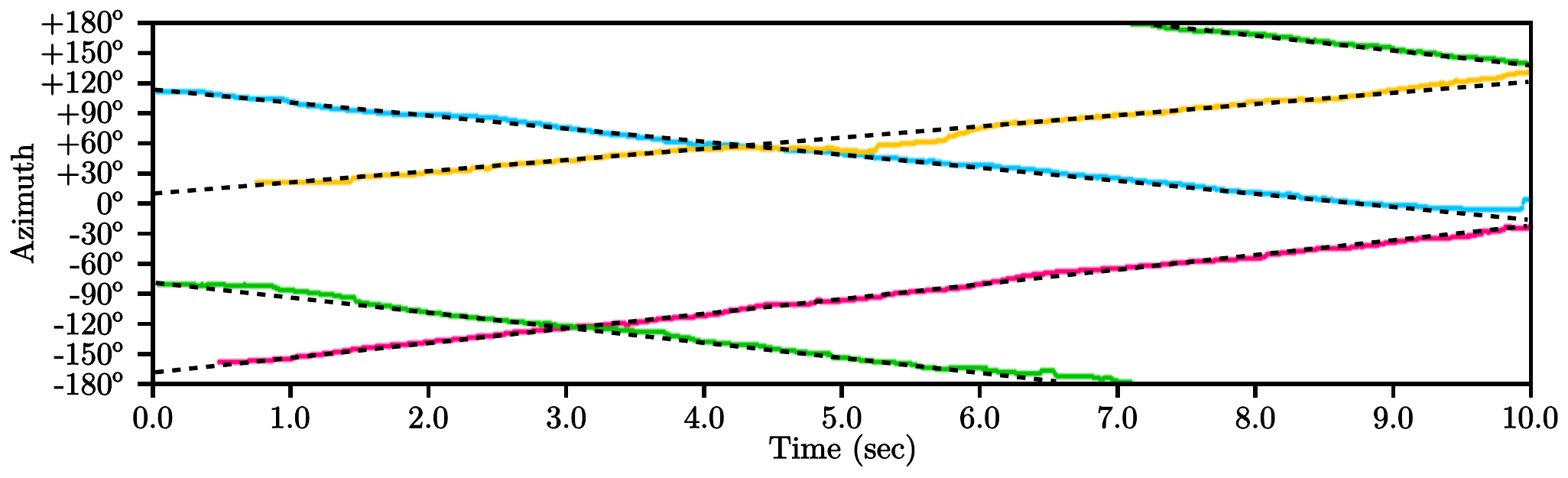}}
\subfloat[Tracked sources with SMC (CMA)]{\label{fig:results_sst_moving_crossing_cma_tracked_particle}\includegraphics[width=0.5\textwidth]{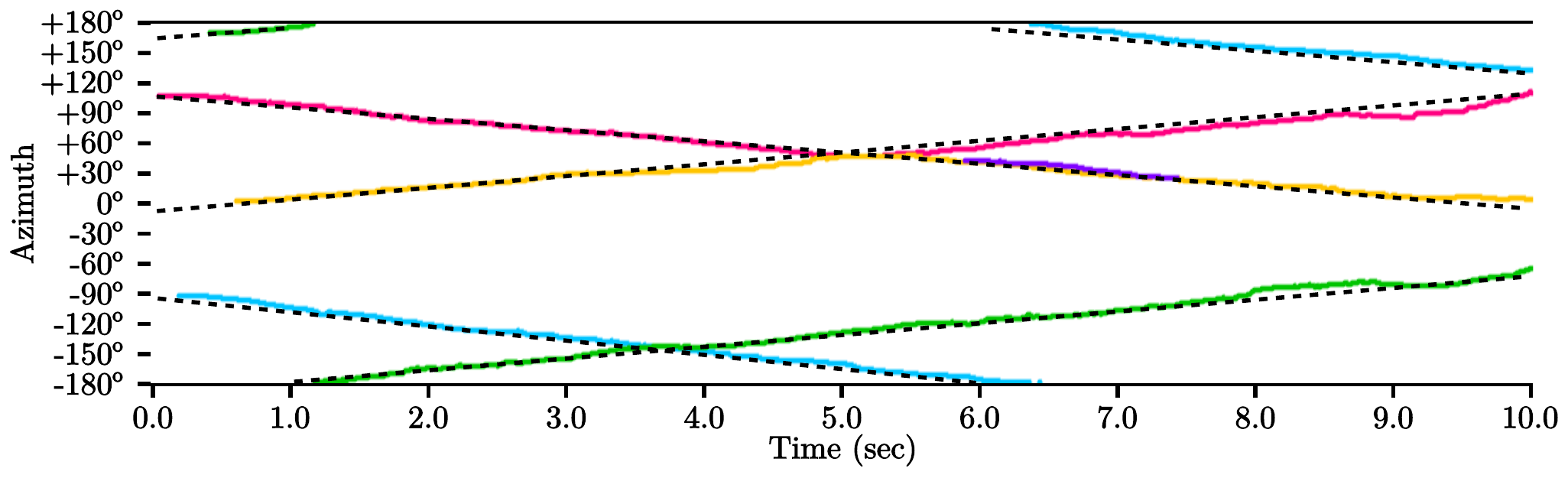}}\\
\subfloat[Tracked sources with M3K (OMA)]{\label{fig:results_sst_moving_crossing_oma_tracked_kalman}\includegraphics[width=0.5\textwidth]{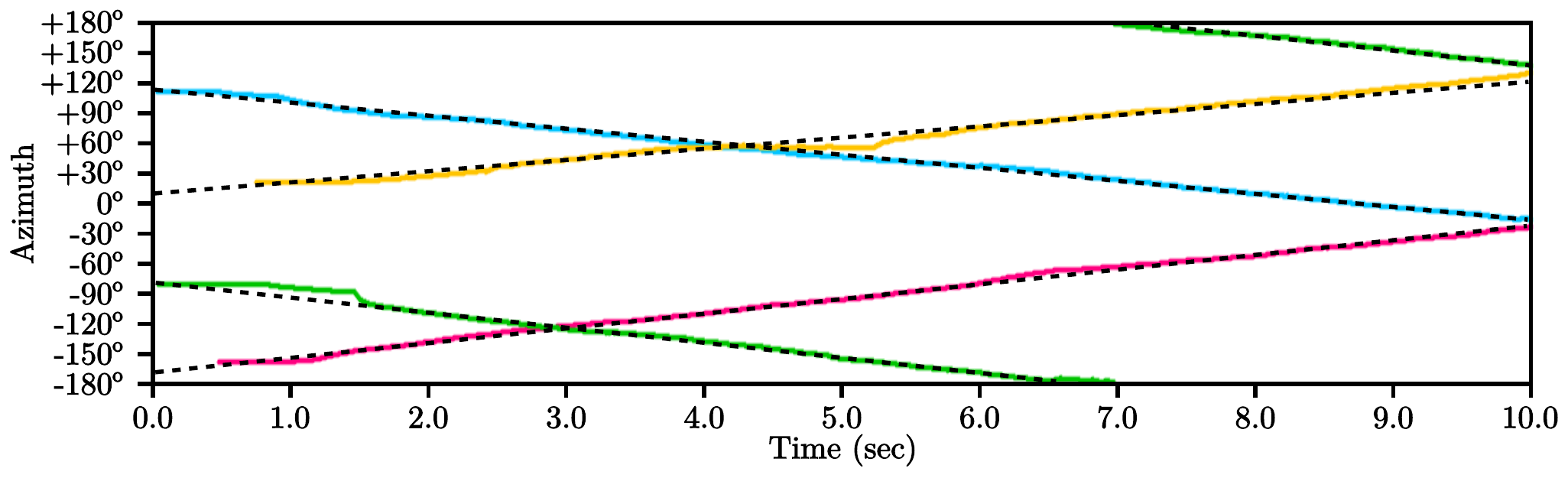}}
\subfloat[Tracked sources with M3K (CMA)]{\label{fig:results_sst_moving_crossing_cma_tracked_kalman}\includegraphics[width=0.5\textwidth]{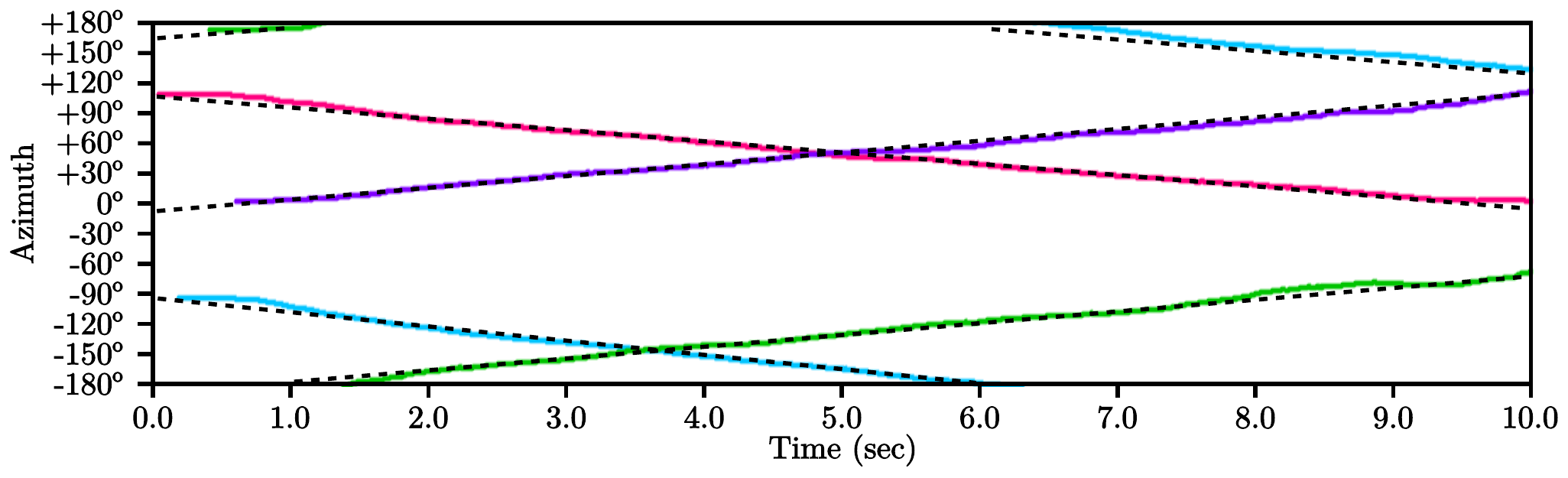}}
\caption{Azimuths of four crossing sources} 
\label{fig:results_sst_moving_crossing}
\end{figure*}

Figure \ref{fig:results_sst_moving_crossing} presents tracking results using both methods, with the OMA and CMA configurations.
Results demonstrate that M3K performs as well as SMC with OMA, and M3K performs better than SMC with CMA.
In fact, sources crossing at 5 sec are permuted with SMC, while they keep their respective trajectories with M3K.
This is caused by the model dynamics that provides more inertia with M3K than with SMC.
There is also a false detection at 6 sec, with the same azimuth and a different elevation, which is due to reverberation from the floor.

\begin{figure*}[!ht]
\centering
\subfloat[Potential sources (OMA)]{\label{fig:results_sst_moving_parallel_oma_pots}\includegraphics[width=0.5\textwidth]{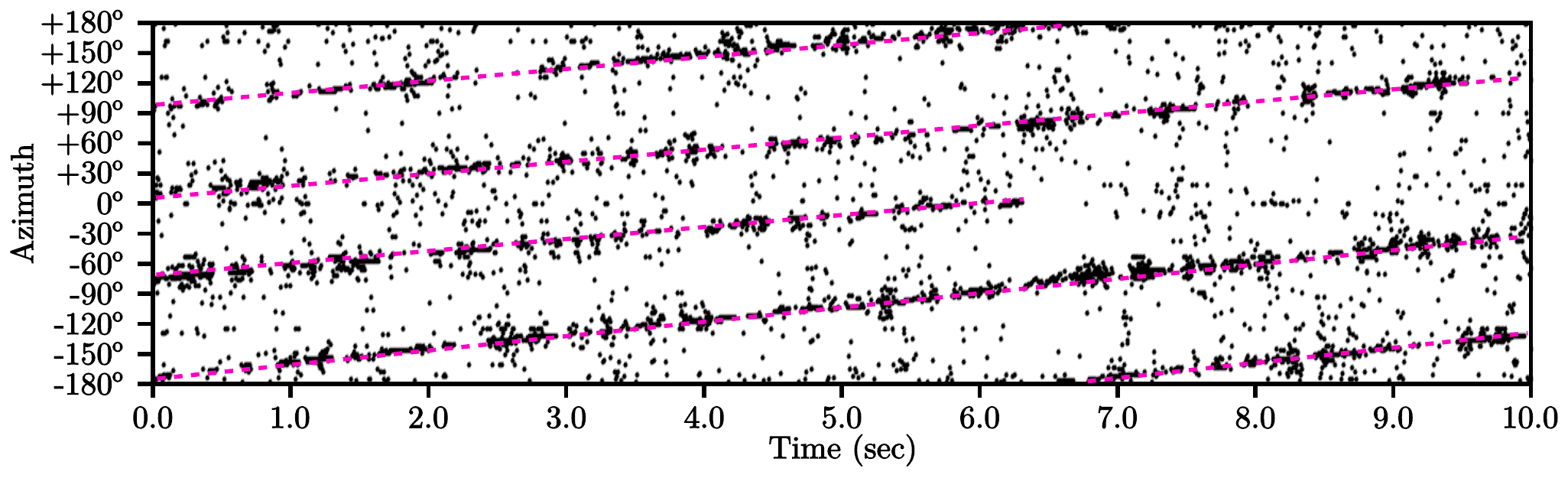}}
\subfloat[Potential sources (CMA)]{\label{fig:results_sst_moving_parallel_cma_pots}\includegraphics[width=0.5\textwidth]{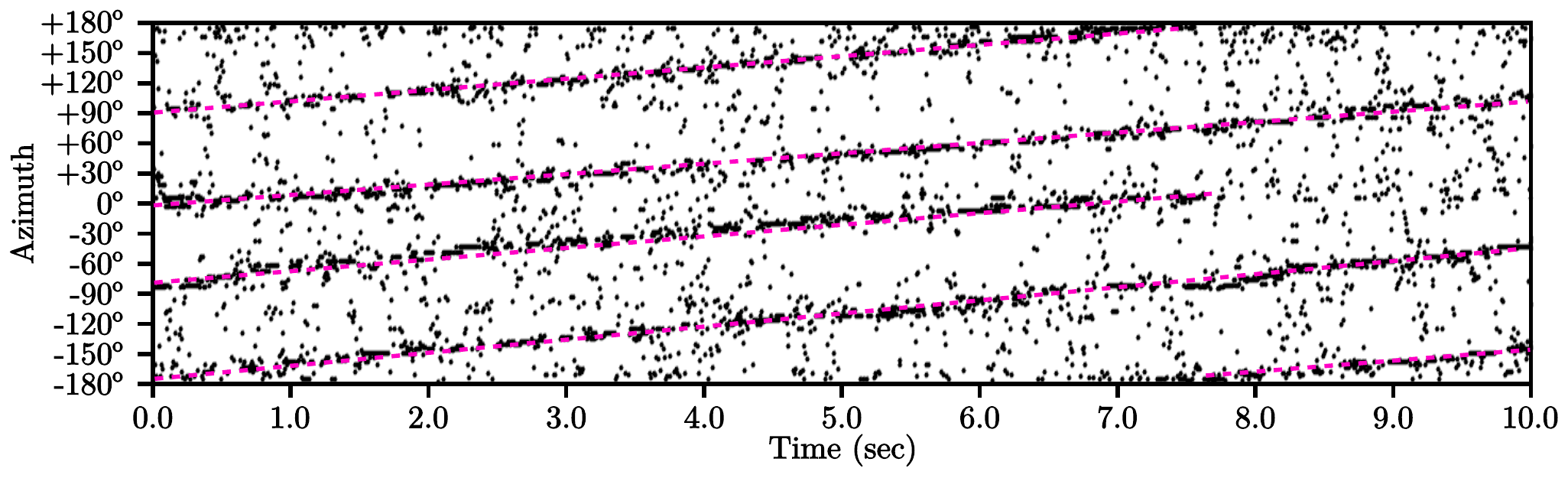}}\\
\subfloat[Tracked sources with SMC (OMA)]{\label{fig:results_sst_moving_parallel_oma_tracked_particle}\includegraphics[width=0.5\textwidth]{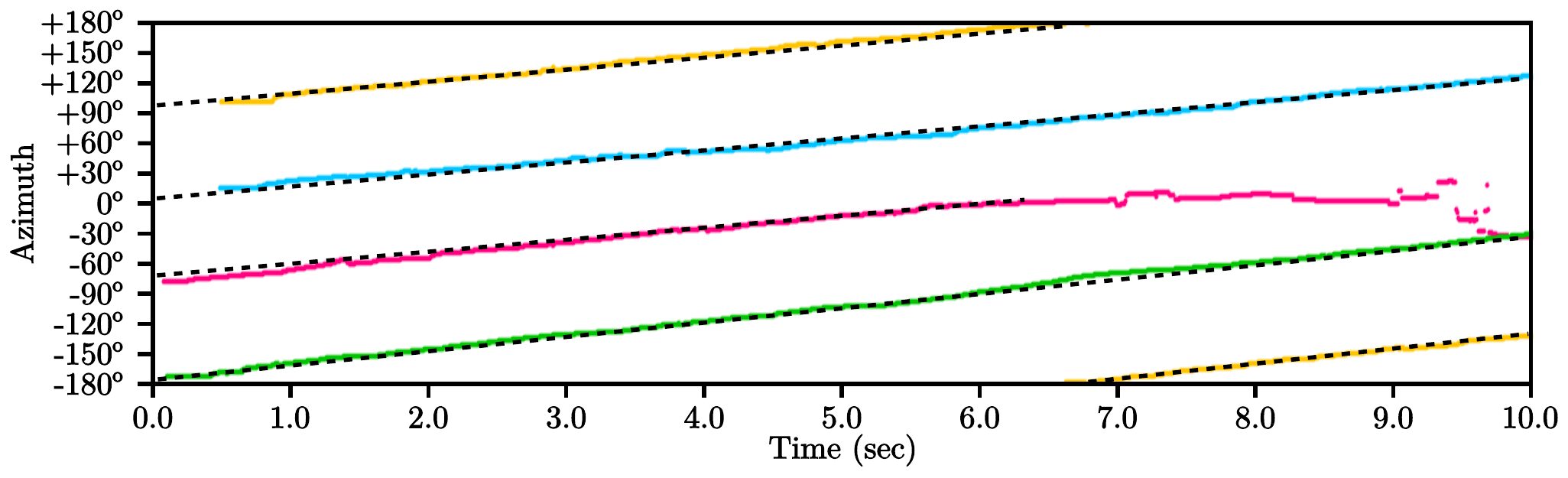}}
\subfloat[Tracked sources with SMC (CMA)]{\label{fig:results_sst_moving_parallel_cma_tracked_particle}\includegraphics[width=0.5\textwidth]{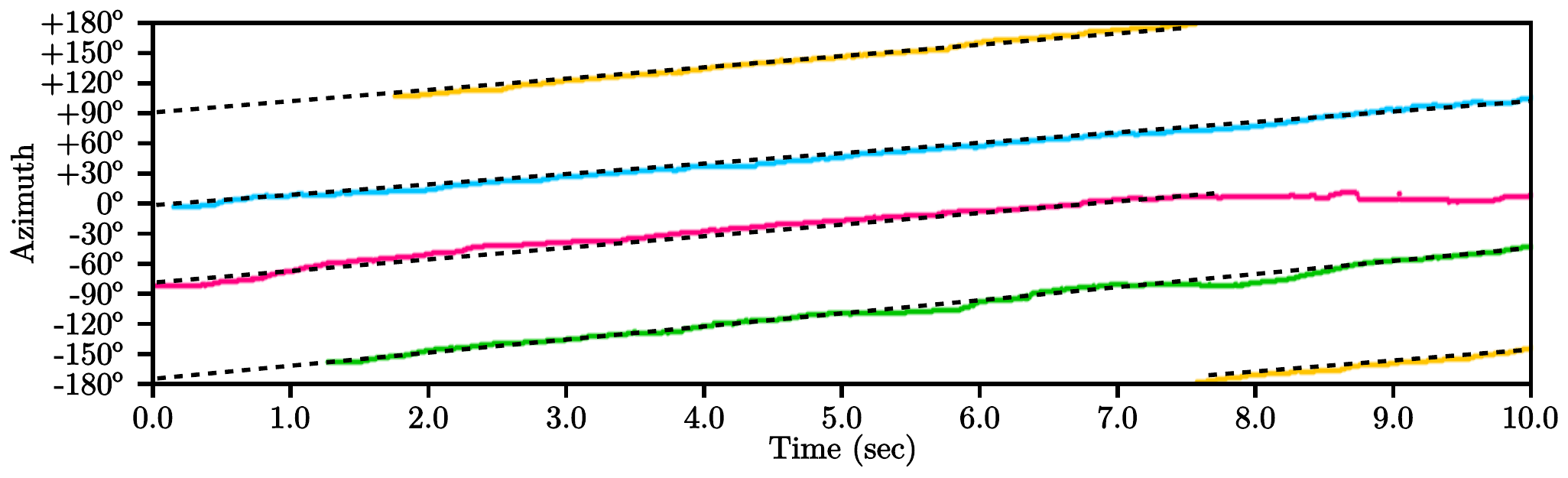}}\\
\subfloat[Tracked sources with M3K (OMA)]{\label{fig:results_sst_moving_parallel_oma_tracked_kalman}\includegraphics[width=0.5\textwidth]{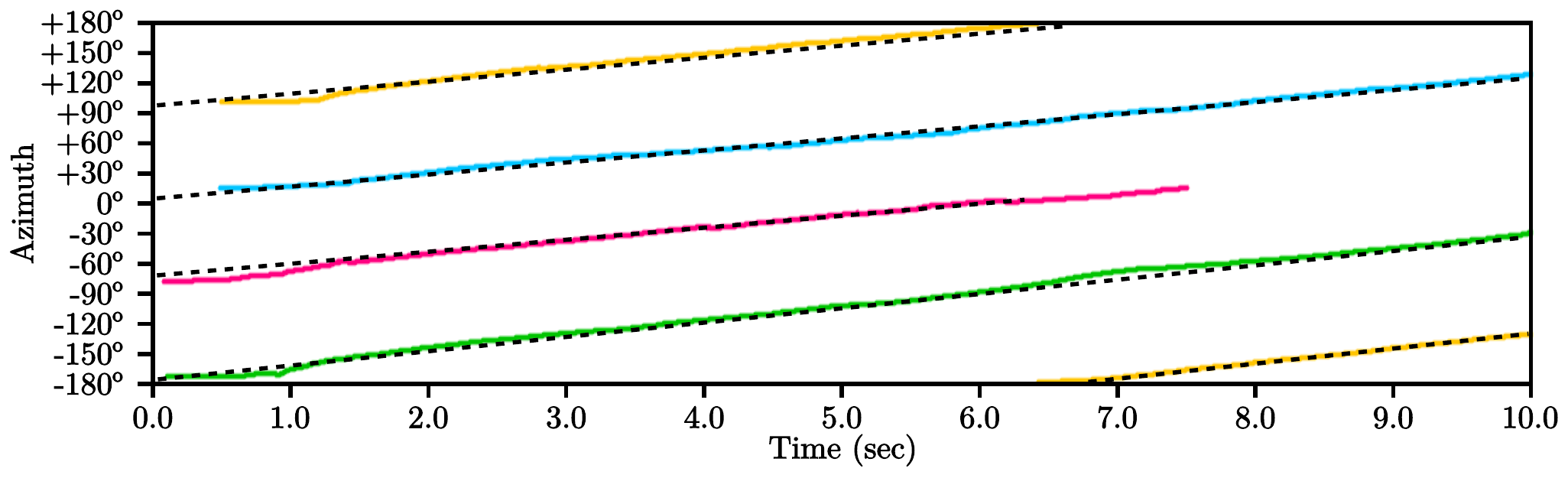}}
\subfloat[Tracked sources with M3K (CMA)]{\label{fig:results_sst_moving_parallel_cma_tracked_kalman}\includegraphics[width=0.5\textwidth]{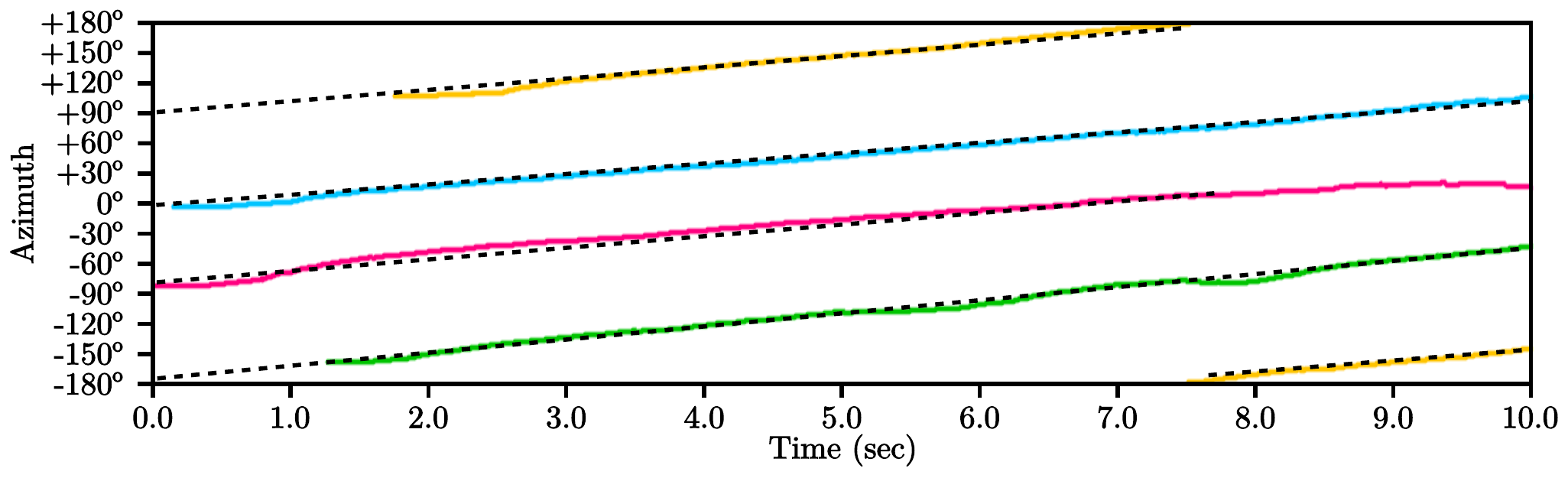}}
\caption{Four speech sources following each other}
\label{fig:results_sst_moving_parallel}
\end{figure*}

Figure \ref{fig:results_sst_moving_parallel} shows the potential sources and tracking results with the OMA and CMA configurations.
With OMA, both tracking 
%% FM: Changea systems pour methods %% FG: OK
methods perform well, except when one source becomes inactive around 6 sec: M3K removes the source in Fig. \ref{fig:results_sst_moving_parallel_oma_tracked_kalman} as required, but SMC keeps tracking it and eventually diverges in the wrong direction, as the particle filter with the parameters proposed in \cite{valin2007robust} is more sensitive to noisy observations.
Similarly, with CMA, M3K and SMC track the sources accurately, except that both keep tracking the source that becomes inactive around 7 sec.
To remove inactive source quickly, the parameter $\mu_{\mathcal{A}}$ defined in Table \ref{tab:setup_sst_parameters} could be increased, at the cost of detecting new sources with more latency. 

\section{Conclusion}
\label{sec:conclusion}

%% FM: Vérifier ma nouvelle phrase %% FG: Oui c'est bon
This paper introduces novel SSL and SST methods designed to improve robustness to noise by allowing to increase the number of microphones used while reducing computational load.
For SSL, SRP-PHAT-HSDA scans the 3D space more efficiently using two grids of coarse and fine resolution.
A microphone directivity model also reduces the amount of computations and reduces false detections with Closed Microphone Array (CMA) configurations.
The TDOA uncertainty model optimizes the MSW sizes according to the array geometry and the uncertainties in the speed of sound and the positions of the microphones.
%% FM: Cette affirmation me semble déjà confirmé pour SST, plutôt spéculative pour SSS et traité dans la dernière phrase de la conclusion 
%Using SRP-PHAT-HSDA should provide clear benefit when combined with sound source tracking and separation methods to provide a complete approach for distant speech recognition and processing. %% FG: Ok on peut la retirer
M3K provides efficient tracking of sound sources in various conditions (simultaneous static sources, simultaneous moving sources crossing and simultaneous moving sources following each other),
with accuracy comparable or better compared to SMC, and reduces by up to 30 times the amount of computations.
This efficiency makes the method more appropriate for implementing SST on low-cost embedded hardware.

In future work, the SRP-PHAT-HSDA method could include a model that optimizes the Maximum Sliding Window size for each individual point to scan and pair of microphones (instead of only each pair of microphones as it is currently the case).
M3K relies on a single dynamic model with a constant velocity for the sound sources.
Particle filtering provides multiple dynamic models (accelerating sources, sources with constant velocity, and stationary sources \cite{valin2007robust}), which may improve tracking performance.
As future work, it would be interesting to replace the single Kalman filter with multiple Kalman filters that obey different dynamic models.
The next step is to include sound source separation to implement a complete pre-filtering system for distant speech recognition.

\section*{Acknowledgment}

This work was supported in part by the Natural Sciences and Engineering Research Council of Canada (NSERC) and the Fonds de recherche du Qu\'{e}bec -- Nature et technologies (FRQNT).
The authors would like to thank Dominic L\'{e}tourneau, C\'{e}dric Godin and Vincent-Philippe Rh\'{e}aume for their help in the experimental setup.

\bibliographystyle{model1-num-names}
\bibliography{references}

\end{document}